\documentclass[10pt,twoside,a4paper]{article}

\usepackage[numbers]{natbib}
\usepackage{graphicx} 
\usepackage[a4paper, total={6.5in, 9.3in}]{geometry}
\usepackage{color}

\usepackage{amsfonts,amssymb}
\usepackage{amsmath,mathtools}
\usepackage{comment}
\usepackage{booktabs}
\usepackage{caption}
\usepackage{subcaption}

\newcommand{\beq}{\begin{equation}}
\newcommand{\eeq}{\end{equation}}

\newcommand {\e}  {\varepsilon}
\newcommand{\pd}{\partial}
\newcommand{\dt}{\pd_{t}}

\newcommand{\dy}{\pd_{y}}

\newcommand{\dyy}{\pd^{2}_{yy}}

\newcommand{\na}{n_{1}}
\newcommand{\nb}{n_{2}}
\newcommand{\nae}{n_{1\e}}
\newcommand{\nbe}{n_{2\e}}
\newcommand{\Ia}{I_{1}}
\newcommand{\Ib}{I_{2}}
\newcommand{\Ii}{I_{i}}
\newcommand{\Iae}{I_{1\e}}
\newcommand{\Ibe}{I_{2\e}}
\newcommand{\Iie}{I_{i\e}}
\newcommand{\nie}{n_{i\e}}
\newcommand{\nuab}{\nu_{1,2}}
\newcommand{\nuba}{\nu_{2,1}}

\newcommand{\uae}{u_{1\e}}
\newcommand{\ube}{u_{2\e}}
\newcommand{\uie}{u_{i\e}}

\DeclareMathOperator*{\argmax}{arg\,max}
\DeclareMathOperator*{\argmin}{arg\,min}
\DeclarePairedDelimiter{\Diagfences}{\Big(}{\Big)}
\DeclarePairedDelimiter{\diagfences}{\big(}{\big)}
\newcommand{\diag}{\operatorname{diag}\Diagfences}
\newcommand{\supp}{\operatorname{supp}\diagfences}

\usepackage{footmisc}

\newcounter{row}

\usepackage[shortlabels]{enumitem}
\usepackage{hyperref}

\usepackage{blindtext}
\title{The development of drug resistance in metastatic tumours under chemotherapy: an evolutionary perspective}
\author{Federica Padovano \footnote{Sorbonne Universit\'e, CNRS, Universit\'e de Paris, Laboratoire Jacques-Louis Lions UMR 7598, 75005 Paris, France. (federica.padovano@sorbonne-universite.fr)}
\and Chiara Villa\footnote{Sorbonne Universit\'e, CNRS, Universit\'e de Paris, Inria, Laboratoire Jacques-Louis Lions UMR 7598, 75005 Paris, France.  (chiara.villa.1@sorbonne-universite.fr)}}

\date{}
\begin{document}

\maketitle


\begin{abstract}
    We present a mathematical model of the evolutionary dynamics of a metastatic tumour under chemotherapy, comprising non-local partial differential equations for the phenotype-structured cell populations in the primary tumour and its metastasis.  These equations are coupled with a 
 physiologically-based pharmacokinetic model of drug delivery,  implementing a realistic delivery schedule. 
 The model is carefully calibrated from the literature, focusing on BRAF-mutated melanoma treated with Dabrafenib as a case study.  
    By means of long-time asymptotic analysis, global sensitivity analysis and numerical simulations, we explore the impact of cell migration from the primary to the metastatic site, physiological aspects of the tumour sites and drug dose on the development of drug resistance and treatment efficacy. Our findings provide a possible explanation for empirical evidence indicating that chemotherapy may foster metastatic spread and that metastatic sites may be less impacted by chemotherapy.
\end{abstract}

\section{Introduction}\label{sec:intro}

\subsection{Biological context and motivation}
In $2022$ around $9.7$ million of people died for cancer worldwide~\cite{gco2022}, accounting for approximately $15\%$ of total deaths, making it one of the primary health problems globally. 
In particular, metastases are responsible for approximately $90\%$ of cancer-related mortality~\cite{gupta2006cancermetastasis,lambert2017emerging}. 
These form following a multi-step process comprising the migration of cancer cells from the original site to regional or distant organs and lymph nodes, \textit{e.g.} by accessing the lymphatic or blood vessels. 
In this latter case the metastatic steps include local invasion of regions surrounding the primary tumour,  the cancer-induced formation of new blood vessels to access more nutrients and foster cancer growth, process known as \textit{angiogenesis}, intravasation, survival in the circulatory system and extravasation through vascular walls of distant sites~\cite{lambert2017emerging}. 
The process of primary tumour cells colonization of other tissues is known as \textit{metastatic spread}, or \textit{primary seeding}, meanwhile when cells return to the original site or an existing metastasis they contribute to \textit{self-seeding}~\cite{nguyen2007genetic,pantel2016thebiology}. Moreover, if the population in the metastatic site grows and acquires itself the ability to metastasise, it can spread to other sites, phenomenon known as \textit{secondary seeding}~\cite{pantel2016thebiology}.

Chemotherapy, \textit{i.e.} the use of cytotoxic drugs to kill cancer cells,  is to this day considered the most effective, and thus most widely used, modality of cancer treatment. 
The development of drug resistance confers a selective advantage upon the cancer cell population, as cancer cells exhibit reduced sensitivity to cytotoxic compounds, and represents a significant challenge in cancer therapies as it often contributes to disease relapse \citep{dagogo2018tumorheterogeneity,gillies2012evolutionary,gottesman2022mechanismsresistance,turner2012henetic,wang2019cancerresistance}.  
Despite the efforts put into the development of treatment strategies evading or reverting the development of drug resistance~\citep{bukowski2020mechanisms}, this process becomes significantly more complex in metastatic cancers,  given the exposure of tumours in distinct organs to disparate concentrations of therapeutic agents and environmental factors~\citep{wu2013druggradient}.  
In fact, it has been reported that even if standard chemotherapy effectively controls disease progression at the primary tumour site, it often fails to influence metastatic populations~\cite{dalterio2020paradoxical,parker2022currentchallenges}.  
Moreover, many studies suggest that there is a strong correlation between treatment resistance and metastatic ability, \textit{i.e.} cancer cells with lower sensitivity to cytotoxic agents usually also manifest enhanced disseminating properties~\cite{dalterio2020paradoxical,karagiannis2018chemotherapy,lambert2017emerging,luebker2019brafinhibitor,shibue2017drugresistance}. 
Therefore the development of drug resistance in the primary tumour during chemotherapy, even in reversible cases,  may favour the persistence of its metastases.  
Nevertheless, it is still unclear whether metastases are intrinsically more resistant than the primary tumour or if their reduced sensitivity arises because they originate from particularly aggressive cell subpopulations or due to subsequent evolution after dissemination~\cite{lambert2017emerging}.  
It may therefore be beneficial to test existing hypotheses on the development of drug resistance in metastatic tumours under chemotherapy using proof-of-concept theoretical frameworks. 

\subsection{Mathematical modelling background}
Mathematical modelling is a valuable tool for gaining insights into the mechanisms driving cancer evolution and the emergence of intratumour and intertumour heterogeneity, for simulating complex and long-term dynamics that may take years to observe in real-life settings, and for designing optimised therapeutic strategies.  
All three objectives share a common aim of paving the way towards precision medicine, \textit{i.e. }the innovative idea of targeting therapies on patients in order to increase their efficacy.

Many mathematical models have been previously employed to describe the different steps of the metastatic cascade, see the review article~\cite{scott2013mathematical} and references therein. 
These may focus, for instance, on the development of metastatic phenotypes~\cite{michor2006dynamics, michor2006stochastic}, their intravasation and extravasation~\cite{brodland2012mechanics,franssen2019modelmetastases,ramis2009multi}, or extensive cell dissemination and potential evolution of cells from the primary to the metastatic site~\cite{bulai2023modeling,diego2013modeling,hartung2014mathematical,newton2013spreaders, newton2015spatiotemporal,scott2013self-seeding}. 
Moreover, some of them also investigate the effect of therapeutic treatments, such as chemotherapy, radiotherapy and immunotherapy, alone or combined, on metastatic cancers~\cite{ghaffari2016mixed,ocana2024mathematical,schlicke2021mathematical,sun2016mathematical}.
Among these, Sun \textit{et al.}~\cite{sun2016mathematical} accounted for the existence of a sensitive and a drug-resistant subpopulation in the primary site. Nevertheless, drug-resistance levels may be better captured on a continuum~\cite{lavi2014simplifying}, and phenotypic heterogeneity in the metastatic site may also play an important role in the failure of treatment.


Many studies in literature investigate the adaptive processes that underlie the development of intratumour phenotypic heterogeneity and the emergence of resistance to chemotherapeutic agents, see for instance~\citep{clairambault2019survey,yin2019areview} and references therein. 
Among these, many works consider models comprising non-local partial differential equations (PDEs) modelling the adaptive dynamics of cancer cell populations structured by a continuous phenotypic trait linked with the cell sensitivity to cytotoxic agents~\cite{chisholm2016cellpopulation,chisholm2015emergence,cho2017modeling,cho2018modeling,cho2018modeling2,cho2020impact,clairambault2019evolutionary,delitalia2011colorectalcarcinoma,lavi2013roleofcell,lorenzi2016tracking,villa2021evolutionchemo}. 
These models enabled the study of the progressive development of increasing levels of drug resistance during treatment and the role of intratumour heterogeneity in cancer persistence after therapy. This was also possible thanks to their higher amenability to analytical investigations compared to their stochastic counterparts and the mathematical theory supporting such analysis~\cite{barles1989wavefront,diekmann2005dynamics,evans1989pde,fleming1986pde,perthame2006transport}.   
Some of these studies further applied optimal control methods in order to explore the most effective treatment strategies~\cite{almeida2019optimizationchemo,olivier2018combination,pouchol2018asymptotic}. 
While this mathematical framework had not been applied to metastatic tumours prior to this manuscript,  Mirrahimi~\cite{mirrahimi2012adaptation} considered the adaptive dynamics of phenotype-structured populations in communicating patches, further developed with applications to speciation in ecology or host-pathogen interactions~\citep{alfaro2023adaptation,boussange2022eco,hamel2021adaptation,lion2022multimorph,mirrahimi2020evolution}, which is easily comparable with the scenario of connected primary tumour and metastatic sites. 

Pharmacokinetic (PK) modelling effectively captures the time course of the drug concentration according to various administration modes and allows to quantify the relationship between the dose and the \textit{in situ} drug concentration~\cite{ahmed2015pkofdrugs,jones2013pbpk,zou2020biophase}. In these models, the body is divided into compartments, also referred to as building blocks, each of which is associated to a variable, or quantity, describing the drug concentration in the block.  The compartments are then linked to each other through the drug exchange.  
This kind of models are usually coupled with a pharmacodynamic (PD) one, which captures the effect of the drug on the considered disease, and employed to find suitable drug dosing schedules to achieve optimal drug concentrations in the target tissues~\cite{gallo2005pkmodel,moss2014optimizing,zhou2007predicting}.  
Moreover, they can also be expanded to include physiological aspects of the considered tissues, such as as organ blood flow and size,  in order to explore their influence on the drug exposure~\cite{gallo2005pkmodel,himstedt2020quantitativemechanistic}. 

\subsection{Synopsis and paper structure}
We propose a mathematical model for the evolutionary dynamics of metastatic tumours under chemotherapy,  where physiological differences of the different tumour sites are integrated within a PK model of drug delivery.  
In particular, the model comprises a system of coupled non-local PDEs for the phenotypic distribution of the cancer cells in the primary tumour and the metastatic site, structured by their level of resistance to the chemotherapeutic agent,  and ordinary differential equations (ODEs) for the pharmacokinetics.  
We focus on a biological framework including a primary tumour that already faced angiogenesis and a newly-formed yet growing metastasis, where cancer cells characterised by higher levels of drug resistance are assumed to be more aggressive and thus able to migrate to distant sites at higher rates.  
We restrict our attention to a highly perfused primary tumour and a metastatic site where cells are proliferating but not yet able to disseminate, as they chose in~\cite{sun2016mathematical}, exploring different levels of tumour vascularisation for the metastasis.  

We specifically consider BRAF-mutated melanoma,  which is a form of skin cancer that develops in melanocytes, \textit{i.e.} the cells responsible for melanin production,  from the uncontrolled proliferation of cells induced by a mutation of the BRAF gene.  
Despite its lower incidence, BRAF-mutated melanoma is the most aggressive and lethal among the skin cancers~\cite{cabaco2022thedarkside,sundararajan2022melanoma}, particularly due to its high metastatic rate~\cite{budczies2015landscape}. 
These tumours can spread locally, regionally and distantly, with the most common metastatic sites being skin and subcutaneous tissue, followed by lungs, liver, bones, and brain~\cite{sundararajan2022melanoma}. 
Metastatic melanomas are often treated with the chemotherapeutic agent Dabrafenib, a kinase inhibitor of mutated BRAF. Despite the rapid response, with a median time around $6$ weeks, and short-term increase in patient survival, resistance to Dabrafenib persists with a median progression-free survival of approximately $6–8$ months~\cite{bowyer2015dabrafenib}. 
For these reasons, a metastatic BRAF-mutated melanoma under Dabrafenib treatment constitutes the ideal case study to adopt for our model, which we carefully calibrate from the literature employing PK parameter values estimated from \textit{in vivo} and \textit{ex vivo} data from patients.  

The paper is organised as follows. In Section~\ref{sec:model} we introduce the model assumptions and equations. In Section~\ref{sec:analysis} we carry out a formal asymptotic analysis of evolutionary dynamics. In Section~\ref{sec:numres} we perform global sensitivity analysis and conduct further numerical investigations, to check the analytical results and explore the role of evolutionary and physiological parameters on the outcome of treatment and the timescale of development of drug resistance.  Section~\ref{sec:conclusion} concludes the paper and provides a brief overview of the model limitations and possible research perspectives.

\section{Description of the model}\label{sec:model}

We present a mathematical model of evolutionary dynamics of a metastatic tumour under chemotherapy, comprising non-local phenotype-structured PDEs for the primary tumour and its metastasis. In order to effectively capture the time course of the drug concentration in each tumour site depending on the administration mode and dose, as well as physiological aspects of each tumour site, we complement the evolutionary dynamics model with a physiologically-based PK model for the drug delivery.

\subsection{Evolutionary dynamics model of metastatic cancer}
We model the evolution of two tumour cell populations, \textit{i.e.} the primary tumour and its metastasis, exposed to a chemotherapeutic agent. 
In order to consider a metastasised tumour,  we assume that the primary tumour is vascularised and the sites are sufficiently well connected so that the cancer cells in the primary tumour can intravasate and, subsequently, extravasate in the secondary site. The opposite process, known as secondary self-seeding~\citep{scott2013self-seeding} is also allowed. 
In particular, the metastatisation process is modeled by allowing tumour cells to transition from one site to another~\citep{mirrahimi2015asymptotic,scott2013self-seeding}. 
Building upon the ideas presented in~\citep{almeida2019optimizationchemo,mirrahimi2012adaptation,villa2021evolutionchemo}, we introduce the variable $y\in [0,1]$, which represents the cell phenotypic state linked to its level of chemoresistance. We assume that the phenotypic variant $y=1$ endows cells with the highest level of cytotoxic-drug resistance and the greatest migratory abilities. This is motivated by evidence suggesting phenotypes with higher drug resistance are more aggressive, and they are often associated with a higher invasive potential~\citep{shibue2017drugresistance}. On the contrary, the state $y=0$ corresponds to cells with the lowest level of cytotoxic-drug resistance and the lowest migratory abilities. \\
From now on we will make use of the index $i\in\{1,2\}$, to represent each tumour site. In particular, $i=1$ corresponds to the primary tumour, while $i=2$ is the metastasis. The phenotypic distribution of tumour cells at time $t \in [0,\infty)$ and site $i$ is described by the function $n_i(t,y)$. Moreover, at each time $t$, we define the density of tumour cells in site $i$ as
\begin{equation} \label{distribu_phenoI}
    I_i(t): = \int_0^1 n_i(t,y) dy \,,
\end{equation}
with the corresponding local mean phenotypic state and related variance defined, respectively, as
\begin{equation}
    \mu_i(t): = \frac{1}{I_i(t)}\int_0^1 y n_i(t,y) dy \quad \mbox{and} \quad \sigma^2_i(t): = \frac{1}{I_i(t)}\int_0^1 y^2 n_i(t,y) dy -\mu_i^2(t) \,.
    \label{distribu_pheno}
\end{equation}
The phenotypic distribution of tumour cells in each site, $n_i(t,y)$, is governed by the following non-local PDE with given initial and Neumann boundary conditions:
\begin{equation}
    \begin{cases}
        \partial_t n_i - \beta_i \partial_{yy}^2 n_i =  R(y,I_i, C_{i})n_i + \nu_{j,i}(y) n_j - \nu_{i,j}(y) n_i, \quad i \neq j, \quad \text{in} \quad (0,\infty) \times [0,1] \\[5pt]
        I_i(t): = {\displaystyle \int_0^1} n_i(t,y) dy ,\\[5pt]
        n_i(0,y)=n_{i,0}(y)  ,\\[5pt]
        \partial_y n_i (t,0) = \partial_y n_i (t,1) = 0\,,
    \end{cases}
    \qquad i=1,2\,.
    \label{pdedensity_final}
\end{equation}
The diffusion term in~$\eqref{pdedensity_final}_1$ models the effects of spontaneous epimutations, which occur at rate $\beta_i \in \mathbb{R}_{>0} $ \citep{chisholm2015emergence,lorenzi2016tracking}. The non-local reaction term takes into account the effects of cell proliferation, natural death, death due to competition for resources, and the cytotoxic action of the drug. The functional $R_i(y,I_i,C_{i})\equiv R_i(y,I_i(t), C_{i}(t))$, models the fitness of tumour cells in site $i$ in the phenotypic state $y$ and under the local environmental conditions at time $t$, characterised by the cell density $I_i \equiv I_i(t)$, and the chemotherapeutic agent concentration $C_{i}\equiv C_{i}(t)$. Building upon the modelling strategies presented in \citep{almeida2019optimizationchemo,villa2021evolutionchemo}, the fitness function $R_i$ is defined as follow
\begin{equation}
    R_i(y,I_i,C_{i}) := p_i(y)-d_i I_i - k_i(y,C_{i}) .
    \label{fitnessfunction}
\end{equation}
In definition~\eqref{fitnessfunction} the term $d_i I_i$ translates into mathematical terms the idea that a higher total cell number corresponds to less available resources and space in the system, hence a higher rate of death due to intrapopulation competition. The parameter $d_i\in\mathbb{R}_{>0}$ is related to the local carrying capacity of the tumour. The function $p_i(y)$, also referred to as the background fitness in the absence of treatment, stands for the net proliferation rate of cancer cells in the phenotypic state $y$ and site $i$, and based on the ideas proposed in \citep{villa2021evolutionchemo} we define it as
\begin{equation}
    p_i(y):=\delta_i (1-y^2) + \varphi_i (1-(1-y)^2) \,,
    \label{proliferation}
\end{equation}
where $\delta_i \in \mathbb{R}_{>0}$ represents the maximum background fitness for highly proliferating and drug sensitive cells in site $i$, while $\varphi_i \in \mathbb{R}_{>0}$ represents the maximum background fitness for weakly proliferating and fully chemoresistant cells in tumour $i$, and we further assume $\delta_i \gg \varphi_i$. Under definition~\eqref{proliferation}, $p_i(y)$ reaches its minimum at $y=1$, \textit{i.e.} the phenotypic trait characterised by the highest level of cytotoxic-drug resistance and the slowest proliferation in the absence of a drug, since we assume that slowly proliferating cells are less susceptible to chemotherapy and thus more likely to develop resistance~\cite{chu2004chemotherapy,corrie2011chemotherapy}. Meanwhile $k_i(y,C_{i})$ is the rate of death induced by the cytotoxic drug, and based on the ideas proposed in \citep{villa2021evolutionchemo}, we define
\begin{equation}
    k_i(y,C_{i}):= \frac{\eta_i \cdot C_{i}}{\alpha_{i}+C_{i}}(1-y)^2 \,,
    \label{cytodeath}
\end{equation}
where $\eta_i\in \mathbb{R}_{>0}$ is the maximal death rate of highly drug sensitive phenotypic variants due to the cytotoxic action of the chemotherapeutic agent, and $\alpha_{i}\in \mathbb{R}_{>0}$ is the Michaelis-Menten constant of the chemotherapeutic agent.  
Under definition~\eqref{cytodeath},  $k_i(y,C_{i})$ is a decreasing function of $y$, \textit{i.e.} the rate of drug-induced death decreases as the level of chemoresistance of the cells increases, and it is null for $y=1$, consistently with the assumption that such a phenotype is completely resistant to the chemotherapeutic agent. With these definitions, after a little algebra, the fitness function in~\eqref{fitnessfunction} can be rewritten as 
\begin{equation}
\label{genericfitness}
    R_i(y,I_i,C_i):=a_i(C_i)-b_i(C_i)\big(y-h_i(C_i)\big)^2-d_iI_i \,,
\end{equation}
where 
\begin{equation}
    a_i(C_i) = \delta_i-\frac{\eta_i \cdot C_{i}}{\alpha_{i}+C_{i}}+ \frac{\big(\varphi_i+\frac{\eta_i \cdot C_{i}}{\alpha_{i}+C_{i}}\big)^2}{\delta_i + \varphi_i +\frac{\eta_i \cdot C_{i}}{\alpha_{i}+C_{i}}} \,, \quad  b_i(C_i) =\delta_i + \varphi_i+ \frac{\eta_i \cdot C_{i}}{\alpha_{i}+C_{i}}  \quad \text{and}  \quad h_i(C_i) = \frac{\varphi_i+\frac{\eta_i \cdot C_{i}}{\alpha_{i}+C_{i}}}{\delta_i + \varphi_i+ \frac{\eta_i \cdot C_{i}}{\alpha_{i}+C_{i}}} \,.
\label{eq:fitness_factors}
\end{equation}
Here $a_i\equiv a_i(C_i)$ is the maximum fitness, $b_i\equiv b_i(C_i)$ is the non-linear selection gradient, and $h_i\equiv h_i(C_i)$ is the phenotypic trait associated to the maximum fitness corresponding to the chemotherapeutic agent concentration $C_i(t)$. We observe that a higher drug concentration $C_i$ results in a lower maximum fitness $a_i$ because of the greater cytotoxic activity due to the compound, a greater fittest phenotypic trait $h_i$ and a stronger selective pressure on tumour cells, \textit{i.e.} a greater $b_i$.\\

\noindent 
The terms $\nu_{i,j}(y)$ and $\nu_{j,i}(y)$ in equation~\eqref{pdedensity_final} are non-negative functions representing the rate of transition of cells in the phenotypic state $y$ from site $i$ to site $j \neq i$, and vice versa, which depend on the ability of the phenotype to intravasate, survive in the circulation and extravasate, given the connectivity of sites $i$ and $j$. 
Motivated by \citep{scott2013self-seeding}, we assume that all cells with phenotypic state $y>0$ can access the bloodstream and reach other sites, and we assume that when cancer cells migrate from one site to the other they maintain their original phenotypic trait~\cite{lambert2017emerging}. Specifically, we define the migration rates as
\begin{equation}
\label{migrationratefunction}
    \nu_{j,i}(y)= \hat{\nu}_{j,i}y^2 ,\quad i\neq j,
\end{equation}
where $\hat{\nu}_{j,i}\in \mathbb{R}_{>0}$ is the maximum transition rate from site $j$ to site $i$, with $i \neq j$.  
Under definition~\eqref{migrationratefunction}, migration is almost absent as $y\rightarrow 0$, while the migration rate approaches its maximum velocity $\hat{\nu}_{j,i}$ as $y\rightarrow1$, \textit{i.e.} cells with higher levels of cytotoxic-drug resistance migrate at more significant rates, consistently with our assumptions~\citep{shibue2017drugresistance}. 

\subsection{PK model for anticancer drug}
\begin{figure}[h]
    \centering
    \includegraphics[scale=0.4]{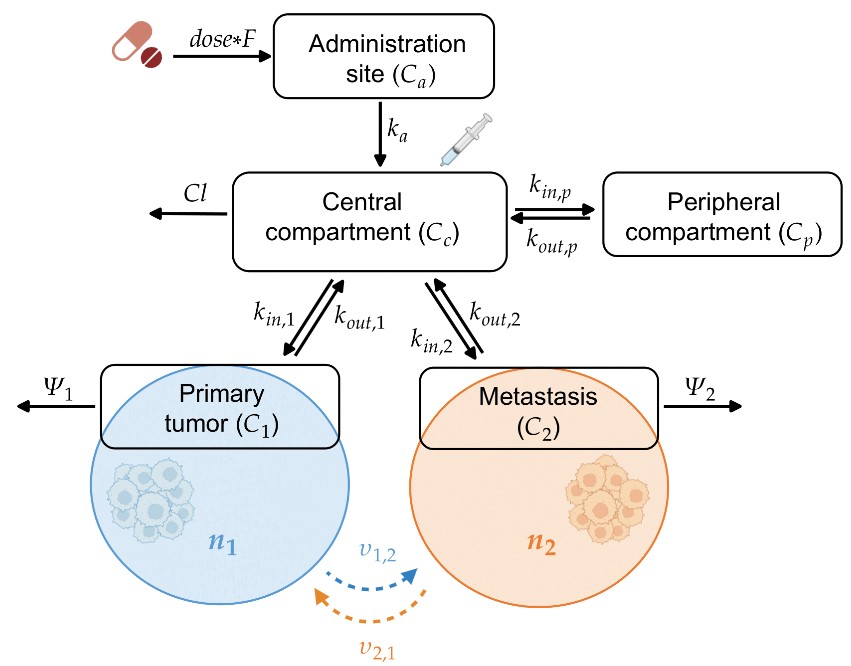}
    \caption{\textbf{Schematic of the model.} The five boxes represent the building blocks of the PK model, while the two circles depict the local environment in primary tumour (blue) and metastatic (orange) sites. 
As illustrated by the overlapping of the boxes and circles, the  local environmental conditions in each tumour site will be affected by the drug concentration in the respective PK compartment.  The red pills and the grey syringe represent oral administration and intravenous injection, respectively. 
The straight arrows represent the drug exchange between compartments, while the dashed ones indicate migration of cancer cells. 
The variables used in the model for the drug concentration in each compartment ($C_i$) and the phenotypic distribution in each tumour site ($n_i$) are indicated in the respective box/circle. The parameters and/or functions used in the model to represent the rate of drug or cell flow are indicated next to the respective arrow. 
}
    \label{fig:compartmentstructure}
\end{figure}

We employ a PK model in order to better predict the drug concentration in two physiologically different tumour sites, given distinct administration methods and doses. To achieve this, we take into account a five-compartments structure, as shown in Figure~\ref{fig:compartmentstructure}, comprising the administration site, the central and peripheral compartments, and the primary tumour and metastasis blocks. The administration site allows us to take into account the scenarios of extravascular drug injection, \textit{i.e.} when the compound does not enter directly into the circulating system, \textit{e.g.} \textit{per os}. The central compartment represents plasma and well-perfused tissues, \textit{i.e.} tissues that receive a rich blood supply relative to their organ weight~\citep{cuenod2013perfusivity}, such as kidneys, liver, heart, and brain. Meanwhile, the peripheral compartment represents poorly-perfused tissues, such as muscle and skin. The primary tumour and metastasis blocks are included in the model structure for two main purposes: to capture the delay between the drug injection and its therapeutic effect, and to introduce in the model some of the physiological differences that can affect the compound distribution in different tumour sites, specifically the tumour size and its vascularization.

\noindent
We denote by $C_a(t)\geq0$ the quantity of administered anticancer drug in the administration site, and let $C_c(t),C_p(t),C_1(t),C_2(t)\geq0$ denote the drug concentration at time $t$ in the central compartment, peripheral block, primary tumour and metastatic site, respectively. Building upon the ideas of \cite{zou2020biophase} we adopt a first-order kinetics strategy to describe the drug exchange between compartments. Then the mass balance for the first compartment in the model reads as
\begin{equation}
\begin{cases}
    \frac{dC_a}{dt}= r F - k_a  C_a \,, \\
    C_a(0)= 0 \,, 
\end{cases}
\label{adm_ode}
\end{equation}
where $F \in (0,1]$ is the bioavailability, \textit{i.e.} the fraction of the administered dose that reaches the systemic circulation, $k_a \in \mathbb{R}_{>0}$ is the absorption rate, and $r \equiv r(t) \geq 0$ is the quantity of administered drug over time. The evolution in time of the drug concentration in the peripheral compartment is given by
\begin{equation}
\begin{cases}
    \frac{dC_p}{dt} = k_{in,p}  C_c - k_{out,p}  C_p\,, \\
    C_p(0)=0 \,,
\end{cases}
\label{periph_ode}
\end{equation}
where $k_{in,p} \in \mathbb{R}_{>0}$ and $k_{out,p} \in \mathbb{R}_{>0}$ are the  first-order constant rates of drug concentration respectively entering and exiting the peripheral compartment.  
The two physiological aspects we want to consider, \textit{i.e.} tumour size and vascularization, are captured by quantities such as tumour volume and \textit{in situ} blood flow. We thus introduce the parameters $V_{1}, V_{2}\in\mathbb{R}_{>0}$ representing the volumes of the primary tumour and the metastasis, respectively, and $Q_{1}, Q_{2} \in\mathbb{R}_{\geq0}$ representing the blood flows through the primary tumour and metastasis, respectively. We assume the primary tumour site to be more vascularised and thus take $Q_{1}/V_1>Q_{2}/V_2$. In order to set the dependency of the first-order distribution constant rates of the two tumour sites on their physiological features, we take inspiration from the physiologically-based PK models and adopt the \textit{perfusion rate-limited} strategy \citep{jones2013pbpk}. This latter assumes that the drug distributes freely and instantly across membranes, hence blood perfusion becomes the only limiting process for the drug distribution in the two tumour sites. Therefore, the ODEs describing the evolution in time of the compound concentration in the primary tumour and its metastasis, are given by
\begin{equation}
\begin{cases}
    \frac{dC_{1}}{dt} = \frac{Q_{1} R}{V_{1}}  C_c - \frac{Q_{1} R}{V_{1}  K_{1}}  C_{1} - \Psi_1 \,, \\
    C_{1}(0)=0 \,,  
\end{cases}
\quad \text{and} \qquad
\begin{cases}
    \frac{dC_{2}}{dt} = \frac{Q_{2} R}{V_{2}} C_c - \frac{Q_{2} R}{V_{2}  K_{2}}  C_{2} - \Psi_2 \,, \\
    C_{2}(0)=0 \,, 
\end{cases}
\label{metast_ode}
\end{equation}
where the constant $R\in (0,1]$ is the blood-to-plasma drug partition coefficient, and $K_{1}, K_{2}\in (0,1]$ are the tumour-to-plasma partition coefficients for primary tumour and metastasis, respectively. These will affect how the chemical distributes throughout the tissues, w.r.t. the plasma concentration, and are an important part of any pharmacokinetic study. The terms $\Psi_1\equiv \Psi_1(C_{1},n_1)$ and $\Psi_2\equiv \Psi_2(C_{2},n_2)$ model consumption of the the cytotoxic drug by the cancer populations and, following the ideas proposed in \citep{villa2021evolutionchemo}, are defined as 
\begin{equation}
     \Psi_i(C_{i},n_i):=\psi_i \cdot \frac{\eta_i \cdot C_{i}(t)}{\alpha_{i}+C_{i}(t)}\int_0^1(1-y)^2 n_i(t,y) dy   \quad i=1,2 \,,
     \label{drug_consuption}
\end{equation}
where $\alpha_i$ and $\eta_i$ were introduced in~\eqref{cytodeath}, while $\psi_i\in \mathbb{R}_{>0}$ is a conversion factor.  
Finally,  the mass balance equation describing the evolution in time of the drug concentration in the central compartment is given by
\begin{equation}
\begin{cases}
    \frac{dC_c}{dt}= \frac{k_a}{V_c}\ C_a  + \frac{Q_{1}}{V_b K_{1}} C_{1}+ \frac{Q_{2}}{V_b  K_{2}} C_{2} + k_{out,p} C_p  - \bigg(\frac{Cl}{V_c} 
+ \frac{Q_{1}+Q_{2}}{V_b}+k_{in,p}\bigg) C_c\,, \\
    C_c(0)=0 \,,
\end{cases}
\label{plasma_ode}
\end{equation}
where $V_c \in \mathbb{R}_{>0}$ is the volume of the central compartment and $V_b \in \mathbb{R}_{>0}$ is the blood volume. This latter parameter is introduced in the ODE because the tumour sites uptake drug from the circulating system and not from the entire central compartment. Moreover, $Cl \in \mathbb{R}_{>0}$ represents the clearance, \textit{i.e.} the volume of plasma cleared of a drug over a specified time period. Under the scenario of drug infusion, the factor $k_aC_a$ in equation~\eqref{plasma_ode}$_1$ will be replaced by the infusion rate.

\section{Analysis of evolutionary dynamics}
\label{sec:analysis}
In this section, we study the long-time behaviour of the system~\eqref{pdedensity_final}, \textit{i.e.} its solution for $t \rightarrow \infty$, with $R_i$ in the form of~\eqref{genericfitness}. 
To achieve this, we focus on a scenario where the concentration of chemotherapeutic agent is constant in time, \textit{i.e.} the function $C_i(t)$ is given and satisfies the following assumption
$$C_{i}(t)\equiv c_i\geq 0\,, \quad i\in\{1,2\} .$$
Consequently, the time dependency of the fitness function is no longer mediated by the drug concentration, and we here make use of the simplified notation $R_i(y,I_i)\equiv R_i(y,I_i,c_i)$, along with $a_i\equiv a_i(c_i)$, $b_i\equiv b_i(c_i)$ and $h_i\equiv h_i(c_i)$ for the factors appearing in~\eqref{genericfitness} and defined in~\eqref{eq:fitness_factors}. 

\subsection{Assumptions}\label{subsec:formal_assumptions}
\paragraph{Fitness functions $R_i$.}
We assume there exist positive constants $I_m$, $I_M$ such that the following hold:

\beq\label{ass:RI1}
\frac{\partial R_i}{\partial I_i}(y,I_i) < 0 \qquad \forall \, y\in [0,1]\,,\quad I_i\in [I_m,I_M] \,, \quad i=1,2 \;;
\eeq

\beq\label{ass:Ry1}
\frac{\partial^2 R_i}{\partial y^2}(y,I_i)<0  \qquad \forall \, y\in [0,1]\,,\quad I_i\in [I_m,I_M] \,, \quad i=1,2 \;.
\eeq 

\noindent 
We have introduced the natural assumption that growth is saturated by an overall higher population density, due to competition for space and resources, which translates mathematically into \eqref{ass:RI1}, \textit{i.e.} each fitness function is a monotonically decreasing function of the local cell density. 
In addition, we have assumed that in each given local environment there is only one fittest phenotypic trait, which translates mathematically into \eqref{ass:Ry1}, \textit{i.e.} each fitness function is strictly concave in $y$ and thus presents only one maximum in $[0,1]$. 
The fittest function definition we give in~\eqref{genericfitness} satisfies assumptions~\eqref{ass:RI1} and \eqref{ass:Ry1}.  

\paragraph{Migration rates $\nu_{i,j}(y)$.}
We consider generic phenotype-dependent migration rates $\nu_{i,j}(y)\geq 0$, and assume there exist positive constants $\nu_m,\nu_M\in \mathbb{R}_{\geq 0}$ so that  $\nu_{i,j}(y)$ satisfy the following properties:
\begin{equation}
\label{ass:nu}
  0 \leq \nu_m \leq \nu_{i,j}(y)\leq \nu_M < \infty \quad \forall y \in [0,1], \quad i,j = 1,2\,, \quad i\neq j , 
\end{equation} 
\begin{equation}
\label{ass:nuy}
    \nu_{i,j}'(y)\geq 0 \quad \forall y \in [0,1], \quad i,j = 1,2\,, \quad i\neq j.
\end{equation}
Property~\eqref{ass:nu} expresses the natural assumption that migration rates from one site to the other are non-negative and bounded for all phenotypic variants.  
Moreover,  we assume that cells with higher levels of cytotoxic-drug resistance migrate to different sites at higher rates~\citep{shibue2017drugresistance}, which translates mathematically into~\eqref{ass:nuy}, \textit{i.e.} the migration rates are increasing functions of the phenotypic state $y$. 
Definition~\eqref{migrationratefunction} for the migration rates satisfies assumptions~\eqref{ass:nu} and~\eqref{ass:nuy}. 

\paragraph{Effective fitness.} Notice that we may rewrite \eqref{pdedensity_final}$_1$ in terms of the effective fitness function $R_i(y,I_i)-\nu_{i,j}(y)$ for $i,j=1,2$ ($i\neq j$). Building on the ideas proposed in~\cite{mirrahimi2012adaptation}, we assume this is such that
\begin{equation}\label{ass:argmaxRmu1}
\argmax_{y\in[0,1]} \big[ R_i(y,I_i)-\nu_{i,j}(y) \big] \quad \text{is a singleton for} \quad i,j=1,2 \quad \text{and} \quad i\neq j \,,
\end{equation}
\begin{equation}\label{ass:argmaxRmu2}
\argmax_{y\in[0,1]} \big[ R_1(y,I_1)-\nu_{1,2}(y) \big] \cap \argmax_{y\in[0,1]} \big[ R_2(y,I_2)-\nu_{2,1}(y) \big] = \emptyset \,,
\end{equation}
that is, the effective fitness function of each population allows only one maximum and the traits corresponding to these maxima are distinct.

\paragraph{Additional technical assumptions.}
Following the ideas proposed in~\cite{mirrahimi2012adaptation}, and given $\nu_m$ and $\nu_M$ introduced in~\eqref{ass:nu}, the analysis will also rely on the additional technical assumption that there exists a constant $\delta>0$ such that

\begin{equation}\label{ass:Im}
\delta \leq \min{\bigg( R_i \Big(y, \frac{\nu_m}{\nu_M} I_m\Big) \,,\, R_i \Big(y,I_m\Big)\bigg)} \qquad \forall \, y\in [0,1]\,, \quad i=1,2 \;,
\end{equation}

\begin{equation}\label{ass:IM}
\max{\bigg( R_i \Big(y, \frac{\nu_m}{\nu_M} I_M\Big) \,,\, R_i \Big(y,I_M\Big)\bigg)} \leq -\delta \qquad \forall \, y\in [0,1]\,, \quad i=1,2 \;.
\end{equation}

\subsection{Hyperbolic time scaling and steady state problem}\label{subsec:hopf_cole}
In accordance with previous studies in literature~\citep{doerfler2006methylation,duesberg2000explaining}, we assume that the rate of spontaneous phenotypic variations $\beta_i$ occur on a slower time scale compared to cell division and death. We therefore introduce a small parameter $\e>0$ and assume both $\beta_i := \e^2$ ($i=1,2$). Following previous studies on the long-time behaviour of non-local PDEs and integro-differential equations modelling the dynamics of continuously structured populations~\cite{barles2009concentration,chisholm2016effects,desvillettes2008selection,diekmann2005dynamics,jabin2016selection,lorz2011dirac,mirrahimi2015asymptotic,perthame2008dirac}, we use the hyperbolic time scaling $t \mapsto \frac{t}{\e}$ in the conservation equation \eqref{pdedensity_final}, and obtain the following system of non-local PDEs for the phenotypic distributions $n_i\left(\frac{t}{\e},y\right)=n_{i\e}(t,y)$ ($i=1,2$):
\begin{equation}
\label{eq:ne:results}
\begin{cases}
\displaystyle{ \e\dt n_{i\e} - \e^2 \, \dyy n_{i\e} = R_i\big(y,I_{i\e}\big) \, n_{i\e}  +\nu_{j,i}(y) \, n_{j\e} - \nu_{i,j}(y) \, n_{i\e}\,, } \quad i \neq j \,,  \\[7pt]
\displaystyle{ I_{i\e}(t) := \int_{0}^1 n_{i\e}(t,y) \, {\rm{d}}y} \\[10pt]
\displaystyle{\dy n_{i\e} (t,0) = \dy n_{i\e} (t,1)= 0 }
\end{cases}
\qquad i=1,2 \;.
\end{equation}
As we are interested in the equilibria of \eqref{pdedensity_final} in the case of rare phenotypic variations, we follow the strategy adopted in~\cite{mirrahimi2012adaptation} and investigate the equilibria of \eqref{eq:ne:results} in the asymptotic regime $\e\to0$. Assume that as $t\to\infty$ we have that $\nie(t,y)\to\nie^{\infty}(y)$ and $\Iie(t)\to\Iie^\infty$ ($i=1,2$). Then, the equilibria of \eqref{eq:ne:results} satisfy the following system of non-local ODEs
\begin{equation}
\label{eq:ne:ss:results}
\begin{cases}
\displaystyle{ R_i\big(y,I_{i\e}^{\infty}\big) \, n_{i\e}^{\infty} + \e^2 \, \left(n_{i\e}^{\infty}\right)''  +\nu_{j,i}(y) \, n_{j\e}^{\infty} - \nu_{i,j}(y) \, n_{i\e}^{\infty} = 0 \,,}\quad i \neq j \,, \\[7pt]
\displaystyle{ I_{i\e}^{\infty} = \int_{0}^1 n_{i\e}^{\infty}(y) \, {\rm{d}}y} \\[10pt]
\displaystyle{n_{i\e}^{\infty}(0) = n_{i\e}^{\infty} (1)= 0}\,
\end{cases} 
\qquad i=1,2\,.
\end{equation}
From now on we will make use of the notation $n_{i\e}(y) \mbox{ and } I_{i\e}$ $(i=1,2)$ to refer to $n_{i\e}^{\infty}(y) \mbox{ and } I_{i\e}^{\infty}$, respectively.

\subsection{Results of formal analysis}\label{subsec:analytical_results}

We here summarise the results of the formal analysis, detailed in appendix~\ref{appendix:analysis}, extending the results of Mirrahimi~\cite{mirrahimi2012adaptation} to the case of phenotype-dependent migration rates between the sites, for our model of evolutionary dynamics of connected metastatic tumours. 

\paragraph{Bounds on $I$.}
It can be shown from \eqref{eq:ne:ss:results} that for all $\e\leq\e_0$, with $\e_0$ small enough, under assumptions
 \eqref{ass:RI1},  \eqref{ass:nu}, \eqref{ass:Im} and \eqref{ass:IM}, 
 we have
\beq\label{bound:Ie:results}
I_m\leq I_{i\e}\leq I_M \qquad i=1,2\,,
\eeq
for all $\e$, where we recall $0<I_m<I_M$. Proof of this follows analogous steps of that in  \cite[Lemma 2.1]{mirrahimi2012adaptation}, and can be found in appendix~\ref{appendix:proofbounds}. As a result, in the asymptotic regime $\e\to 0$ we have $I_{i\e}\to I_i$ with
\beq\label{bound:I:results}
I_m\leq I_{i}\leq I_M \qquad i=1,2\,. 
\eeq

\paragraph{Asymptotic regime $\e\rightarrow 0$.} 
Building on the strategies adopted in~\cite{mirrahimi2012adaptation}, we introduce the Hopf-Cole transformation 
\begin{equation}
    \nie(y) = e^{\uie(y)/\e} \qquad i=1,2\,,
\label{fa:hopf_cole:results}
\end{equation}
with $\uie(y)$ semi-convex (\textit{i.e.} $\dyy \uie \geq -E$, for some constant $E>0$). Then,  we expect that $\Iae\to\Ia$, $\Ibe\to\Ib$, $\uae\to u_1$ and $\ube\to u_2$ in the asymptotic regime $\e\to0$, where $\Ia$, $\Ib$, $u_1$ and $u_2$ are the leading-order terms of the asymptotic expansions for $\Iae$, $\Ibe$, $\uae$ and $\ube$, respectively. 
Assuming that $u_1=u_2=u$ and that the semi-convexity of $\uie$ ($i=1,2$) is preserved in the limit, so that $u=u(y)$ is also semi-convex, we have that $u$ is a viscosity solution to the following constrained Hamilton-Jacobi equation
\beq
\label{fa:hj2:results}
\begin{cases}
\displaystyle{- \bigg(\frac{du}{dy}\bigg)^2 = H(y,\Ia,\Ib)} \,,\\
\displaystyle{ \max_{y \in [0,1]} u(y) = 0 }\,,
\end{cases} 
\eeq
with $H(y,\Ia,\Ib)$ being the largest eigenvalue of the matrix 
$$
\mathcal{A} =
  \begin{pmatrix}
   R_1\big(y,\Ia\big) - \nuab(y)  & \nuba(y)  \\
    \nuab(y) & R_2\big(y,\Ib\big) - \nuba(y)
  \end{pmatrix}  \,.
$$
Given the assumptions introduced in Section~\ref{subsec:formal_assumptions}, from \eqref{bound:Ie:results} and \eqref{bound:I:results} we deduce that $\nae$ and $\nbe$ converge weakly to measures $\na$ and $\nb$. Moreover, from~\eqref{fa:hj2:results},  we have that
\beq
\label{fa:n:results}
\nie(y) \xrightharpoonup[\e  \rightarrow 0]{\scriptstyle\ast}  \sum_{k=1}^{K} \rho_{ik} \, \delta(y - y_k) \qquad i=1,2 \,,
\eeq
\textit{i.e.} the measures $\na$ and $\nb$ to which $\nae$ and $\nbe$ converge weakly in the asymptotic regime $\e\to0$ concentrate as $K$ Dirac masses centered at $y_k$ ($k=1,...,K$), where the weights $\rho_{ik}\geq0$ must be such that
\beq
\label{fa:rho1}
\Ii = \sum_{k=1}^{K} \rho_{ik} \qquad i=1,2 \,,
\eeq
and where the finite number of points $y_k \in \Omega \cap \Gamma$, with $\Omega := \big\{ y \in [0,1] : u(y) = 0 \big\}$ and $\Gamma := \big\{ y \in [0,1] :  H(y,\Ia,\Ib) = 0 \big\}$.
Specifically, this yields
\beq
\label{fa:rho2:results}
\begin{cases}
\big( R_1(y_k,\Ia)-\nuab(y_k)\big) \rho_{1k} +\nuba(y_k)\rho_{2k} =0 \\[5pt]
\big( R_2(y_k,\Ib)-\nuba(y_k)\big) \rho_{2k} +\nuab(y_k)\rho_{1k} =0 
\end{cases}\qquad k=1,..,K \,.
\eeq 
Details of the formal analysis in the asymptotic regime $\e\rightarrow 0$ can be found in appendix~\ref{appendix:prooftheorem}. 

\paragraph{Analytical results for the metastatic spread case: $\nu_{1,2}\not \equiv 0$ and $\nu_{2,1}\equiv 0$.}\label{par:analytical results}
The long-time solution of our system with the fitness function $R_i(y,I_i)$ defined as in \eqref{genericfitness} and the migration rate $\nu_{i,j}(y)$ defined as in \eqref{migrationratefunction}, and obtained by solving~\eqref{fa:rho2:results}, is the following:
\begin{equation}
\label{sol:metastatic:y:n}
\displaystyle{n_1(y) = \Ia \, \delta (y-y_1) \qquad \text{and} \qquad n_2(y) = \rho_{21} \, \delta (y-y_1) +  \rho_{22} \, \delta (y-y_2)}  \,,
\end{equation}
with
\begin{equation}
\label{metastatic:y:y}
\displaystyle{y_1 = \frac{b_1}{b_1+\hat{\nu}_{1,2}}h_1\,, \qquad y_2 = h_2 \,, }
\end{equation}
and
\begin{equation}
\label{metastatic:y:rho}
\displaystyle{ \Ia= \frac{1}{d_1}\bigg[ a_1  - \frac{\hat{\nu}_{1,2} b_1 h_1^2}{b_1+\hat{\nu}_{1,2}} \bigg] \,, \quad \Ib = \frac{a_2}{d_2} \,,  \qquad \rho_{21} = \min \bigg(\frac{\hat{\nu}_{1,2}\,y_1^2}{b_2(y_1-h_2)^2}\Ia, \Ib\bigg) \,, \quad \rho_{22} = \Ib - \rho_{21} \,.  }
\end{equation}
The proof can be found in the appendix~\ref{appendix:proofsolutions}, while the results for the localised tumours ($\nu_{1,2}\equiv 0$ and $\nu_{2,1}\equiv 0$) and secondary seeding ($\nu_{1,2}\not \equiv 0$ and $\nu_{2,1}\not \equiv 0$) scenarios can be found in the supplementary sections S.1.1 and S.1.2, respectively.

\subsection{Biological insights}
\begin{figure}[htb!]
    \centering
    \includegraphics[width=0.7\linewidth]{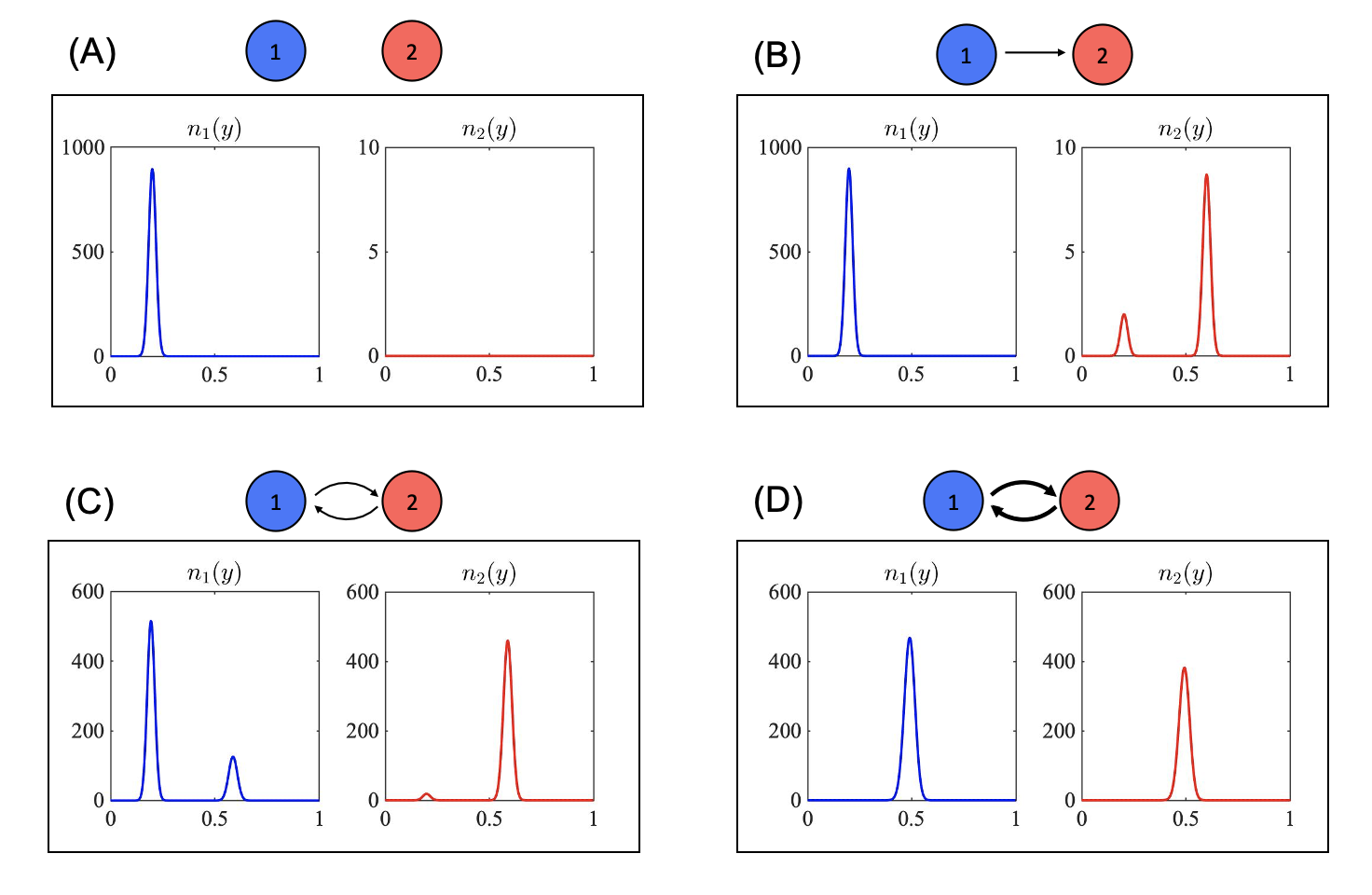}
    \caption{
    \textbf{Model outcome dependency on the migration rates.} Possible model outcomes under different proof-of-concept non-dimensional parameter sets. 
We simulate system~\eqref{pdedensity_final}, under definition~\eqref{genericfitness} for the fitness functions $R_i$ and definition~\eqref{migrationratefunction} for the migration rates $\nu_{i,j}(y)$,  under the initial conditions $n_{0,1}(y)>0$ and $n_{0,2}(y)=0$ $\forall y\in[0,1]$. Each plot displays the equilibrium solution in a different biological scenario:
 (A) Localised tumour, under the parameter set $\beta_1=\beta_2=10^{-7}$, $a_1=8$, $a_2=0.1$, $b_1=1$, $b_2=0.8$, $h_1=0.2$, $h_2=0.6$, $d_1=d_2=0.2$, and $\hat{\nu}_{1,2}=\hat{\nu}_{2,1}=0$; (B) Metastatic spread, under the parameter set of A, except $\hat{\nu}_{1,2}(y)=0.007$; (C) Secondary seeding, under the parameter set $\beta_1=\beta_2=10^{-7}$, $a_1=6$, $a_2=5$, $b_1=1$, $b_2=0.6$, $h_1=0.2$, $h_2=0.6$, $d_1=d_2=0.2$, $\hat{\nu}_{1,2}=0.1$ and $\hat{\nu}_{2,1}=0.05$. (D) Secondary seeding in highly connected sites, under the parameter set of C except $\hat{\nu}_{1,2}=0.6$ and $\hat{\nu}_{2,1}=0.3$.}
    \label{fig:possible_outcomes_results}
\end{figure}

The biological interpretation of the results of the formal analysis summarised in Section~\ref{subsec:analytical_results} for the metastatic spread case yields interesting biological insights. Figure~\ref{fig:possible_outcomes_results} displays the possible model outcomes, based on the constraints imposed on the migration rates, obtained simulating a non-dimensional version of the model.  We particularly focus on the metastatic spread case here, and further illustrate the dependency of the solution on certain parameters in Figure~\ref{analysis:parametersaffectpopulation}, but provide more details on the other biological scenarios in the supplementary material. 

\paragraph{Different biological scenarios.}
From the analysis, and as can be seen in Figure~\ref{fig:possible_outcomes_results}, we note that:
\begin{itemize}[noitemsep,topsep=0pt]
\item[(i)] In the case of a localised tumour, the primary tumour evolves into a monomorphic population, with the selected trait being the fittest locally ($y_1 = h_1$) and the total cell density reaching the carrying capacity of the site ($I_1=\frac{a_1}{d_1}$), 
and no metastasis forms (cf.  Figure~\ref{fig:possible_outcomes_results}A).  
\item[(ii)] In the case of metastatic spread, from \eqref{sol:metastatic:y:n} we have that the population in the primary tumour will evolve into a monomorphic population, while the population in the metastatic site will evolve into a dimorphic population (or a monomorphic one under certain conditions, explained later), as also illustrated in Figure~\ref{fig:possible_outcomes_results}B.  
\item[(iii)] In the case of metastatic spread with secondary seeding, dimorphism may be observed in both tumours (cf.  Figure~\ref{fig:possible_outcomes_results}C) or, in the case of highly connected sites, both tumours may evolve into a monomorphic population adapted to both environments (cf.  Figure~\ref{fig:possible_outcomes_results}D), although this scenario is unlikely to be observed \textit{in vivo}. 
\end{itemize}

\paragraph{Metastatic spread: the primary site.} From \eqref{sol:metastatic:y:n}-\eqref{metastatic:y:rho} we have that the composition and size of the population in the primary tumour not only depend on the local environment but also on the ability of cells to intravasate and eventually metastasise. In particular,  we observe the following:
\begin{enumerate}[noitemsep,topsep=0pt]
\item[(iv)] The selected trait in the primary tumour $y_1$ is smaller than the locally fittest one $h_1$, unlike in the localised tumour case. Hence, if cells have different metastatic abilities depending on their phenotypic state, at equilibrium cells less prone to metastasise are found in the primary tumour, probably since the more metastatic phenotypes have left the site.  
\item[(v)] In particular, $y_1$ is a decreasing function of $\hat{\nu}_{1,2}$, indicating that the magnitude of this phenotypic shift ($|y_1-h_1|$) increases with the migration rate.
\item[(vi)] The total population at equilibrium in the primary tumour is lower than the carrying capacity of the site, \textit{i.e.} $I_1<\frac{a_1}{d_1}$, unlike in the case of a localised tumour. 
Recall however that this comes with the trade off of having established a dimorphic population in the metastatic site, which increases the chances of surviving environmental changes, overall resulting in greater evolutionary advantage.
\end{enumerate}

\paragraph{Metastatic spread: the metastatic site.} From \eqref{sol:metastatic:y:n}-\eqref{metastatic:y:rho},  we observe that the total cell density of the metastatic tumour reaches carrying capacity, \textit{i.e.} $I_2=\frac{a_2}{d_2}$, but the contribution of cells coming in from the primary tumour will affect the composition of the local population. In particular, we observe the following:
\begin{enumerate}[noitemsep,topsep=0pt]
\item[(vii)] We have that $\rho_{21}$ is an increasing function of $I_1$ and $\hat{\nu}_{1,2}$, suggesting that a larger population in the primary site or a higher migration rate will result in a larger subpopulation in the secondary site presenting traits selected in the primary, as also illustrated in Figure~\ref{analysis:parametersaffectpopulation} (third column).  
\item[(viii)] Moreover, if $\hat{\nu}_{1,2}$ is significantly high, this might lead to $\rho_{21}=I_2$ and $\rho_{22}=0$ (cf. bottom plot in the third column of  Figure~\ref{analysis:parametersaffectpopulation}). This means that if the migration rate of cells between sites is particularly high then the subpopulation in the secondary site presenting traits selected in the primary may outnumber the subpopulation with traits adapted to the local environment and drive it to extinction. This scenario can also occur in the case the population of the primary tumour $I_1$ is significantly big w.r.t. to the one of the metastasis $I_2$.
\item[(ix)] We have that $\rho_{21}$ is a decreasing function of the non-linear selection gradient $b_2$, suggesting that a stronger selective pressure from the environment in the secondary site will result in a smaller local subpopulation presenting traits selected in the primary,  as also shown in Figure~\ref{analysis:parametersaffectpopulation} (first column).  
\item[(x)] Moreover, in the limit $b_2 \to \infty$ we have $\rho_{21}\to 0$ (cf. bottom plot in the first column of Figure~\ref{analysis:parametersaffectpopulation}). This means that in the case of extremely strong selective pressure, assuming cells from the primary site already manage to create a local niche in the metastatic site,  this will eventually only evolve into a monomorphic population, as the trait selected in the primary site will not be fit enough to survive. 
\item[(xi)] Finally, $\rho_{21}$ is a decreasing function of the distance between $y_1$ and $h_2$, indicating that the closer the selected trait in the primary site is to the fittest trait of the metastatic site, the higher the subpopulation density in the metastatic site of cells presenting the trait selected in the primary site is.  This may depend on the shift in the phenotypic trait selected in the primary site compared to the fittest one in the local environment $h_1$, but also on the similarity of environmental condition in the two sites, \textit{i.e.} the distance $|h_1-h_2|$, as also illustrated in  Figure~\ref{analysis:parametersaffectpopulation} (second column).  
\end{enumerate}

\begin{figure}[htb!]
    \centering
    \includegraphics[width=1\linewidth]{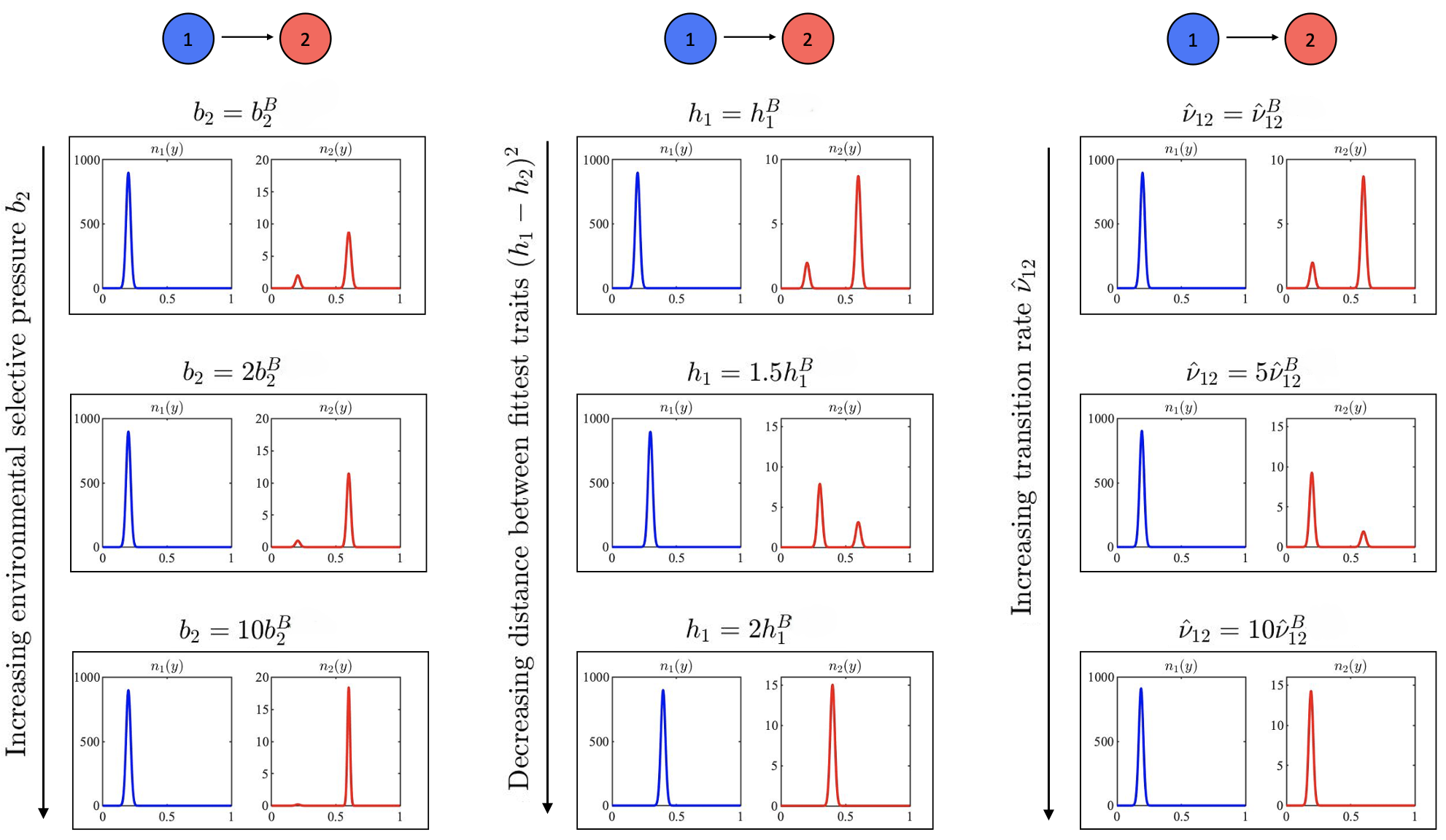}
    \caption{\textbf{Model outcome dependency on input factors in the metastatic spread scenario.}
Illustrative example showing how the selection gradient $b_2$, the fittest trait $h_1$ and the maximum migration rate $\hat{\nu}_{1,2}$ may affect the equilibrium distributions of the cancer cell populations of each site, in the metastatic spread case. 
We simulate system~\eqref{pdedensity_final}, under definition~\eqref{genericfitness} for the fitness functions $R_i$ and definition~\eqref{migrationratefunction} for the migration rates $\nu_{i,j}(y)$,  under the initial conditions $n_{0,1}(y)>0$ and $n_{0,2}(y)=0$, $\forall y\in[0,1]$.  
Each column displays the equilibrium solutions of three simulations obtained by progressively varying the parameters $b_2$ (first column), $h_1$ (second column) and $\hat{\nu}_{1,2}$ (third column) form their baseline values $b_2^{B}=0.8$, $h_1^{B}=0.2$ and $\hat{\nu}_{1,2}^{B}=0.007$. The remaining parameters are set to $\beta_1=\beta_2=10^{-7}$, $a_1=8$, $a_2=0.1$, $b_1=1$, $b_2=b_2^{B}$, $h_1=h_1^{B}$, $h_2=0.6$, $d_1=d_2=0.2$, $\hat{\nu}_{1,2}=\hat{\nu}_{1,2}^{B}$ and $\hat{\nu}_{2,1}=0$. }
    \label{analysis:parametersaffectpopulation}
\end{figure}

\section{Numerical results}\label{sec:numres}
We complement the analytical results presented in Section~\ref{sec:analysis} with numerical solutions of the model equations, focusing on the metastatic spread case and under dynamic drug concentrations as predicted by the PK model, \textit{i.e.} we solve system~\eqref{pdedensity_final}, under definitions~\eqref{fitnessfunction} and \eqref{genericfitness}-\eqref{migrationratefunction},  coupled with equations~\eqref{adm_ode}-\eqref{plasma_ode}. 

In Section~\ref{sec:setup} we present the set-up and parameter values employed for the numerical simulations, along with the numerical method employed to simulate the model. In Section~\ref{sec:gsa} we perform global sensitivity analysis of the model equations, while in Section~\ref{sec:numerical_results} we investigate how the parameters impact the model outcome, that is, the equilibrium solution of the system of equations, and the time it takes for the system to reach steadiness, in view of the relative impact this may have during the course of treatment. 

\subsection{Set-up of numerical simulations and numerical methods.}\label{sec:setup}
\paragraph{Set-up and model parametrization.}
As introduced in Section~\ref{sec:intro},  we consider a metastatic BRAF-mutated melanoma, assuming no self-seeding takes place (\textit{i.e.} $\hat{\nu}_{2,1}=0$), treated with Dabrafenib orally administered twice a day in the form of a $150$ mg tablet~\citep{luebker2019brafinhibitor}. To carry out numerical simulations of the system, we calibrate the model with parameters drawn from the literature. 
For the PK parameters we rely on the reports on Tafinlar, \textit{i.e.} brand name for Dabrafenib, realised by the European Medicine Agency (EMA) \cite{ema2013tafinlar} and the US Food and Drug Administration (FDA) \cite{fda2013dabrafenib}, as reported in Table~\ref{tab:pk_params}. 
\begin{table}[h]
    \centering
    \begin{tabular}{c|l|c|l}
        \multicolumn{4}{c}{\textbf{PK parameters}}  \\
        \midrule
        Parameter & Value  & Unit & Ref \\
        \midrule
        $k_a$ & $1.8$ & h$^{-1}$ & \cite{balak2020dabrafenib}\\
        $F$  & $0.95$ & - & \cite{ema2013tafinlar,fda2013dabrafenib,puszkiel2019dabrafenib} \\
        $Cl$  & $17$ & h$^{-1}$ & \cite{balak2020dabrafenib,fda2013dabrafenib}\\ 
        $R$  & $0.54$ & h$^{-1}$ & \cite{ema2013tafinlar,fda2013dabrafenib} \\
        $V_{c}$  & $37.525$ & l & \cite{balak2020dabrafenib,puszkiel2019dabrafenib}\\ 
        $V_{b}$  & $5$ & l & \cite{sharma2022physiology}\\
        $k_{in,p}$ & $0.0974$ & h$^{-1}$ & \cite{balak2020dabrafenib} \\
        $k_{out,p}$ & $0.196$ & h$^{-1}$ & \cite{balak2020dabrafenib} \\
        \bottomrule
    \end{tabular}
    \caption{Pharmacokinetics parameter values.}
    \label{tab:pk_params}
\end{table}
For the parameters appearing in the cancer evolutionary dynamic model,  we mainly rely on the values reported in~\cite{lorenzi2018spatialvariations,villa2021evolutionchemo}. However, we set the conversion factors $\psi_i$ ($i=1,2$) in equations~\eqref{metast_ode} to zero, since we consider PK parameters obtained from a model that was calibrated with \textit{ex vivo} data and thus the value of the clearance $Cl$ accounts for the whole process of elimination, including drug consumption by tumour cells~\cite{balak2020dabrafenib}. For the migration rates we consider the ratio $\hat{\nu}_{i,j} = (\text{Intravasation rate})\times(\text{Survival in the circulation})\times(\text{Extravasation probability})$, where the intravasation rate takes value in the range $[10^{-11},10^{-2}]$ cells/day~\citep{scott2013mathematical}, the probability of survival in the circulation is in the interval $[5\cdot 10^{-4},2.5\cdot 10^{-2}]$ and the extravasation probabilities are in the range $[0.1986,0.5461]$, as reported in~\citep{franssen2019modelmetastases}. 
Building on these works, we identified upper and lower bounds that each parameter should take, selecting its reference value in this range, as detailed in Table~\ref{tab:bounds_tumour_param}.  
\begin{table}[h]
    \centering
    \begin{tabular}{c|l|l|l|l|l|l}
        \multicolumn{7}{c}{\textbf{Tumour site specific parameters}} \\
        \midrule
        Parameter & LB  & UB & RV $i=1$ & RV $i=2$ & Unit & Ref \\
        \midrule
        $Q_{i}$  & -  &  -  & $0.3$ & $0.01$ & lh$^{-1}$ & \cite{jones2013pbpk}\\
        $V_{i}$ & -  &   - & $0.5$ & $0.05$ & l &  \cite{gallo2005pkmodel} \\
        $K_{i}$ & - & - & $0.8$ & $0.5$ & - & \cite{mittapalli2013mechanisms} \\
        \midrule
        $\beta_i$  & $10^{-13}$  &  $10^{-8}$ & $10^{-9}$ & $10^{-9}$ & s$^{-1}$ & \cite{chisholm2015emergence} \\
        $\delta_i$  & $10^{-5}$  & $10^{-3}$ & $10^{-4}$ & $10^{-4}$ & s$^{-1}$ & \cite{ward1997mathematical}\\
        $\varphi_i$  & $10^{-7}$  & $10^{-5}$ & $10^{-5}$ & $10^{-5}$ & s$^{-1}$ & \cite{gordan2007hif2alpha}\\
        $\eta_i$  & $10^{-5}$  & $10^{-3}$ & $1.8\cdot10^{-4}$ & $1.8\cdot10^{-4}$ & s$^{-1}$ & \cite{ward1997mathematical}\\
        $\alpha_i$  & $10^{-7}$  & $10^{-5}$ & $2\cdot10^{-6}$ & $2\cdot10^{-6}$ & g cm$^{-3}$ & \cite{norris2006modelling}\\
        $d_i$  & $10^{-14}$  & $10^{-12}$ & $2\cdot10^{-13}$ & $2\cdot10^{-13}$ & cm$^{3}$ s$^{-1}$ cells$^{-1}$ & \cite{li1982glucose}\\
        $\hat\nu_{i,j}$  & $10^{-12}$  & $10^{-8}$ & $1.3\cdot10^{-10}$ & $0$ & s$^{-1}$ & \cite{franssen2019modelmetastases,scott2013mathematical}\\
        \bottomrule
    \end{tabular}
    \caption{Parameter lower bounds (LB), upper bounds (UB) and reference values (RV), for the primary tumour site ($i=1$) and its metastasis ($i=2$).}
    \label{tab:bounds_tumour_param}
\end{table}

\paragraph{Numerical methods.} Numerical solutions are constructed using a uniform discretisation of the interval $[0, 1]$, consisting of 101 grid points, as the computational domain of the independent variable $y$. 
We consider $t \in [0, T]$, with $ T > 0$ being the final time of simulations, chosen sufficiently large to reach steady state or to mimic the duration of therapy. 
We discretize the interval $[0, T]$ with a uniform step, sufficiently small to ensure numerical stability under each parameter set used. 
We construct numerical solutions employing an explicit scheme, based on a first order forward difference approximation in time and a second order central difference approximation in space, applied also to the zero-flux boundary conditions.
Following \cite{almeida2019optimizationchemo,villa2021evolutionchemo}, we set the initial phenotypic distribution in each site $n_{i,0}(y)$ to be a weighted normal distribution, centered in $0$ and truncated in $[0,1]$, integrating to the initial cell density $I_{i,0}$, \textit{i.e.} 
$$
n_{i,0}(y) = \frac{I_{i,0}}{\Theta}\exp \left(-\frac{y^2}{8\cdot10^{-6}}\right)\,, \quad \text{with} \quad \Theta = \int_0^1 \exp \left(-\frac{y^2}{8\cdot10^{-6}}\right) dy\,.
$$
We chose this consistently with the assumption that most cancer cells are sensitive to the drug prior to treatment, and we also set as initial population size $I_{i,0} = \frac{\delta_i}{d_i}$. All numerical computations are performed in {\sc Matlab}. 

\paragraph{Steady-state criterion.} We check the time $T_{ss}$ at which the phenotypic distribution $n_i(t,y)$ reaches steadiness. In order to identify this numerically,  we consider a sufficiently small tolerance, $tol>0$, below which the relative difference in the numerical solution at two consecutive time steps is considered to no longer be significant.  
Let $n_{i,j}^{k}$ denote the numerical approximation of the density of cancer cells in site $i$ ($i=1,2$), endowed with phenotypic trait $y_j$ ($j=0,\dots,J$), at time step $t_k$ ($k=0,\dots,N$).  Then, we define the relative difference $D_{i,k}$ between the phenotypic distribution at site $i$ at two consecutive time steps $t_{k-1}$ and $t_{k}$ as:
\begin{equation}
D_{i,k} = \frac{1}{J+1}\sum_{j=0}^J\bigg|\frac{n_{i,j}^{k}-n_{i,j}^{k-1}}{n_{i,j}^k}\bigg|  \quad \text{for} \quad 1\leq k \leq N\,,
  \label{steady-state}
\end{equation}
and denote $T_{ss}$ as the first time step $t_k$ after which the inequality $D_{i,k}<tol$ is always satisfied for $i=1,2$.

\subsection{Global sensitivity analysis}\label{sec:gsa}
We here make use of Global sensibility analysis (GSA) to verify that the insights gained from the analytical results of Section~\ref{sec:analysis} still hold outside the asymptotic regime of rare spontaneous phenotypic changes.  
GSA is the study of how uncertainty in the output of a model can be apportioned to different sources of uncertainty in the model input and, specifically, it offers a wide overview of how the parameter interactions impact the model output.  
The most suitable GSA techniques for systems modelling highly non-linear dynamics, generally the case in biological applications and certainly the case in this work,  are the elementary effect (EE) and Sobol methods~\cite{kiparissides2009GSAbiological,qian2020sensitivityanalysis}. 
    
\paragraph{EE method.} The EE method is of screening type, \textit{i.e.} it aims at identifying the parameters that have negligible impact on the output variability, and is intended for use when dealing with a large number of input parameters, since it has a lower computational cost w.r.t. other methods. Inspired by the \textit{radial design} strategy proposed in \cite{campolongo2010fromscreening}, we consider a vector $Z$ of $p$ independent input parameters $Z_i$ ($i=1,...,p$), varying in their input space. Given the model output $Y=f(Z)$, we compute $r$ evaluations of the elementary effect $EE^l_i$ ($l=1,...,r$) associated to the $i$-th input factor. Given two sample points of $Z$, \textit{e.g.} $a^l=(a^l_1,\dots,a^l_p)$ and $b^l=(b^l_1,\dots,b^l_p)$, the elementary effect $EE^l_i$ is defined as
\begin{equation}\label{EEi}
        EE^l_i =  \frac{f((a^l_1,\dots,a^l_i,\dots,a^l_p))-f((a^l_1,\dots,b^l_i,\dots,a^l_p))}{a^l_i-b^l_i} \,, \quad l=1,...,r \,.
    \end{equation}
We also compute the corresponding measures as 
\begin{equation}\label{EEmesure}
    \overline{EE}_i = \frac{1}{r}\sum_{l=1}^r EE_i^l \,, \quad   \overline{EE}_i^* = \frac{1}{r}\sum_{l=1}^r |EE_i^l|\quad \text{and} \quad SD_i = \bigg(\frac{1}{r}\sum_{l=1}^r (EE_i^l-\overline{EE}_i)^2\bigg)^{1/2} \,.
\end{equation}
In particular:
\begin{itemize}[noitemsep,topsep=0pt]
     \item $\overline{EE}_i$ (\textit{sensitivity measure}) represents the overall influence of the $i$-th parameter on the output. $\overline{EE}_i^*$ has the same meaning but is adopted for avoiding the cancellation effect when dealing with non-monotonic models.
    \item $SD_i$ (\textit{interactions measure}) gives information of the non-linearity and/or interaction effects of the $i$-th input. If it is small it means that the $EE_i$ are similar all along the sample space, suggesting a linear relationship between the $i$-th input and the output. On the contrary, if it is large it means that the $EE_i$ are strongly affected by the sample point at which they are computed, thus the parameter effects are considered to be non-linear and/or due to interaction with other factors.
\end{itemize}
    
\paragraph{Sobol method.} The Sobol method is a variance-based technique, and is both of screening and ranking type, \textit{i.e.} it aims at ordering the inputs based on their impact on the output variability. This approach is computationally demanding, since it requires a large number of model simulations. For this reason, it is usually adopted for a small set of input parameters. Given the input vector $Z$ and the model output $Y$, the Sobol indices are given by
\begin{equation}\label{sobol}
        S_i = \frac{Var(\mathbb{E}[Y|Z_i])}{Var(Y)}, \quad \text{and} \quad  S_{Ti} = S_i + \sum_{u\subseteq \{1,\dots,p\} \backslash\{i\}} \frac{Var(\mathbb{E}[Y|Z_i,Z_u])}{Var(Y)}\,,
\end{equation}
which are both non-negative.  In particular:
\begin{itemize}[noitemsep,topsep=0pt]
    \item $S_i$ provides the first-order contribution of the $i$-th input to the output variance and measures the main effect of $Z_i$.
    \item $S_{Ti}$ measures the total effect of the parameter $Z_i$ on the output. $S_{Ti}=0$ implies that the parameter $Z_i$ is non-influential, while a significant difference between $S_i$ and $S_{Ti}$ suggests that the factor $Z_i$ is involved in important interactions.
\end{itemize}
We compute the primary and total indices adopting the Monte Carlo method, and in particular we make use of the estimators for $S_i$ and $S_{Ti}$ suggested in~\cite{saltelli2002makingthebest}.

\paragraph{Model inputs and outputs.}
We consider $Z = (\beta_1,\beta_2,\varphi_1,\varphi_2,\delta_1,\delta_2,\alpha_1,\alpha_2,\eta_1,\eta_2,d_1,d_2,\hat\nu_{1,2})$ and the input vector of the GSA, while the other parameters are fixed to the reference values of Tables~\ref{tab:pk_params} and \ref{tab:bounds_tumour_param}. We consider each parameter a random variable, independent from the others, distributed with a log-uniform distribution in the ranges defined in Table~\ref{tab:bounds_tumour_param}. We adopt this distribution in order to better explore the different orders of magnitude of the input parameters. Furthermore, we take into account a drug schedule of $150$ mg oral tablets twice a day, and set a final time $T=91$ days. We consider two model outputs:
\begin{itemize}
    \item $Y_{I}=I_1(T)+I_2(T)$, \textit{i.e.} the total tumour mass at time $T$;
    \item $Y_{\mu}=\frac{\mu_1(T)+\mu_2(T)}{2}$, \textit{i.e.} the mean phenotypic trait average between the two tumour sites at time $T$.
\end{itemize}
We have used Sobol’ quasi-random sequences in order to generate our sets of quasi-random points \cite{campolongo2010fromscreening}.

\subsubsection{Results of the EE method for GSA}

\begin{figure}
  \renewcommand{\thesubfigure}{\alph{subfigure}.\arabic{row}}
  \centering
  \setcounter{row}{1}%
  \begin{subfigure}[b]{0.45\textwidth}
    \centering
    \includegraphics[width=0.95\textwidth]{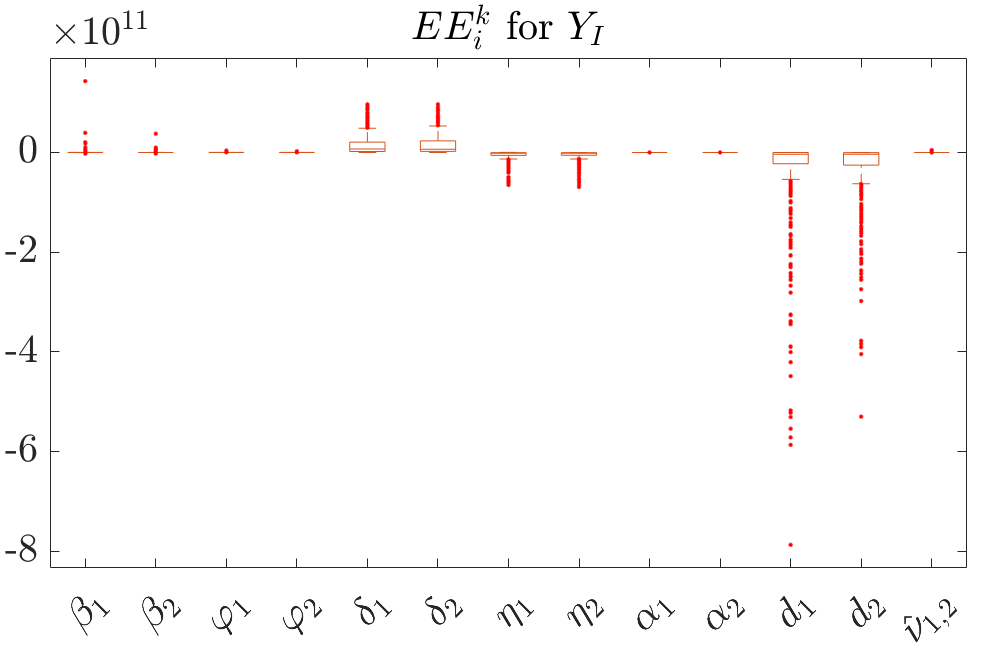}
    \caption{}
    \label{fig:EEi_rho}
  \end{subfigure} \quad
  \begin{subfigure}[b]{0.45\textwidth}
    \centering
    \includegraphics[width=0.95\textwidth]{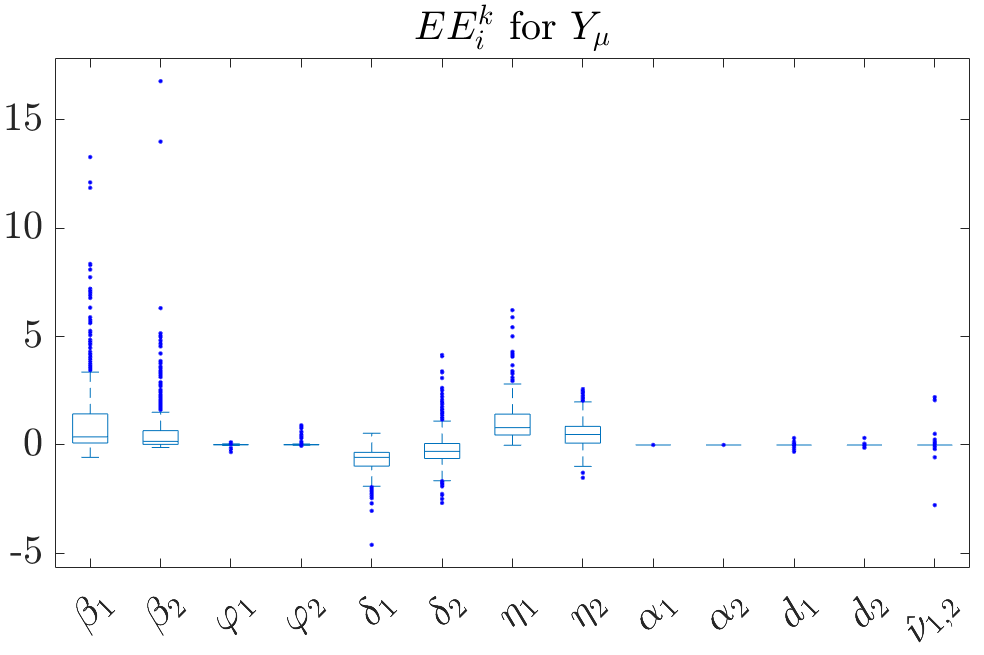}
    \caption{}
    \label{fig:EEi_mu}
  \end{subfigure}
  \vskip\baselineskip
  \stepcounter{row}
  \setcounter{subfigure}{0}
  \begin{subfigure}[b]{0.45\textwidth}
    \centering    
    \includegraphics[width=\textwidth]{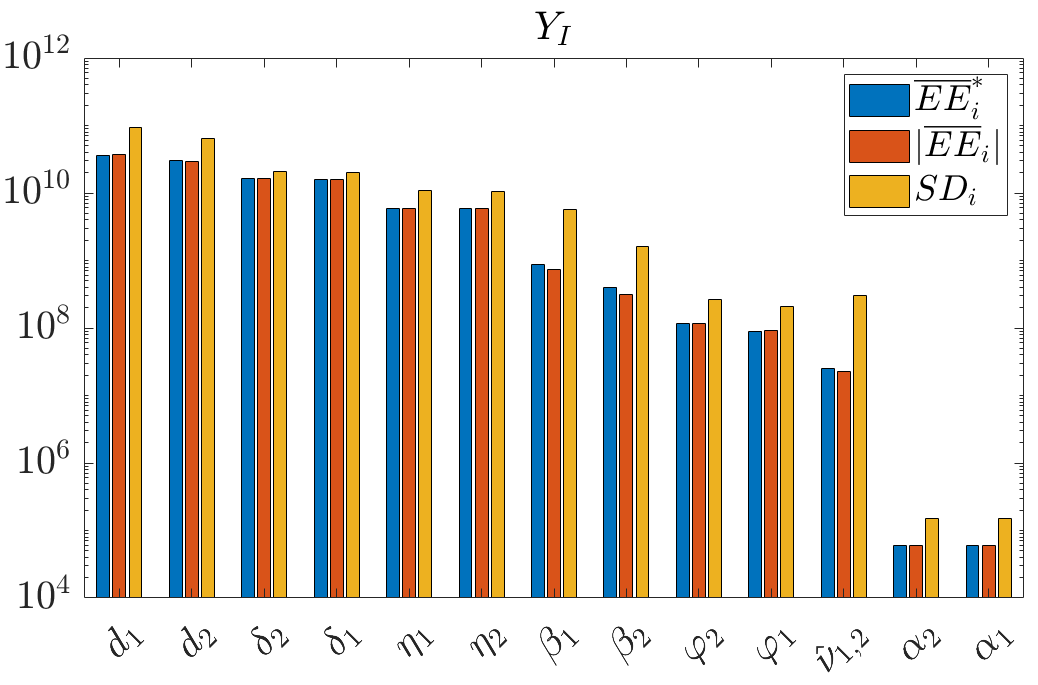}
    \caption{}
    \label{fig:EEmeasures_rho}
  \end{subfigure} \quad
  \begin{subfigure}[b]{0.45\textwidth}
    \centering    
    \includegraphics[width=\textwidth]{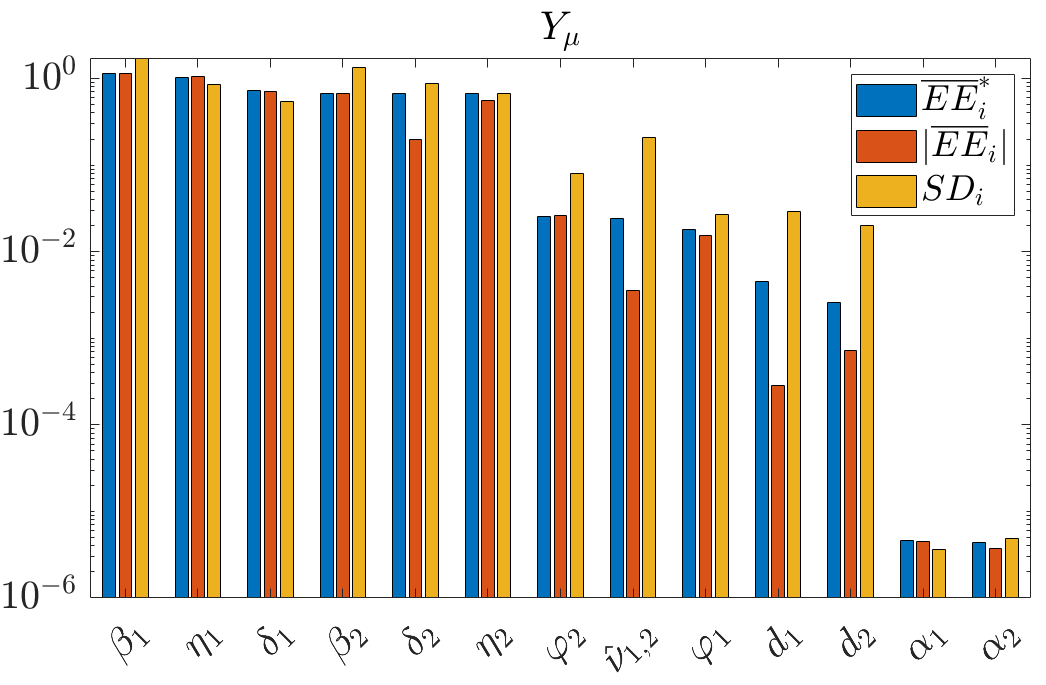}
    \caption{}
    \label{fig:EEmeasures_mu}
  \end{subfigure}
  \vskip\baselineskip
  \stepcounter{row}
  \setcounter{subfigure}{0}
  \begin{subfigure}[b]{0.45\textwidth}
    \centering
    \includegraphics[width=0.95\textwidth]{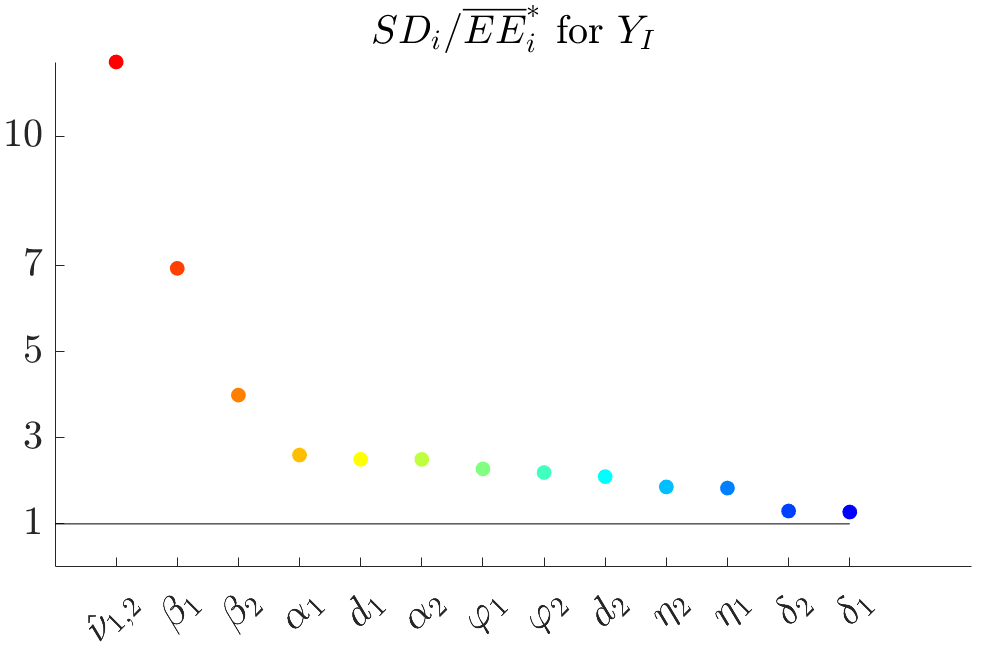}
    \caption{.}
    \label{fig:EEratio_rho}
  \end{subfigure} \quad
  \begin{subfigure}[b]{0.45\textwidth}
    \centering
    \includegraphics[width=0.95\textwidth]{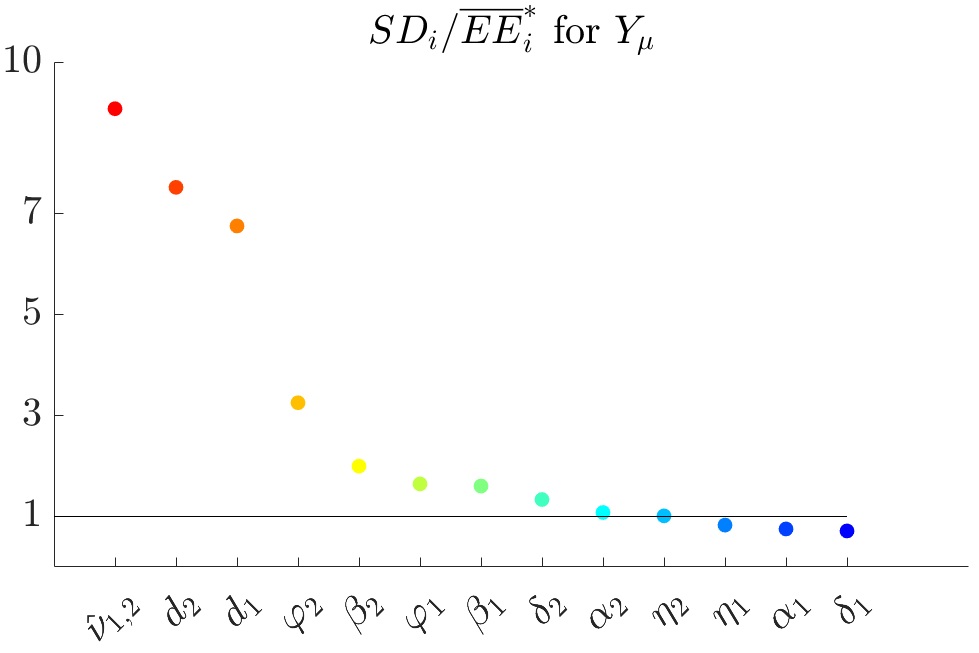}
    \caption{}
    \label{fig:EEratio_mu}
  \end{subfigure}
   
  \caption{\textbf{Results of the Elementary effects method for GSA.} EE measures associated with the model output $Y_{I}$ (a.$1$-$3$) and $Y_{\mu}$ (b.$1$-$3$), \textit{i.e.} the total tumour mass and average mean phenotypic state of the sites.  
  The estimates are computed with $r=500$ evaluations of the elementary effect for each parameter.  The elementary effects $EE_i$, defined in~\eqref{EEi}, calculated for each parameter are displayed in plots (a.$1$) and (b.$1$). The associated measures $\overline{EE}_i^*$, $|\overline{EE}_i^*|$ and $SD_i$, defined in~\eqref{EEmesure}, are shown in plots (a.$2$) and (b.$2$).  Plots (a.$3$) and (b.$3$) present the ratio between $SD$ and $\overline{EE}_i^*$.}
  \label{fig:EEindices}
\end{figure}

Figure~\ref{fig:EEindices} displays the results of EE method for GSA, from which we remark the following:
\begin{enumerate}[1)]
    \item Figures~\ref{fig:EEratio_rho} and \ref{fig:EEratio_mu} display the ratios between $SD$ and $\overline{EE}^*$ for the two model outputs. They both display non-linearity and/or interactions between the parameters, consistently with the non-linear nature of the model. In particular, for both outputs the migration rate $\hat\nu_{1,2}$ seems to have a highly non-linear effect and/or to strongly interact with the other factors, consistently with its key role in connecting they dynamics in the two sites.
    \item For $Y_{I}$ the indices $\overline{EE}^*$ and $\overline{EE}$ are in agreement (Figures~\ref{fig:EEmeasures_rho} and \ref{fig:EEi_rho}), indicating a lack of cancellation effects and thus a monotonic nature of the input-output response. This is not the case for $Y_{\mu}$ (Figures~\ref{fig:EEmeasures_mu} and \ref{fig:EEi_mu}) for which many parameters show both negative and positive $EE_i$, suggesting a non monotonic behaviour. 
    \item Consistently with the analytical results of Section~\ref{par:analytical results}, in Figure~\ref{fig:EEmeasures_rho} we can observe that parameters $\delta_i$, $d_i$ and $\eta_i$ ($i=1,2$) have the largest impact on the total cell number $Y_{I}$. Figure~\ref{fig:EEi_rho} shows that the $EE_i$ evaluations for $d_i$ and $\eta_i$ are negative, while those for $\delta_i$ are positive, in agreement with equations~\eqref{eq:fitness_factors} and \eqref{metastatic:y:rho}.
    \item In Figure~\ref{fig:EEmeasures_mu} we can observe that parameters $\delta_i$, $\beta_i$, and $\eta_i$ ($i=1,2$) exhibit the largest impact on the mean phenotypic trait average $Y_{\mu}$, which is consistent with the analytical results of Section~\ref{par:analytical results}. In particular, in Figure~\ref{fig:EEi_mu} we observe that most of the $EE_i$ evaluations for $\eta_i$ are positive, in agreement with equations~\eqref{eq:fitness_factors} and \eqref{metastatic:y:y} in which $y_i$ is proportional to the maximum cytotoxic death rate.  
    \item It is interesting to notice that $\beta_i$ ($i=1,2$) emerges as a significant input parameter for $Y_{\mu}$, with its $EE_i$ evaluations being all positive (cf. Figure~\ref{fig:EEi_mu}). 
    This finding, which could not emerge from the analytical results due to the asymptotic regime considered in Section~\ref{sec:analysis}, reasons with the fact that higher values of $\beta_i$ should correlate with a larger variance of the phenotypic distributions -- \textit{e.g.} as seen in~\citep{villa2021evolutionchemo} -- which may shift the mean of the distribution in the bounded domain $[0,1]$.  
    \item We note that the migration rate $\hat{\nu}_{1,2}$ has negligible impact on both outputs, likely due to the small value it may take in the admissible range for the GSA.  Considering a wider range of values, \textit{i.e.} $\hat{\nu}_{1,2}\in[10^{-12},10^{-5}]$,  we observed an increase in the importance of this parameter, as expected, particularly for $Y_{\mu}$ (cf. Figure S.4).  Nonetheless, a highly non-linear effect is still observed in both parameter ranges. 
    \item The Michaelis-Menten coefficients $\alpha_i$ ($i=1,2$) exhibit a really low impact on both outputs, likely due to the large dose of drug injected.  In fact, considering a lower drug dose leads to an increase in the importance of $\alpha_i$, along with a decrease in the importance of the maximum death rate $\eta_i$ (cf. Figure S.5).  This suggests that at low drug doses, increasing the cytotoxic efficiency of the compound will not yield significant therapeutic improvements.  
\end{enumerate}

\subsubsection{Results of Sobol method for GSA}
\begin{figure}
    \centering
    \begin{subfigure}[b]{0.45\textwidth}
        \includegraphics[width=\textwidth]{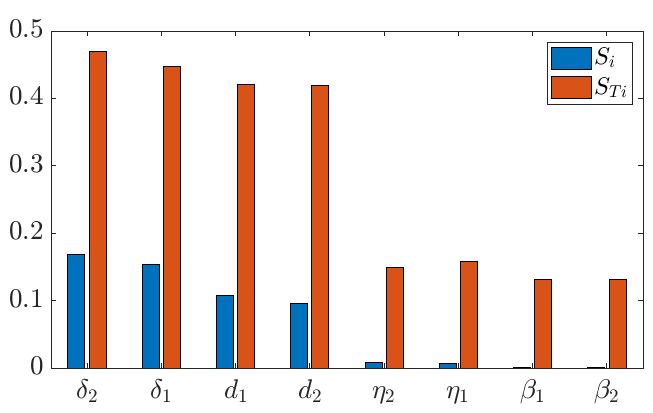}
        \caption{Model output $Y_{I}$.}
    \end{subfigure} \qquad
    \begin{subfigure}[b]{0.45\textwidth}
        \includegraphics[width=\textwidth]{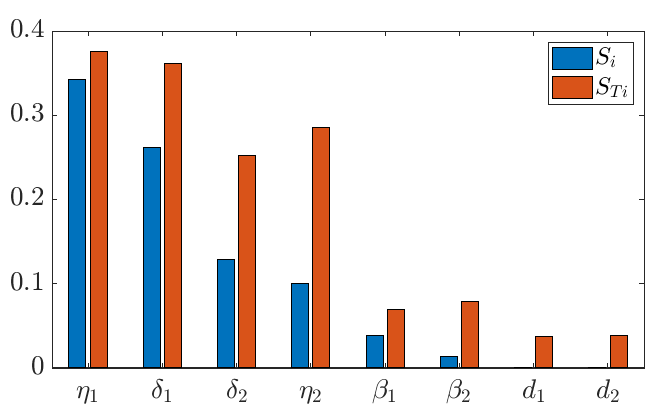}
        \caption{Model output $Y_{\mu}$.}
    \end{subfigure}
    \caption{\textbf{Results of the Sobol method for GSA.}  Sobol measures $S_i$ and $S_{Ti}$, defined in~\eqref{sobol}, associated with the model output $Y_{I}$ (a) and $Y_{\mu}$ (b),  \textit{i.e.} the total tumour mass and average mean phenotypic state of the sites.  The estimates are computed from $5000$ sample points in the input space.}
    \label{fig:SBmeasures}
\end{figure}

We conduct further GSA following the Sobol strategy on the most influential parameters highlighted by the results obtained with the EE method.  
We thus consider a shorter input vector $Z = (\beta_1,\beta_2,\delta_1,\delta_2,\eta_1,\eta_2,d_1,d_2)$ and fix the rest of the parameters to the values of Table~\ref{tab:bounds_tumour_param}. 
The obtained results, illustrated in Figure~\ref{fig:SBmeasures}, are generally consistent with the findings of the GSA following the EE method. In addition, severable notable observations can be made:
\begin{itemize}
    \item[8)] In contrast to the results obtained from the EE method, $\beta_i$ ($i=1,2$) appear to have relatively low importance for $Y_{\mu}$. Moreover their $S_{Ti}$ are significantly bigger than the $S_i$, implying that they mainly act in interaction with the other parameters. This suggests that higher rates of phenotypic changes, and thus intra-population heterogeneity, imply greater complexity and non-linearity.
    \item[9)] Parameters associated with the primary tumour exhibit higher values of $S_i$. This may be due to the fact that we consider a metastatic spread scenario, where the primary tumour affects the metastasis, but not the other way around.
    \item[10)] For $Y_{I}$, the significantly higher values of $S_{Ti}$ compared to those of $S_i$ provide further evidence of the importance of interactions among these parameters. Meanwhile, for $Y_{\mu}$ the $S_i$ values occupy a considerable proportion of $S_{Ti}$, showing that the  main effects of $\eta_i$ and $\delta_i$ are more important than their interactions. 
\end{itemize}

\subsection{Numerical simulations}\label{sec:numerical_results}
Given the findings of the GSA, we further explore how the most impactful parameters affect the model outcome. 
Specifically, we investigate how different tumour environments may influence possible treatment outcomes, and how different factors or biological processes may influence the evolutionary timeline of the tumours under treatment.
        
\subsubsection{Highly communicating sites with different tissue-to-plasma partition coefficients}

In this section we take into account two different scenarios, referred to as the baseline and non-baseline scenarios,  which take different values of the maximum migration rate $\hat\nu_{1,2}$ and the partition coefficients of the two sites $K_i$ ($i=1,2$), as summarised in Table~\ref{tab:baseline&nonscenarios}.  
\begin{table}[h]
    \centering
    \begin{tabular}{c|l|l|l|l|l|l}
      & \multicolumn{2}{c|}{\textbf{Baseline scenario}} & \multicolumn{2}{c|}{\textbf{Non-baseline scenario}} & \multicolumn{2}{c}{\textbf{}}\\
    \midrule
    Parameter & $i=1$  & $i=2$  & $i=1$ & $i=2$ & Unit & Ref \\
    \midrule
    $K_{i}$ & $0.8$ & $0.5$ & $1$ & $10^{-4}$  & - & \cite{mittapalli2013mechanisms} \\
    $\hat\nu_{i,j}$  & $1.3\cdot10^{-10}$  & $0$ & $1.3\cdot10^{-5}$ & $0$ & s$^{-1}$ & \cite{franssen2019modelmetastases,scott2013mathematical}\\
    \bottomrule
    \end{tabular}
    \caption{Cancer physiological and evolutionary dynamics parameter values for the baseline and non-baseline scenarios. $K_i$ represents the partition coefficient of site $i$, and $\hat \nu_{i,j}$ the maximum migration rate from site $i$ to site $j$, where $i=1$ is the primary tumour and $i=2$ the metastatic site.}
    \label{tab:baseline&nonscenarios}
\end{table}
Specifically, the non-baseline one is an extreme scenario in which we significantly increase the migration rate from primary tumour to the metastasis to a value out of the range reported in Table~\ref{tab:bounds_tumour_param}. 
We do this for illustrative purposes, as from the results of the GSA we expect to be able to more easily observe the potential impact of the communication between sites by considering parameter values outside this range. 
In addition, we vary the partition coefficients of the tissues in the two sites, in order to consider how significantly different tumour environments in different sites may affect the treatment outcome in metastatic tumours. 
In fact, the parameter $K_{i}$ indicates the degree of tissue drug accumulation, attributed to phenomena such as protein binding, lysosomal trapping, and lipid dissolution \cite{jones2013pbpk}, and a smaller value of $K_2$ compared to that of $K_1$ implies a lower drug concentration will accumulate in the metastatic site compared to the primary.  
In order to mainly focus on the impact of these biological factors on the treatment outcome, the other parameters are fixed to their reference values of Tables~\ref{tab:pk_params} and \ref{tab:bounds_tumour_param}, and are thus the same in both scenarios.

\begin{figure}[h] 
    \centering
    \begin{subfigure}[b]{0.45\textwidth}
        \centering
        \includegraphics[width=\textwidth]{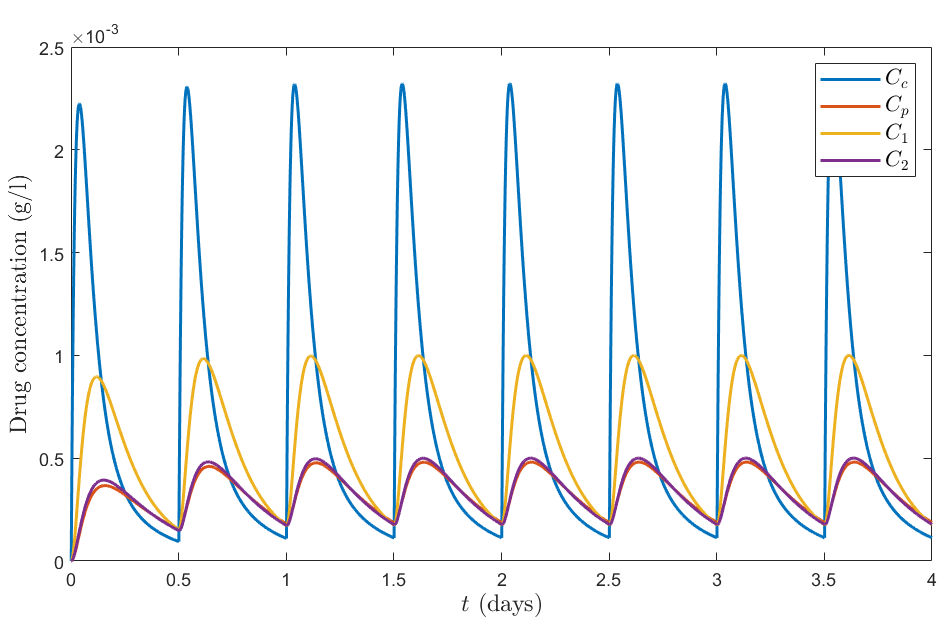}
        \caption{Baseline scenario.}
    \end{subfigure} \qquad 
    \begin{subfigure}[b]{0.45\textwidth}
        \centering
        \includegraphics[width=\textwidth]{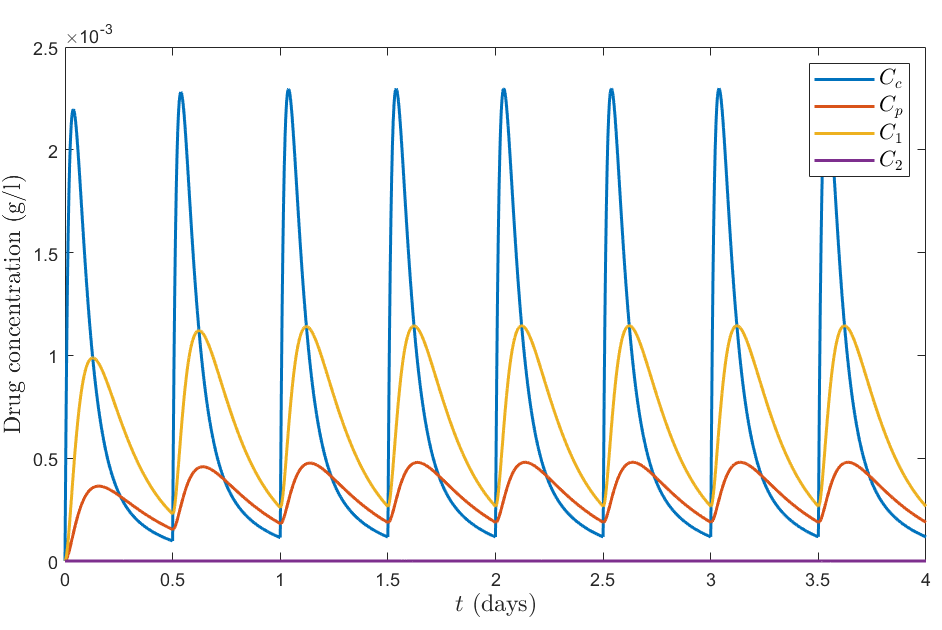}
        \caption{Non-baseline scenario.}
    \end{subfigure}
    \caption{\textbf{Drug concentrations under baseline and non-baseline scenarios.}
    Pharmacokinetics results of the numerical simulations under the baseline (a) and  non-baseline (b) scenarios with the drug schedule of $150$ mg orally administered twice a day, and a final time $T=4$ days.  For each scenario we plot the drug concentration in the central (blue), peripheral (orange), primary tumour (yellow) and metastatic tumour (purple) compartments.  
    More details on the simulation set-up and numerical methods can be found in Section~\ref{sec:setup}. 
       }
    \label{fig:pk_variables}
\end{figure}

\paragraph{Different \textit{in situ} drug concentrations.} 
Figure~\ref{fig:pk_variables} displays the drug concentration in four compartments, excluding the administration site for visualisation purposes. 
First of all, we remark that the peak of the plasma compound concentration ($C_c$) is reached at the time $t_{max} =0.93$ h, and with a value of $C_{c,max}=0.0022$ g/l, of the same order of magnitude of the values found in literature~\cite{ema2013tafinlar,fda2013dabrafenib}.
Moreover, the drug concentration in the metastatic site ($C_2$) is significantly lower in the non-baseline scenario compared to the baseline one, reaching a maximum value of the order of $10^{-7}$ g/l while in the baseline scenario it reaches higher peaks comparable to those reached by the drug concentration in the primary site ($C_1$). 
This is consistent with the biological meaning of $K_i$, and highlights how different tissue-to-plasma partition coefficients may result in different drug concentrations in the two tumour sites, creating substantial discrepancies between the two tumour environments. 

\begin{figure}
        \centering
        \begin{subfigure}[b]{\textwidth}
            \centering
            \includegraphics[width=0.88\textwidth]{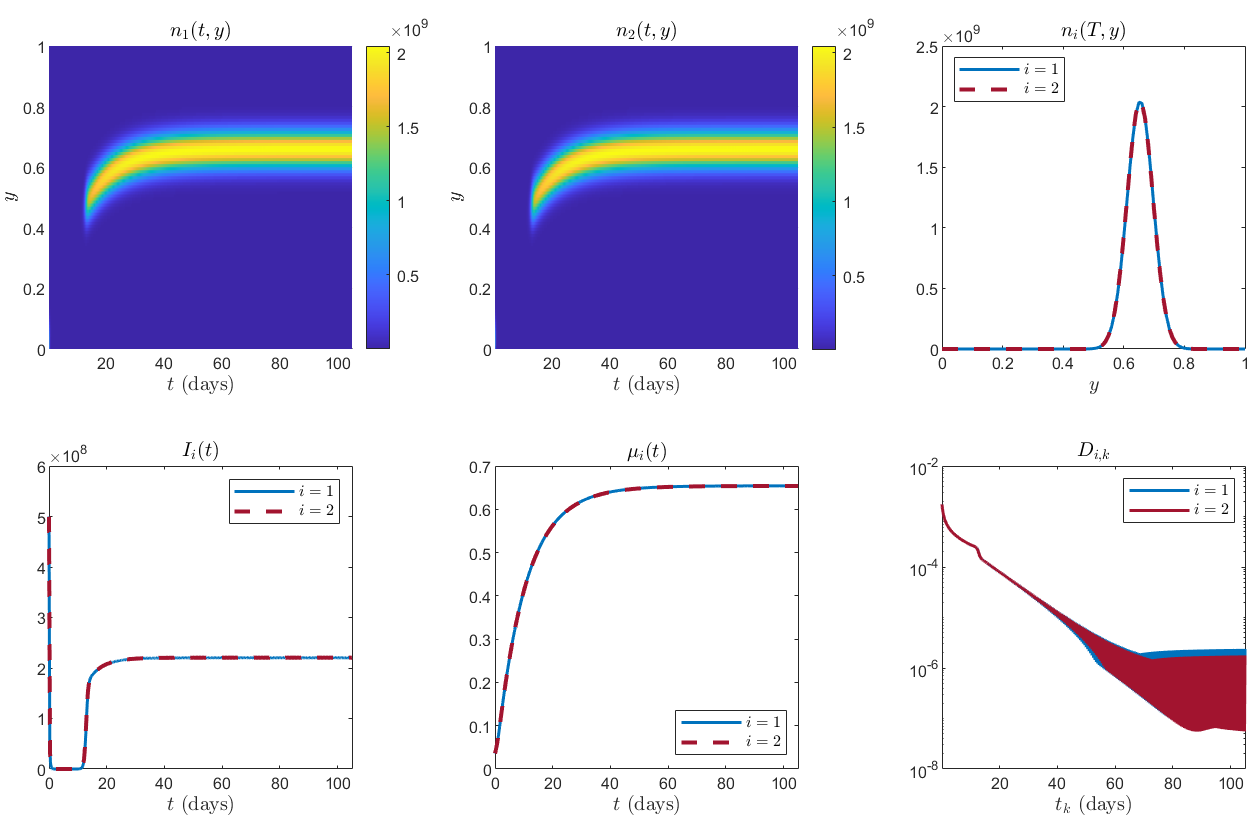} 
        \caption{Baseline scenario}
        \label{fig:evol_dynam_baseline}
        \end{subfigure}
        \begin{subfigure}[b]{\textwidth}
            \centering
            \includegraphics[width=0.88\textwidth]{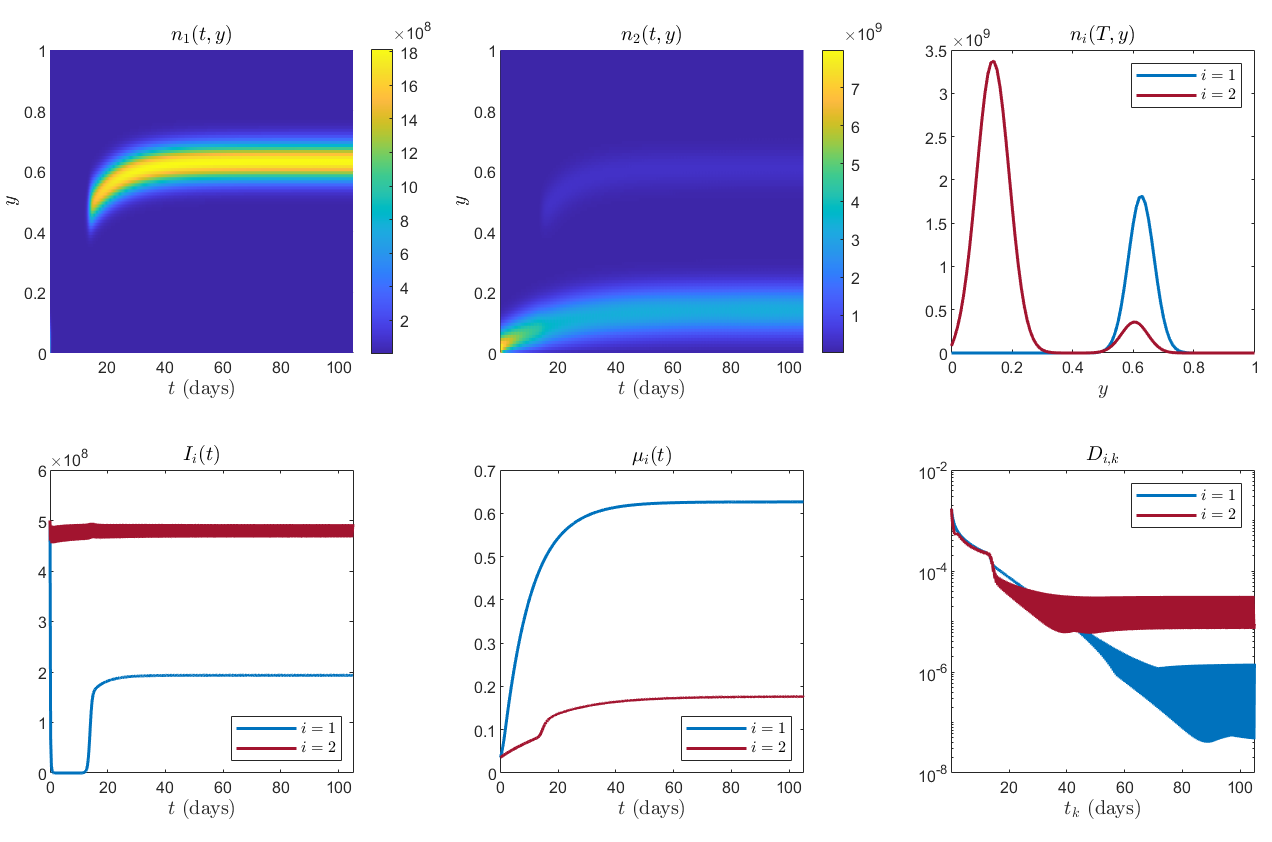} 
        \caption{Non-baseline scenario}
        \label{fig:evol_dynam_nonbaseline}
        \end{subfigure}
        \caption{\textbf{Cancer evolutionary dynamics under baseline and non-baseline scenarios.} Cancer evolutionary dynamics results of the numerical simulations under the baseline (a) and  non-baseline (b) scenarios with the drug schedule of $150$ mg orally administered twice a day, and a final time $T=105$ days. 
For each scenario we plot in the first row the phenotypic distribution over time in the primary tumour (first panel) and the metastatic tumour (second panel), and compare them at the final time $T$ (third panel), and we plot in the second row the population size (first panel), mean phenotypic trait (second panel) and step difference $D_{i,k}$ (third panel), as respectively defined in equations~\eqref{distribu_phenoI}, \eqref{distribu_pheno} and~\eqref{steady-state}. More details on the simulation set-up and numerical methods can be found in Section~\ref{sec:setup}. }
        \label{fig:baseline&non_scenarios_results}
\end{figure}

\paragraph{Different evolutionary outcomes of the tumours.} 
Such a difference in drug concentration in the two sites results in different evolutionary dynamics of the two tumour populations,  unlike in the baseline scenario where no significant difference is observed between the phenotypic distributions in the two sites, as illustrated in Figure~\ref{fig:baseline&non_scenarios_results} comparing the long-term evolution of the tumour populations in the two scenarios. 
In particular, in the non-baseline scenario we observe that the lower drug concentration accumulating in the metastatic site results in a population mostly characterised by lower levels of drug resistance compared to that in the primary site, exposed to a higher compound concentration.  Nonetheless,  the high migration rate results in a small subpopulation in the metastatic site presenting levels of resistance comparable to those developed in the primary tumour.  
In the baseline scenario, cell migration has no meaningful impact on the phenotypic distribution given the similarity of the tumour environments, consistently with the analytical results illustrated in Figure~\ref{analysis:parametersaffectpopulation} (second column).

\paragraph{Consistency with analytical results.}  The obtained results are consistent with the analytical study of Section~\ref{sec:analysis}. In fact, from definitions~\eqref{eq:fitness_factors}, by significantly decreasing $C_2$ we obtain a lower fittest phenotypic trait $h_2$ and selective gradient $b_2$, and a higher maximum fitness $a_2$. Following \eqref{metastatic:y:y}, this results in a lower selected trait $y_2$ and a higher metastatic population size $I_2$, which is exactly what we detect in Figure~\ref{fig:baseline&non_scenarios_results}. Given the considerably different selected traits $y_1$ and $y_2$, and the increased migration rate, the population density of the metastatic site shows polymorphism, in accord with equations~\eqref{sol:metastatic:y:n} and \eqref{metastatic:y:rho}.  Moreover, we detect a slight decrease in both the primary tumour population size and selected trait,  in agreement with equation~\eqref{metastatic:y:rho} considering that the order of magnitude of the migration rate is lower than that of $b_1$.

\paragraph{Sensitivity of cancer evolutionary dynamics to temporal oscillations in drug concentrations.} 
Indeed temporal oscillations in the \textit{in situ} compound concentration, due to the drug administration schedule, imply temporal oscillations in the moments of the phenotypic distribution during treatment, as can be better observed by plotting the solutions of Figure~\ref{fig:baseline&non_scenarios_results} over just two days (cf. Figure S.6).  
As already clear from Figure~\ref{fig:baseline&non_scenarios_results}, the great variability of the $I_2$ curve in the non-baseline scenario indicated that the metastatic population size is considerably susceptible to the drug concentration variation.  
In fact, it can be detected from the zoomed-in plots that under both scenarios $I_i$ is the most sensitive to the drug concentration variations, while both $\mu_i$ and $\sigma^2_i$ are substantially less susceptible.  
This may be explained by noticing that while the pharmacokinetics are in the scale of hours, the cancer evolutionary dynamics, \textit{i.e.} the phenotypic adaptation of the cancer cell populations to the local environments, are in the scale of days.  
Nonetheless, we remark that the fluctuations of these quantities are negligible in the baseline scenario and the primary site in the non-baseline scenario, where their amplitude is small compared to the average value of the respective quantity.  This is not as much the case in the metastatic site of the non-baseline scenario,  characterised by a lower average drug concentration and thus higher sensitivity to small variations of this quantity.  We thus conclude that a higher administered drug dose may result in a lower evolutionary dynamics sensibility to the compound variability.  

\subsubsection{Oral administration vs intravenous injection}

In the previous section we remarked that under the drug administration regimen of a $150$ mg oral dose twice a day, in the baseline scenario, the impact of the drug concentration oscillations on the evolutionary dynamics and overall therapeutic outcome is negligible.  We thus wish to conduct further numerical investigations considering the equilibrium phenotypic distribution to which the system converges at the end of treatment, as well as the time it takes to reach this.  
However, considering oral drug administration the system does not reach numerical steadiness for either scenario, as can be seen from the lower-right panels of Figure~\ref{fig:baseline&non_scenarios_results} displaying the step-difference $D_{i,k}$ introduced in equation~\eqref{steady-state}, which becomes approximately periodic with a $12$ h period (cf. zoom-in plot in Figure S.6), \textit{i.e.} the time interval between two consecutive drug administrations.  
This would not be the case if we considered a constant intravenous drug injection. 

\begin{figure}[h]
    \centering
        \includegraphics[width=\textwidth]{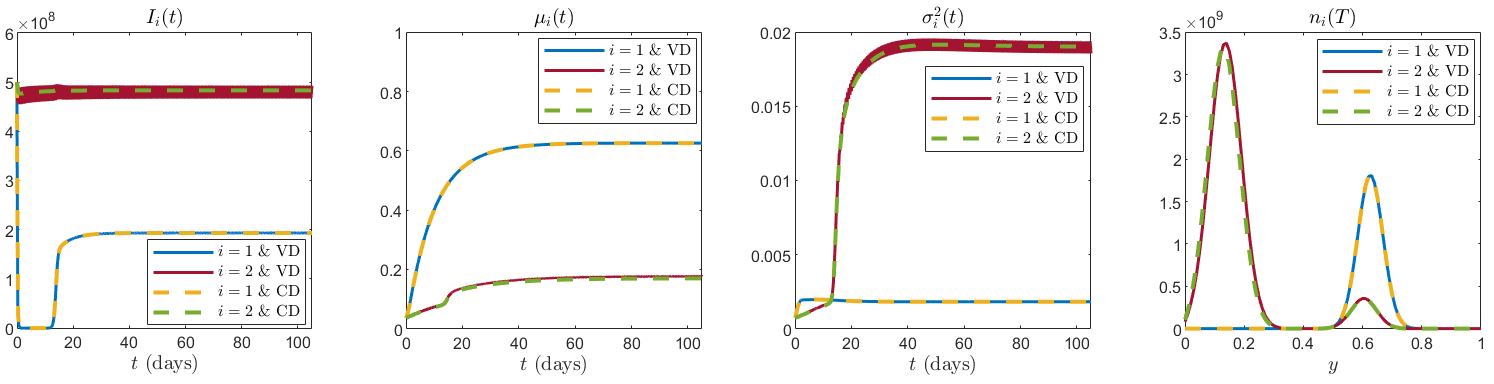}
    \caption{\textbf{Oral administration vs intravenous injection under non-baseline scenario.} Cancer evolutionary dynamics results under the non-baseline scenario and a drug schedule of $150$ mg orally administered twice a day (VD) or constant intravenous injection of $2.6915$ $\mu$g/s (CD), and final time $T=105$ days. From left to right, the panels display the cancer population sizes, the mean phenotypic traits and the corresponding variances over time for the two tumour cell populations, respectively defined in equations~\eqref{distribu_phenoI}, and~\eqref{distribu_pheno}, and the phenotypic distributions at time $T$. More details on the simulation set-up and numerical methods can be found in Section~\ref{sec:setup}. }
    \label{fig:constant_drug_dose}
\end{figure}

 Therefore, we aim at finding the intravenously-injected constant drug dose leading to cancer evolutionary dynamics analogous to those obtained with an orally-administered drug dose of $150$ mg twice a day, which we will use in the next section for convenience, acknowledging that Dabrafenib is administered \textit{per os} to patients.
 Given the final distribution obtained with the dose administered twice a day $n_{i}^*(T)$ and the final distribution obtained with a constant intravenously-injected dose $n_{i}^{dose}(T)$, we define the relative error between the two outcomes as
\begin{equation}
            err_{i,dose} = \frac{1}{J}\sum_{j=1}^J\bigg|\frac{n_{i,j}^*(T)-n_{i,j}^{dose}(T)}{n_{i,j}^*(T)}\bigg| \,.
\end{equation}
Minimizing the function $f(dose)=0.5\left( err_{1,dose}+err_{2,dose}\right)$ in the baseline scenario, we obtain that the intravenously-injected drug dose that best reproduces the evolutionary outcomes of the oral administration regimen is of $2.6915$ $\mu$g/s. 
We verified that this optimal dose also provides a good approximation of the dynamics obtained in the non-baseline scenario, as displayed in Figure~\ref{fig:constant_drug_dose}.  Moreover, we report the times $T_{ss,i}$ at which the phenotypic distributions in the two sites reach steadiness, which are larger than the median time of response and below the median Dabrafenib treatment durations reported in~\citep{martin2019effectiveness}. 

\begin{table}[h]
        \centering
        \begin{tabular}{c|c|c}
             & $T_{ss,1}$ & $T_{ss,2}$ \\  \hline
            Baseline & $66.98$ days & $67.02$ days \\ \hline
            Non-baseline & $65.24$ days & $59.16$ days \\
        \end{tabular}
        \caption{Steady state time of the phenotypic distribution of the primary tumour ($T_{ss,1}$) and the metastasis ($T_{ss,2}$) in the baseline and non-baseline scenarios under an intravenously-injected constant drug dose of $2.6915$ $\mu$g/s. The steady state time is the first time at which $D_{i,k}<tol$, with $D_{i,k}$ defined in~\eqref{steady-state} and a tolerance of  $tol=10^{-6}$. }
        \label{tab:Tss_constant_dose}
\end{table}

\subsubsection{Steady-state study of treatment outcomes}

Having identified the dose of constant intravenous drug injection that yields analogous therapeutic outcomes to the oral administration regimen used in the clinic, we now investigate how the therapeutic outcome may be affected by some key factors.  
Specifically, we consider the level of vascularisation of the metastatic tumour, \textit{i.e.} a physiological factor which will regulate the distribution of the compound in the metastatic site, and the most important parameters emerged from the GSA focussing on $Y_{\mu}$ as the model output , \textit{i.e.} $\delta_i$, $\eta_i$ ($i=1,2$) and $\beta_i$. 
This choice is motivated by the fact that the mean phenotypic trait $\mu_i$ in the two tumours gives us an indication of the development of drug resistance during treatment. Moreover, as highlighted in the previous section,  $\mu_i$ is the least affected by the drug concentration variability,  and we thus expect conclusions drawn in this section to hold in both drug delivery regiments with higher certainty.  
We also remark that from the analytical results of Section~\ref{sec:analysis} we expect that the total population in the sites ($I_i$) depends on the selected traits ($y_1$ and $y_2$), but not viceversa.

To test the influence of these factors on the treatment outcome, for each parameter we vary its value in the range defined in Table~\ref{tab:bounds_tumour_param}, and compare the model outputs, \textit{i.e.} the cancer population sizes $I_i$ and mean phenotypic traits $\mu_i$ in the two sites, after $30$ weeks of continuous intravenous constant drug administration, \textit{i.e.} the average length of Dabrafenib treatment~\citep{martin2019effectiveness}. 
For each simulation we also report on when the drug concentration and the phenotypic distributions in the two sites reach steady state, as it gives insights in how the different input factors affect the speed of the cancer cell adaptation to the drug regimen.  This also suggests reasonable timeframes in which to assess the advancement of the therapeutic intervention. 

 \begin{figure}
          \centering
         
          \begin{subfigure}[b]{0.85\textwidth}
            \centering
            \includegraphics[width=\textwidth]{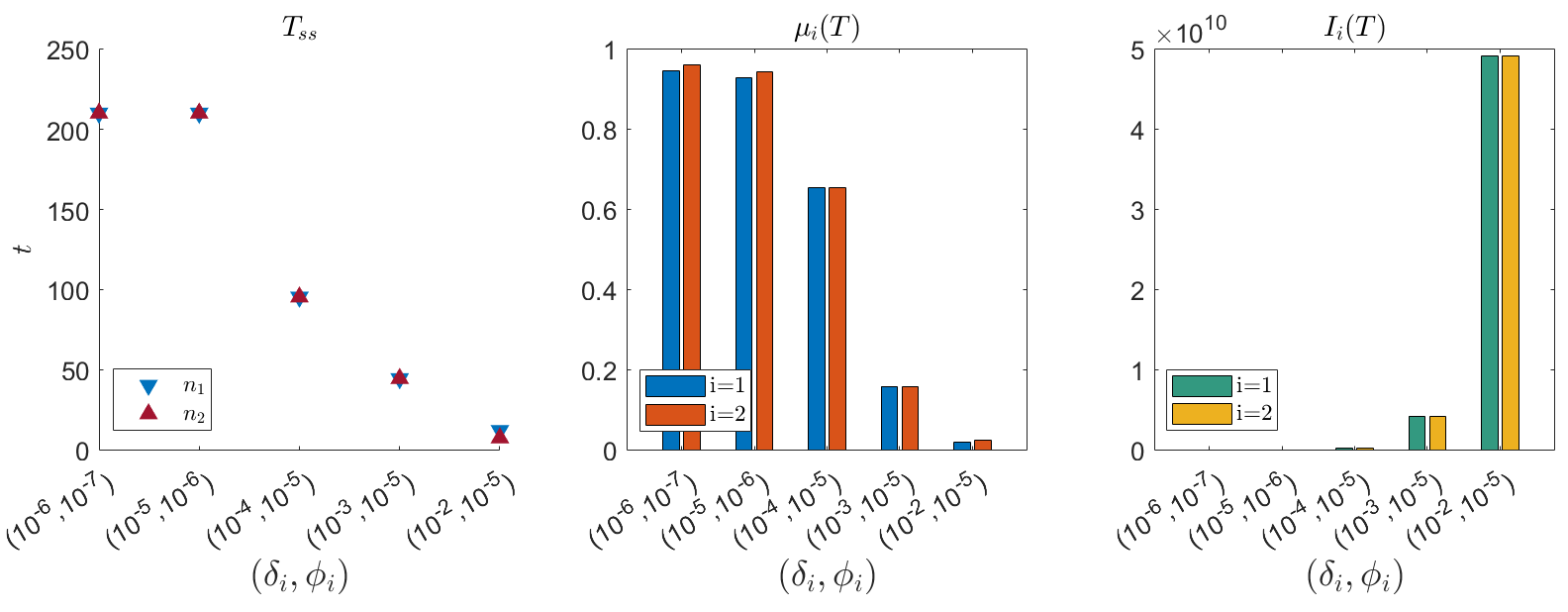}
            \caption{Maximal background fitness}
            \label{fig:sensitivity_delta}
          \end{subfigure}
          
          \begin{subfigure}[b]{0.85\textwidth}
            \centering
            \includegraphics[width=\textwidth]{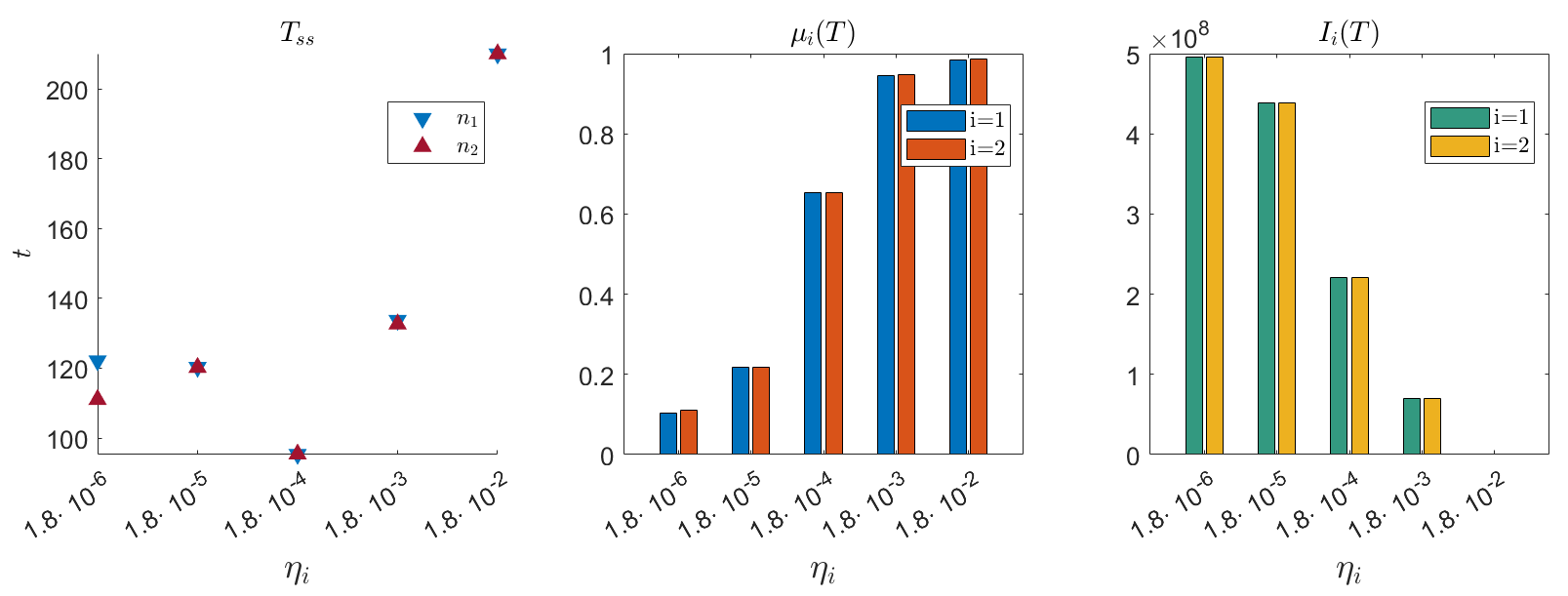}
            \caption{Maximum cytotoxic death rate}
            \label{fig:sensitivity_eta}
          \end{subfigure}
          
           \begin{subfigure}[b]{0.85\textwidth}
            \centering
            \includegraphics[width=\textwidth]{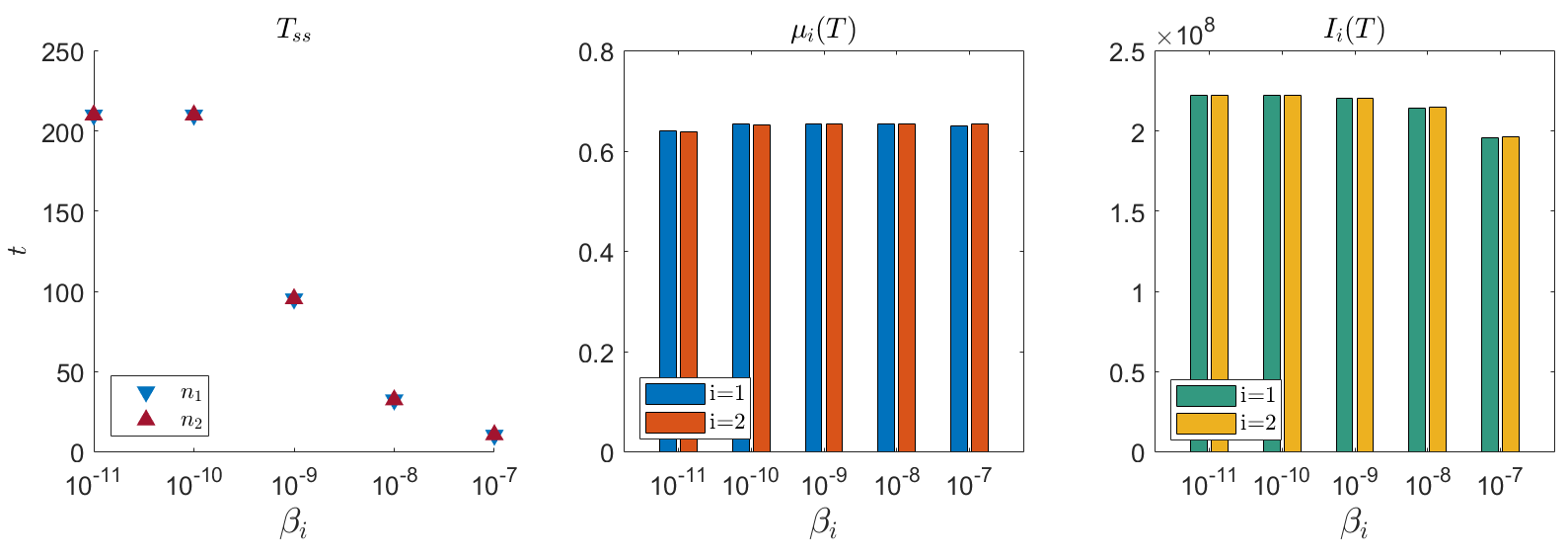}
            \caption{Epimutation rate}
            \label{fig:sensitivity_beta}
          \end{subfigure}
    \caption{\textbf{Treatment outcome dependency on $\delta_i$, $\eta_i$ and  $\beta_i$.} 
Treatment outcomes obtained from numerical simulations of the model under the baseline scenario, for an intravenously-injected drug dose of $2.6915$ $\mu$g/s and a final time of $T=210$ days. The graphs display the steady-state times (first panel), mean phenotypic traits (second panel) and tumour sizes (third panel) at steady state, varying: a) the maximal background fitness $\delta_i$, and $\phi_i$ to ensure the assumption $\delta_i\gg\phi_i$ still holds; b) the maximum cytotoxic death rate $\eta_i$; c) the epimutation rate $\beta_i$.  
The steady state time is the first time at which $D_{i,k}<tol$, with $D_{i,k}$ defined in~\eqref{steady-state} and a tolerance of  $tol=10^{-6}$. 
 More details on the simulation set-up and numerical methods can be found in Section~\ref{sec:setup}.    
}
    \label{fig:sensitivity}
    \end{figure}

\paragraph{Fitness-related parameters.} 
Figure~\ref{fig:sensitivity_delta} and Figure~\ref{fig:sensitivity_eta} display the model sensitivity to the maximum background fitness for highly proliferating yet sensitive cells, $\delta_i$ ($i=1,2$), and the maximal death rate of sensitive cells, $\eta_i$ ($i=1,2$), respectively. 
We notice that incrementing $\delta_i$ speeds up the adaptive dynamics, increases the cancer cell population sizes, while decreases the mean phenotypic traits.  
Meanwhile, consistently with its biological meaning, we detect that incrementing $\eta_i$ decreases the cancer cell population sizes, while increases the mean phenotypic traits, but has no consistent impact on the adaptation speed. All these findings are in line with previous analytical results in the literature~\cite{almeida2019optimizationchemo,villa2021evolutionchemo}.

\paragraph{Epimutation rate.} Since the effects of the epimutation rates $\beta_i$ ($i=1,2$) could not emerge from the asymptotic analysis of Section 3, and the results of the GSA yield contrasting information on the importance of these parameters, we consider the results of the numerical steady-state investigation, reported in Figure~\ref{fig:sensitivity_beta}.  
We immediately notice that as $\beta_i$ take greater values the phenotypic distributions $n_i$ reach steadiness faster, implying a greater speed of adaptation.  
On the other hand, in the parameter range considered $\beta_i$ do not appear to affect $I_i$ and $\mu_i$, as suggested by the results of the Sobol method for GSA, although increasing the epimutation rates results in a higher variance of the phenotypic distribution (cf. Figure S.7), resulting in a more heterogeneous population.  
These findings are consistent with analytical results in the literature~\cite{almeida2019optimizationchemo,villa2021evolutionchemo}, highlighting that higher epimutation rates increase phenotypic diversity, in turn speeding up natural selection.

\begin{figure}[h]
        \centering
        \begin{subfigure}[b]{0.45\textwidth}
            \includegraphics[width=\textwidth]{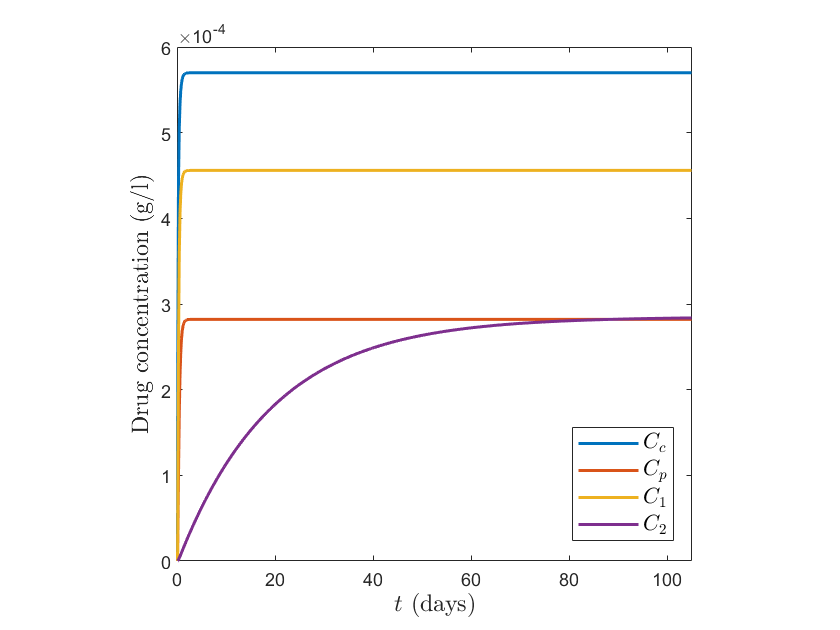}
            \caption{Drug concentrations for $Q_{2}/V_2= 2 \cdot 10^{-3}$ 1/h.}
            \label{fig:Pk_var_perfusivity}
        \end{subfigure}
        \begin{subfigure}[b]{0.45\textwidth}
            \includegraphics[width=\textwidth]{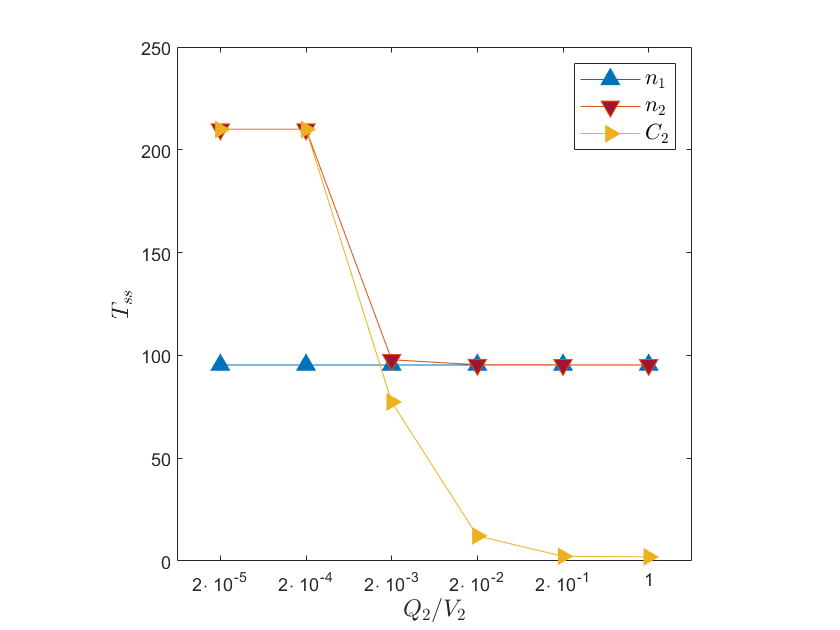}
            \caption{Steady-state times for varying $Q_{2}/V_2$.}
            \label{fig:Tss_perfusivity}
        \end{subfigure}
    \caption{\textbf{Evolutionary timescale dependency on the vascularisation of the metastatic site. } 
    Time evolution of the drug concentration and steady-state times obtained from numerical simulations of the model under the baseline scenario, for an intravenously-injected drug dose of $2.6915$ $\mu$g/s, and final time $T=210$ days, varying the blood flow-to-volume ratio in the metastatic site $Q_{2}/V_2$.  
 In (a) we set $Q_{2}/V_2= 2 \cdot 10^{-3}$ 1/h,  leaving the other parameters at their reference value, and display the drug concentration in the central (blue), peripheral (orange), primary tumour (yellow) and metastatic tumour (purple) compartments, up to day 105. In (b) we display the steady-state times for $n_1$ (blue), $n_2$ (red) and $C_2$ (yellow) for different values of the ratio $Q_{2}/V_2$, specifically $Q_{2}/V_2\in\{2\cdot10^{-5},2\cdot10^{-4},2\cdot10^{-3},2\cdot10^{-2},2\cdot10^{-1},1\}$ 1/h. 
 The steady state time is the first time at which $D_{i,k}<tol$, with $D_{i,k}$ defined in~\eqref{steady-state} and a tolerance of  $tol=10^{-6}$.  
 More details on the simulation set-up and numerical methods can be found in Section~\ref{sec:setup}.    
 }
    \label{fig:perfusivity}
    \end{figure}

\paragraph{Vascularisation of the metastatic site.} Lastly, we study the effect of the level of vascularization of the metastatic site on the evolutionary timeline during treatment by varying the blood flow-to-volume ratio, \textit{i.e.} $Q_2/V_2$.  
From Figure~\ref{fig:perfusivity} it is evident that a lower blood flow-to-volume ratio delays the time at which the \textit{in situ} compound concentration reaches its equilibrium value.  
This can be seen in Figure~\ref{fig:Pk_var_perfusivity} for $C_2$ compared to the drug concentration in the other model compartments, and it is consistent with the biological interpretation of these factors: given that there is less blood entering the metastatic site, and consequently less compound, it takes much more time for the tissue to fill up.  
We remark that for $Q_2/V_2\geq 2 \cdot 10^{-2}$ 1/h, \textit{i.e.} for particularly well vascularised metastatic tumours, the steady-state time of the local phenotypic distribution remains constant, as can be observed in Figure~\ref{fig:Tss_perfusivity}. Nonetheless,  for lower values of the ratio $Q_2/V_2$ we observe that the steady-state time for $n_2$  significantly increases, indicating that for particularly low levels of tumour vascularisation the slower drug perfusion delays the cancer evolutionary dynamics. 

\section{Conclusions}\label{sec:conclusion}

\subsection{Summary of major results}
In this paper, we conducted a mathematical study of the evolutionary dynamics of metastatic tumours under chemotherapy. Specifically, our approach involves formal asymptotic analysis and numerical simulations of a system of non-local PDEs describing the phenotypic evolution of the tumour cells coupled with a system of ODEs modelling the cytotoxic drug pharmacokinetics. 

Focussing on a biological scenario allowing for metastatic spread but no secondary or self-seeding, the analytical results indicate that, while the primary tumour will evolve into a monomorphic population with trait selected according to local environmental conditions, the metastasis may display polymorphism if local environmental conditions differ from those in the primary site, with a subpopulation of cells presenting traits selected in the primary tumour.  
Moreover, the size of this subpopulation increases as the connectivity of the sites and the migratory ability of cells selected in the primary tumour increases, or as the selective pressure in the metastatic site decreases.  

We performed global sensitivity analysis and long-term numerical simulations of the full system, to verify the analytical results outside the asymptotic regime of rare phenotypic changes due to epimutations, and to better understand the consequences of cancer adaptive dynamics on the outcome of treatment.
To do so we adopted a case study of metastatic melanoma under Dabrafenib monotherapy, through which we could gain deeper insights into the impact of tissue-specific physiological parameters and drug kinetics on the evolution of the metastatic cancer.  
We found that the tumour stage and its location may influence \textit{in situ} drug distribution during the course of treatment~\cite{ziemys2018progression}. In particular, the tumour angiogenic switch and the organ characteristics may affect the overall level of tumour vascularisation, regulating the timeline of local drug delivery. Moreover, the tissue drug affinity of the tumour, \textit{i.e.} the tissue-to-plasma partition coefficient, regulates the maximum local concentration of cytotoxic compound. 
As a consequence,  discrepancies in drug concentrations between the two tumour sites may arise during treatment, resulting in differing local environments to which cells will adapt. 
In line with the results of the asymptotic analysis,  this fosters phenotypic diversity among tumours at different locations and the emergence of distinct subclones~\cite{tsuruo1981differences,yancovitz2012intra}. 

In particular, assuming that the primary tumour has significantly increased its vascular supply prior to undergoing metastatic spread, the higher local concentration of the cytotoxic drug will result in the selection of more resistant phenotypic traits in the primary tumour. 
Conversely, assuming the metastasis is yet to undergo substantial angiogenesis for further dispersal, the expected lower vascular supply of the secondary tumour should result in a lower drug concentration at this site. This correlates with a lower selective pressure in the metastatic site which, together with the selection of cells with higher migratory abilities in the primary tumour,  favours the emergence of a subpopulation of highly drug-resistant cells in the metastasis, in agreement with the results of the asymptotic analysis.  
Overall, this confers further evolutionary advantage to the tumour population, as it will have a higher chance of survival to sudden environmental changes, a big trade off feature of biological dispersal despite leaving the primary site may be a risk for tumour cells. 
This is coherent with empirical studies suggesting that chemotherapy fosters metastatisation and may provide an explanation for the evidence indicating that the metastatic site is often less impacted by chemotherapy~\citep{dalterio2020paradoxical,massague2016metastatic,obenauf2015therapy,parker2022currentchallenges}.

Finally, investigating the impact of drug kinetics on cancer evolution under the twice-a-day oral administration protocol used in the clinic, we observed that the pharmacokinetics and the cancer adaptive dynamics evolve on different timescales, \textit{i.e.} hours vs. days.  This discrepancy implies that the long-term evolutionary outcome of treatment may not be significantly influenced by short-term PK dynamics and fluctuations of the \textit{in situ} drug concentration, but rather by average cytotoxic levels throughout the duration of chemotherapy, especially when the drug dose is particularly high.


\subsection{Limitations and perspectives}
 
  \paragraph{Cell migration and metastatic spread.}
 We chose the migration rate $\nu_{1,2}(y)$ to significantly increase only as the phenotypic state of a cell gets closer to the highest levels of drug resistance, but it would be relevant to consider alternative definitions, in particular non-smooth ones, as they might better reflect the switch in the intravasation ability of cancer cells as they become more aggressive. 
The use of different definitions for $\nu_{1,2}(y)$ in the PDE for $n_1$ and that for $n_2$ would also be relevant in view of the fact that successful intravasation of cancer cells does not necessarily correlate with survival in the circulation and successful extravasation to the metastatic site. 
In fact, circulating tumour cells in the bloodstream may be attacked by the immune system and face the challenge of adhesion to the vascular walls to exit the circulation, and may potentially endow themselves with more resistant traits~\citep{fidler2003pathogenesis}.  
The addition of a PDE describing the evolutionary dynamics of circulating tumour cells may therefore provide a more realistic extension of our modelling framework, which currently relies on the limiting assumption that cells exiting the primary tumour will immediately appear in the metastatic site maintaining their original phenotypic state.  
In addition, the extension of this model to the one including motion between several patches, \textit{e.g.} as in~\cite{mirrahimi2012adaptation},  could be used to investigate more complex dynamics arising in cancers presenting multiple metastatic sites.  
This could allow for the theoretical exploration of the \textit{seed and soil} hypothesis~\citep{fidler2003pathogenesis} of metastatic spread, accounting for different extravasation probabilities to different sites, \textit{e.g.} as in~\cite{franssen2019modelmetastases}. 
Moreover, this mathematical framework may help explain why secondary seeding is much more efficient at spreading cells than dispersal from the primary site~\cite{scott2013self-seeding,vitos2022model},  thanks to promising results of previous studies on host-pathogen interactions~\citep{alfaro2023adaptation,hamel2021adaptation} analysing how pathogen persistence is mediated by the presence of a middle host. 

\paragraph{Cell proliferation and tumour burden.}
 Building upon the strategies presented in \citep{almeida2019optimizationchemo,villa2021evolutionchemo}, in our model we took into account a one-dimensional phenotypic trait whereby higher levels of drug resistance correlate with lower proliferation rates of cancer cells~\citep{chu2004chemotherapy,corrie2011chemotherapy}. 
 Nevertheless, subclones simultaneously exhibiting high proliferation rates and high levels of drug resistance may exist~\cite{sharma2010chromatin}.  
 It would therefore be biologically significant to consider a two-dimensional phenotypic state, as previously done in~\cite{chisholm2015emergence},  to capture the phenotypic evolution of tumours without any \textit{a priori} assumption of a link between the level of drug tolerance and proliferation potential of the cells. 
 This may affect the relation, emergent from the model, between the development of resistance and the evolution of the tumour burden during therapy. 
 On this note, it is important to remark that we included the tumour volume $V_i$ as a physiological parameter in the PK model, but kept this quantity constant throughout the simulations, for the sake of simplicity.  Indeed the tumour volume is dependent on the number of cancer cells composing it, and it would therefore be significant to allow $V_i$ to vary in time as a function of the cell number $I_i$ in future work. This would allow to account for the impact on the drug delivery of changes in the tumour burden expected to occur during chemotherapy.

\paragraph{Therapeutic strategy and personalised treatment.} 
 Given the novelty of this work, we here restricted our attention to a case study of BRAF-mutated melanoma under Dabrafenib monotherapy.  In the clinic, however, this is usually combined with Trametinib~\cite{long2017dabrafenib}, a kinase inhibitor of MEK protein adopted to prevent the development of resistance to the BRAF inhibitor~\citep{luebker2019brafinhibitor}.  
Therefore, future work investigating the development of drug resistance in metastatic melanoma under treatment should extend this modelling framework to include a  PK model for Trametinib, as they did in~\cite{sun2016mathematical}, and its drug resistance-inhibiting effect on cancer cells, \textit{e.g.} modelled via a drift in phenotypic space,  being aware that compounds interact with each other when administered simultaneously. 
Moreover, while in this theoretical study we adopted parameter values collected from the literature, in order to employ this model for treatment optimisation purposes, it should rely on patient-specific data to improve reliability.  
To achieve this it is necessary to collect \textit{ex vivo} data on plasma and tissue compound concentrations, and use them for model calibration using a two-stage approach, as done in~\cite{himstedt2020quantitativemechanistic}.  
First, one should estimate the systemic PK parameters, \textit{i.e.} the ones related to central and peripheral compartments, based on the plasma concentration-time profile.  Secondly, tissue concentration–time profiles should be used to estimate the remaining tissue-specific parameters.  We note, however, that experimental methods to measure the partition coefficient parameters, independently of the mathematical model, do exist~\cite{holt2019prediction}.


\section*{Acknowledgements and Funding}
\paragraph{Acknowledgments.} 
The authors thank Luis Almeida (LJLL, Sorbonne Université, FR), Tommaso Lorenzi (Politecnico di Torino, IT) and Marco Picasso (EPFL, CH) for the insightful discussions in the initial stages of the project. 

\paragraph{Funding.}
This project has received funding from the European Union’s Horizon 2020 research and innovation programme under the Marie Sk\l{}odowska-Curie grant agreement No 945298-ParisRegionFP. 
C.V. is a Fellow of the Paris Region Fellowship Programme, supported by the Paris Region.

\bibliographystyle{siam}

\newpage
\appendix
\renewcommand{\theequation}{\thesection.\arabic{equation}}
\setcounter{equation}{0}
\renewcommand{\thefigure}{\thesection.\arabic{figure}} 
\setcounter{figure}{0}
\section{Proofs of analytical results}\label{appendix:analysis}

\subsection{Proof of the bounds on $I_{i\e}$} \label{appendix:proofbounds}
We prove the result by contradiction. Suppose that $0<I_M<I_{1\e}$. We integrate $\eqref{eq:ne:ss:results}_1$ and adopt the boundary conditions $\eqref{eq:ne:ss:results}_3$ to obtain
\begin{equation}
\label{proof1:eq}
        \int_0^1 R_1(y,I_{1\e}) n_{1\e}(y) dy  + \int_0^1 \nu_{2,1}(y) n_{1\e}(y) dy  - \int_0^1 \nu_{1,2}(y) n_{1\e}(y) dy  = 0 \,.
\end{equation} 
For the first term of equation~\eqref{proof1:eq} we employ assumptions~\eqref{ass:RI1} and \eqref{ass:IM}, along with the inequality $I_M<I_{1\e}$, to deduce
$$
        R_{1}(y,I_{1\e})<R_{1}(y,I_{M})\leq \max_{y\in[0,1]} \big(R_{1}(y,I_M)\big)\leq -\delta\,.
$$
Multiplying both sides of this inequality by $n_{1\e}(y)$ and integrating, we obtain
$$
        \int_0^1 R_{1}(y,I_{1\e}) n_{1\e}(y) dy<   \int_0^1 -\delta n_{1\e}(y) dy= -\delta I_{1\e}\,.
$$
For what concerns the second and third terms of equation~\eqref{proof1:eq}, from assumption~\eqref{ass:nu} we can deduce
$$
        \int_0^1 \nu_{2,1}(y)n_{2\e}(y)dy\leq \nu_M I_{2\e} \quad \mbox{and} \quad - \int_0^1 \nu_{1,2}(y)n_{1\e}(y)dy\leq - \nu_m I_{1\e}\,.
$$
Combining these results we obtain
$$
    0 =\int_0^1 R_1(y,I_{1\e}) n_{1\e}(y) dy  + \int_0^1 \nu_{2,1}(y) n_{1\e}(y) dy  - \int_0^1 \nu_{1,2}(y) n_{1\e}(y) dy  < -\delta I_{1\e} + \nu_M I_{2\e}  - \nu_m I_{1\e}\,,
$$
and thus
\begin{equation}
\label{proof1:Iineq}
    I_{2\e} > \frac{(\nu_m+\delta)}{\nu_M} I_{1\e} > \frac{\nu_m}{\nu_M} I_{1\e} > \frac{\nu_m}{\nu_M} I_{M}\,.
\end{equation}
Now we integrate equation~$\eqref{eq:ne:ss:results}_1$, both for $i=1$ and $i=2$, apply boundary conditions~$\eqref{eq:ne:ss:results}_3$, and then sum them to obtain
\begin{equation}
\label{proof1:sumR}
     \int_0^1 R_1(y,I_{1\e}) n_{1\e}(y) dy +  \int_0^1 R_2(y,I_{2\e}) n_{2\e}(y) dy =0.
\end{equation}
From {assumption~\eqref{ass:RI1}, inequality~\eqref{proof1:Iineq} and} the technical assumption~\eqref{ass:IM}, we have that
$$
    R_1(y,I_{1\e})<\max_{y\in[0,1]}\bigg(R_1(y,I_M)\bigg)\leq -\delta 
    \quad \mbox{and} \quad 
    R_2(y,I_{2\e})<\max_{y\in[0,1]}\bigg(R_2\big(y,\frac{\nu_m}{\nu_M}I_M\big)\bigg)\leq -\delta \,.
$$
Together with equation~\eqref{proof1:sumR} and, again, the previous result~\eqref{proof1:Iineq}, this gives
$$
    0 = \int_0^1 R_1(y,I_{1\e}) n_{1\e}(y) dy +  \int_0^1 R_2(y,I_{2\e}) n_{2\e}(y) dy < -\delta (I_{1\e} + I_{2\e}) < -\delta \big(I_M + \frac{\nu_m}{\nu_M}I_M\big) <0\,,
$$
a contradiction. We thus conclude that $I_{1\e}\leq I_M$.  
The proof for the case with $0<I_M<I_{2\e}$, and for the lower bounds of $I_{i\e}$ ($i=1,2$) follow analogous arguments, relying on assumption~\eqref{ass:Im}. 

\subsection{Proof of the results for the asymptotic regime $\e\to0$}\label{appendix:prooftheorem}

We formally extend the analysis carried out in~\cite{mirrahimi2012adaptation} for the case of constant migration rates to the case of phenotype-dependent migration rates.

\paragraph{Hopf-Cole transformation.} 
Following the work of~\cite{mirrahimi2012adaptation}, we introduce the Hopf-Cole transformation $\nie(y) = e^{\uie(y)/\e}$~\cite{barles1989wavefront,evans1989pde,fleming1986pde}, with $\uie(y)$ semi-convex (\textit{i.e.} $\dyy \uie \geq -E$, for some constant $E>0$), for $i=1,2$. 
Then we can evaluate
$$\frac{d^2\nie}{dy^2} =\Big( \e^{-1} \frac{d^2\uie}{dy^2} + \e^{-2} \Big(\frac{d\uie}{dy}\Big)^2 \,\Big) \, \nie \,,$$
and equation $\eqref{eq:ne:ss:results}_1$ becomes
$$R_i\big(y,I_{i\e}\big) \, n_{i\e} + \e^2 \, \Big( \e^{-1} \frac{d^2\uie}{dy^2} + \e^{-2} \Big(\frac{d\uie}{dy}\Big)^2 \,\Big) \, \nie +\nu_{j,i}(y) \, n_{j\e} - \nu_{i,j}(y) \, n_{i\e} = 0, \qquad y\in(0,1). $$
We can rewrite the system \eqref{eq:ne:ss:results}$_1$ for $\nie$ ($i=1,2$) as the following steady-state system for $u_{i\e}$ ($i=1,2$)  in matrix-vector form:
\begin{equation}
\label{fa:ss:ue}
\mathcal{A}_\e \, \mathcal{N}_\e = \mathcal{L}_\e \, \mathcal{N}_\e
\end{equation}
where
\begin{align}
&\mathcal{A}_\e =
  \begin{pmatrix}
   R_1\big(y,\Iae\big) - \nuab(y)  & \nuba(y)  \\
    \nuab(y) & R_2\big(y,\Ibe\big) - \nuba(y)
  \end{pmatrix} \,, \\
&\mathcal{N}_\e = 
  \begin{pmatrix}
   \nae  \\
   \nbe
  \end{pmatrix} = 
  \begin{pmatrix}
   e^{\uae/\e} \\
   e^{\ube/\e}
  \end{pmatrix} \,, \\
&\mathcal{L}_\e = \diag{- \e \frac{d^2\uae}{dy^2} - \left(\frac{d\uae}{dy}\right)^2,- \e \frac{d^2\ube}{dy^2} - \left(\frac{d\ube}{dy}\right)^2} \,.
\end{align}
Then given the eigenvalue $H_\e=H_\e(y,\Iae,\Ibe)$ of $\mathcal{A}_\e$ with corresponding eigenvector $\mathcal{N}_\e$, from \eqref{fa:ss:ue} we have
\beq
\label{fa:hj:e}
\mathcal{L}_\e \, \mathcal{N}_\e = H_\e \, \mathcal{N}_\e \,,
\eeq
which corresponds to the following equations for $\uie$
\beq
\label{fa:hj:e2}
- \e \frac{d^2\uie}{dy^2} - \Big(\frac{d\uie}{dy}\Big)^2 = H_\e(y,\Iae,\Ibe) \qquad i=1,2 \,.
\eeq

\paragraph{Hamilton-Jacobi equation.} Firstly, given the assumptions introduced in Section~\ref{subsec:formal_assumptions}, from \eqref{bound:Ie:results} and \eqref{bound:I:results} we deduce that $\nae$ and $\nbe$ converge weakly to measures $\na$ and $\nb$, \textit{i.e.}
\beq\label{fa:weak:results}
\nie(y) \xrightharpoonup[\e  \rightarrow 0]{\scriptstyle\ast} n_{i}(y) \qquad i=1,2 \,.
\eeq 
On the basis of the analysis carried out in~\cite{mirrahimi2012adaptation} for the case of constant migration rates, we expect that in the asymptotic regime $\e\to0$ both sequences $\uae$ and $\ube$ converge uniformly in $[0,1]$ to a continuous function $u\in C([0,1])$ and $(\Iae,\Ibe)$ converges to $(\Ia,\Ib)$. We further assume that the semi-convexity of $\uie$ ($i=1,2$) is preserved in the limit such that $u=u(y)$ is also semi-convex. 
If that is the case, then in the limit $\e\to0$ equation \eqref{fa:ss:ue} yields
\begin{equation}\label{fa:ss:u}
\mathcal{A} \, \mathcal{N} = \mathcal{L} \, \mathcal{N}
\end{equation}
\noindent where
\begin{align}
&\mathcal{A} =
  \begin{pmatrix}
   R_1\big(y,\Ia\big) - \nuab(y)  & \nuba(y)  \\
    \nuab(y) & R_2\big(y,\Ib\big) - \nuba(y)
  \end{pmatrix}  \,, \\
&\mathcal{N} = 
  \begin{pmatrix}
   \na  \\
   \nb
  \end{pmatrix}  \,, \\
&\mathcal{L} = \diag{- \left(\frac{d u}{dy}\right)^2, - \left(\frac{d u}{dy}\right)^2 } \,.
\end{align}
Then as $\e\to0$ equation~\eqref{fa:hj:e2} and the bounds \eqref{bound:I:results} lead to the following constrained Hamilton-Jacobi equation for $u(y)$: 
\beq
\label{fa:hj2}
\begin{cases}
\displaystyle{- \bigg(\frac{du}{dy}\bigg)^2 = H(y,\Ia,\Ib)} \\
\displaystyle{ \max_{y \in [0,1]} u(y) = 0 }
\end{cases} \,,
\eeq
 where the constraint is required to ensure bounds on $\Ii$ ($i=1,2$) in~\eqref{bound:I:results} are satisfied. Note that the right-hand-side of \eqref{fa:hj2}$_1$ is zero at points $y_k\in\argmax{ u(y)}$ and negative for all $y\in [0,1]$ that are not stationary points of $u(y)$.
This implies
\beq
\label{fa:Hmax}
 \begin{cases}
H(y_k,\Ia,\Ib)= 0 \qquad\qquad \forall\; y_k \in\argmax{u(y)} , \\[3pt]
 \displaystyle{\max_{y \in [0,1]}  H(y,\Ia,\Ib) = 0},
\end{cases}
 \eeq 
 where $H=H(y,\Ia,\Ib)$ is the largest eigenvalue of $\mathcal{A}$ with corresponding eigenvector $\mathcal{N}$. Hence $H$ is given by
\beq\label{fa:H}
H(y,\Ia,\Ib) = F + \sqrt{F^2-4G} \;,
\eeq
with
\begin{align}
F &= R_1(y,\Ia)-\nuab(y) + R_2(y,\Ib)-\nuba(y) \,,\label{fa:F} \\
G &= \big( R_1(y,\Ia)-\nuab(y) \big) \big( R_2(y,\Ib)-\nuba(y) \big) - \nuab(y)\nuba(y) \label{fa:G} \,.
\end{align}

\noindent Note from \eqref{fa:H} that for \eqref{fa:Hmax}$_1$ to be satisfied we must have that $F$ is negative at each ${y}_k\in\argmax{u(y)}$. In such case, \eqref{fa:Hmax} implies
\beq\label{fa:minG}
 \begin{cases}
G(y_k,\Ia,\Ib)= 0 \qquad\qquad \forall\; y_k \in\argmax{u(y)}  \\[3pt]
 \displaystyle{\min_{y \in [0,1]} G(y,\Ia,\Ib) = 0} \;.
\end{cases}
\eeq
Then from \eqref{fa:hj2} and \eqref{fa:Hmax} we have
  \beq
\label{fa:supp}
\supp{n_i}  \subset  \Gamma \qquad i=1,2 \,,
\eeq
 \noindent where
\beq
\label{fa:supp:gamma}
\Gamma := \argmax_{y\in[0,1]} H(y) \equiv \big\{ y \in [0,1] :  H(y,\Ia,\Ib) = 0 \big\}\,,
\eeq 
 
\noindent which -- under assumption \eqref{ass:argmaxRmu1} -- we expect to be discrete and finite, with cardinality $K\in\mathbb{N}$. \\ 

\noindent \textit{\textbf{Remark 1:} As we have shown that \eqref{fa:Hmax}, together with definitions \eqref{fa:H} and \eqref{fa:F}, implies \eqref{fa:minG}, we could alternatively define $\Gamma$ in \eqref{fa:supp} as
\begin{equation*}
\Gamma := \argmin_{y\in[0,1]} G(y) \equiv \big\{ y \in [0,1] :  G(y,\Ia,\Ib) = 0 \big\}\,.
\end{equation*}}

\noindent \textit{\textbf{Remark 2:} The support of $n_i$ is only a proper subset of $\Gamma$ as defined in \eqref{fa:supp:gamma}, since $\Gamma$ includes both the points at which $u$ attains its maxima and minima (from \eqref{fa:hj2}$_1$), while $n_i$ will be zero at the points in $\argmin u$ (from \eqref{fa:hj2}$_2$). Hence \eqref{fa:supp} may be more precisely written as
\begin{equation*}
\supp{n_i}\subseteq \Omega \subset \Gamma \qquad i=1,2 \,,
\end{equation*}
\noindent where $\Gamma$ is defined as in \eqref{fa:supp:gamma} (or as in Remark 1) and $\Omega$ is defined as
\begin{equation*}
\Omega := \argmax_{y\in[0,1]} u(y) \equiv \big\{ y \in [0,1] : u(y) = 0 \big\} \,. 
\end{equation*}}

\paragraph{Concentration as Dirac masses.}
From \eqref{bound:Ie:results}, we deduce that, along subsequences and for $i=1,2$, $\nie$ converges weakly in the asymptotic regime $\e\to0$ to a measure $n_i$.
In view of the results so far obtained on the support of $n_i$, we have that the measures $\na$ and $\nb$ to which $\nae$ and $\nbe$ converge concentrate as Dirac masses, \textit{i.e.}
\beq
\label{fa:n}
\nie(y) \xrightharpoonup[\e  \rightarrow 0]{\scriptstyle\ast}  \sum_{k=1}^{K} \rho_{ik} \, \delta(y - y_k) \qquad i=1,2 \,,
\eeq
\noindent where the points $y_k\in[0,1]$ can be found by solving \eqref{fa:Hmax} -- or alternatively \eqref{fa:minG} -- while the weights $\rho_{ik}\geq0$ must be such that
\beq
\label{fa:rho1}
\Ii = \sum_{k=1}^{K} \rho_{ik} \qquad i=1,2 \,,
\eeq
 and can be found by integrating the system \eqref{fa:ss:u} 
over a ball $B_r(y_k):=\big\{ y\in[0,1] : |y-y_k|\leq r \big\}$ centred at $y_k$, with radius $r$ small enough that no other point in the support of $n_i$ ($i=1,2$) is contained in $B_r(y_k)$. This yields
\beq
\label{fa:rho2}
\begin{cases}
\big( R_1(y_k,\Ia)-\nuab(y_k)\big) \rho_{1k} +\nuba(y_k)\rho_{2k} =0 \\[5pt]
\big( R_2(y_k,\Ib)-\nuba(y_k)\big) \rho_{2k} +\nuab(y_k)\rho_{1k} =0 .
\end{cases}\qquad k=1,..,K \,
\eeq

\subsection{Proof the analytical results for the metastatic spread case}\label{appendix:proofsolutions}
If $\nu_{1,2}(y)\geq 0$,  with $\nu_{1,2}(y)> 0$ for some $y\in[0,1]$, and $\nu_{2,1}(y)\equiv0$ for all $y\in[0,1]$, then solving \eqref{fa:Hmax} leads to the following systems:
\begin{equation}
\label{fa:localised}
\begin{cases}
\displaystyle{R_1({y}_1,\Ia) -\nu_{1,2}({y}_1)= 0} \\[5pt]
\displaystyle{\partial_y R_1({y}_1,\Ia) -\partial_y\nu_{1,2}({y}_1)= 0}
\end{cases}
\qquad \text{and} \qquad
\begin{cases}
\displaystyle{R_2({y}_2,\Ib) = 0 }\\[5pt]
\displaystyle{\partial_y R_2({y}_2,\Ib) = 0 } \;.
\end{cases} 
\end{equation}
Notice that \eqref{fa:localised} together with assumption \eqref{ass:argmaxRmu2} ensure that $F$ defined in \eqref{fa:F} is negative at $y_1$ and $y_2$, as required by \eqref{fa:Hmax} and \eqref{fa:H}.
From the second equation of each system in~\eqref{fa:localised} we obtain $y_1$ and $y_2$, while from the first one of each system we get $I_1$ and $I_2$. Moreover, equations~\eqref{fa:rho1} and~\eqref{fa:rho2} result in 
\beq
\rho_{11}=\Ia \,,  \quad \rho_{12}=0\,  \quad  \rho_{21}=\min \bigg(-\frac{\nu_{1,2}(y_1)}{R_2(y_1,\Ib)}\rho_{11}, \Ib\bigg) \,,  \quad  \rho_{22} = \Ib - \rho_{21} \,.
\eeq
Solving \eqref{fa:localised} for $R_i(y,I_i)$ ($i=1,2$) given by \eqref{genericfitness} and for $\nu_{1,2}(y)$ given by \eqref{migrationratefunction}, we have that~\eqref{fa:weak:results} and~\eqref{fa:n} corresponds to the following asymptotic solution:
\begin{equation}
\displaystyle{n_1(y) = \Ia \, \delta (y-y_1) \qquad \text{and} \qquad n_2(y) = \rho_{21} \, \delta (y-y_1) +  \rho_{22} \, \delta (y-y_2) }\;,
\end{equation}
with
\begin{equation}
\displaystyle{y_1 = \frac{b_1}{b_1+\hat{\nu}_{1,2}}h_1\,, \qquad y_2 = h_2 \,, }
\end{equation}
and
\begin{equation}
\label{sol:localised:y:rho}
\displaystyle{ \Ia= \frac{1}{d_1}\bigg[ a_1  - \frac{\hat{\nu}_{1,2} b_1 h_1^2}{b_1+\hat{\nu}_{1,2}} \bigg] \,, \quad \Ib = \frac{a_2}{d_2} \,,  \qquad \rho_{21} = \min \bigg(\frac{\hat{\nu}_{1,2}\,y_1^2}{b_2(y_1-h_2)^2}\Ia, \Ib\bigg) \,, \quad \rho_{22} = \Ib - \rho_{21} \,.  }
\end{equation}

\end{document}


\maketitle

\section{Supplementary analytical results}

The following analytical results extend those in Section 3.3 of the Main Manuscript to the case of localised tumour and secondary seeding. Therefore we refer to equations appearing in Section 3.3 and Appendix A.2 of the Main Manuscript. 

\subsection{Analytical results for the localised tumors case: $\nu_{1,2} \equiv 0$ and $\nu_{2,1}\equiv 0$}
If both $\nu_{1,2}(y)\equiv0$ and $\nu_{2,1}(y)\equiv0$ for all $y\in[0,1]$, then solving (A.16) 
leads to the following systems:
\begin{equation}
\begin{cases}
\displaystyle{R_i({y}_i,I_i) = 0} \\[5pt]
\displaystyle{\partial_y R_i({y}_i,I_i) = 0 } 
\end{cases} 
\quad i=1,2 \;.  \label{fa:separate}\\
\end{equation}
 Notice that \eqref{fa:separate} together with assumption (20) 
ensure that $F$ defined in (A.18) 
is negative at $y_1$ and $y_2$, as required by (A.16) and (A.17). 
From \eqref{fa:separate}$_2$ we obtain $y_1$ and $y_2$, and from \eqref{fa:separate}$_1$ we get $I_1$ and $I_2$, while (A.25) 
leads to having $\rho_{21}=\rho_{12}=0$ which together with (A.24) 
imply $\rho_{11}=\Ia$ and $\rho_{22}=\Ib$. 
Doing this for $R_i(y,I_i)$ ($i=1,2$) given by~(7) 
leads to the following solution:
\begin{equation}
\displaystyle{n_i(y) = I_i \, \delta (y-y_i) \qquad \text{with} \qquad y_i = h_i\,, \quad  I_i= \frac{a_i}{d_i}} \qquad \text{for} \qquad i=1,2 \;. \label{sol:separate}
\end{equation} 
These results, in agreement with previous results in the literature 
 focusing on a single localised tumour, provide a formal description of the fact that populations living in separate, non-communicating sites evolve independently of each other. Each population presents monomorphism at equilibrium, and the trait selected corresponds to the fittest trait $h_i$ ($i=1,2$) determined by the local environment. That is, of course, if cells were present in both sites initially: if the secondary site is representing a metastatic site, in this scenario no cells ever reach it and we expect $n_2\equiv 0$ at equilibrium.

\subsection{Analytical results for the secondary seeding case: $\nu_{1,2}\not \equiv 0$ and $\nu_{2,1}\not \equiv 0$}
Consider now the case in which $\nu_{1,2}\not\equiv0$ and $\nu_{2,1}\not\equiv0$.
In particular we focus on the case in which the selected traits $y_k$ are such that $\nu_{i,j}(y_k)>0$ ($i,j=1,2 \;\; i\neq j$).
In this case, for each $k=1,..,K$ we have from~(A.25)$_1$ 
that $\rho_{2k}>0$ implies $\rho_{1k}>0$ and $ R_1(y_k,\Ia)-\nuab(y_k)<0$. Similarly, from~(A.25)$_2$ 
we have that $\rho_{1k}>0$ implies $\rho_{2k}>0$ and $ R_2(y_k,\Ib)-\nuba(y_k)<0$. Therefore we either have both $\rho_{1k}=\rho_{2k}=0$ (trivial case) or $\rho_{1k}>0$, $\rho_{2k}>0$ and
\beq
\label{fa:neg}
R_i(y_k,I_i)-\nu_{i,j}(y_k)<0 \,, \quad i,j=1,2 \;\; i\neq j\,.
\eeq
%
%
\noindent Under definition~(A.18), 
\eqref{fa:neg} ensures that $F$ is negative at every $y_k$ ($k=1,..,K$) as required by (A.16) and (A.17). 
In addition, \eqref{fa:neg} and (A.25) 
imply
\beq
\rho_{1k} = \Bigg( \frac{\nuba(y_k)}{\nuab(y_k)-R_1\big(y_k,\Ia\big)}\Bigg) \rho_{2k} =  \Bigg( \frac{\nuba(y_k)-R_2\big(y_k,\Ib\big)}{\nuab(y_k)}\Bigg) \rho_{2k} \quad \text{for} \quad k=1,...,K \,.
\eeq
For a more precise characterisation of the phenotypic distributions at equilibrium, we must solve (A.20), 
which is equivalent to simultaneously solving $G=0$, $\partial_y G=0$ and $\partial^2_{yy}G\geq 0$. Due to analytical complexity, we here do not solve this explicitly. Instead we investigate the number $K$ of discrete points that we might find solving~(A.20), 
under the specific fitness functions and transition rates, to shed light on the conditions under which we may expect the population to go extinct, present monomorphism or polymorphism at equilibrium. 

\begin{figure}[h!]
\centering
 \includegraphics[width=1\textwidth]{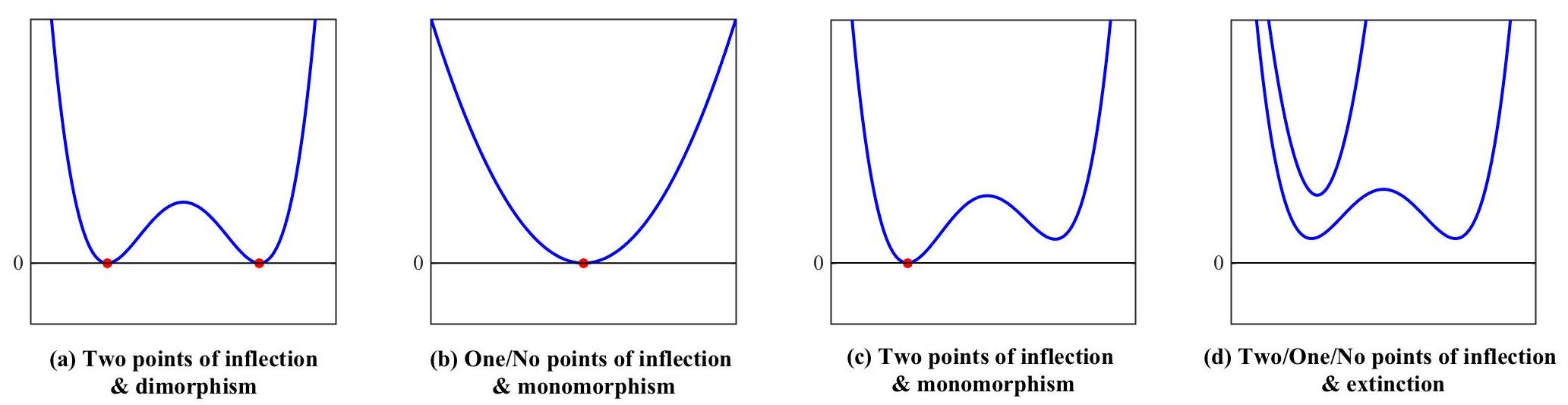}
 \caption{\textbf{Four different cases for the function $G$.} Possible plots of $G(y,I_1,I_2)$ as a function of $y$ displaying how $G$ having two inflection points is necessary -- {cf.} (a) and (b) -- but not sufficient -- \textit{vid} (c) and (d) -- for the population to present dimorphism at equilibrium.}\label{fig:conditions}
\end{figure}

Let $R_i(y,I_i)$ ($i=1,2$) be given by (7) 
and $\nu_{i,j}(y)$ ($i,j=1,2 \;\;\; i\neq j$) be given by (9). 
Note first of all that, under these specific definitions, $G(y,\Ia,\Ib)$ in (A.19) 
is a polynomial of order 4 in $y$. Therefore $\partial^2_{yy}G$ is quadratic in $y$, suggesting that $G$ can have at most 2 inflection points, determined by solving $\partial^2_{yy}G=0$ in $y$. Figure \ref{fig:conditions} illustrates why in order for the population to be dimorphic we must have 2 inflections points: this will give us a necessary condition for dimorphism and, in the opposite case (\textit{i.e.} if we have one or no inflection points), a sufficient condition for monomorphism -- always assuming that the population does not go extinct. Said condition is determined by applying the well-known binomial formula to $\partial^2_{yy}G=0$ and investigating its discriminant $D(\Ia,\Ib)$. We here omit simple calculations to derive this and focus on the obtained necessary condition for dimorphism, which is given by
\beq\label{cond}
D(\Ia,\Ib)>0\,.
\eeq

\noindent Under definitions (7) and (9), 
condition \eqref{cond} can be rewritten to give the following condition:
\begin{equation*}\label{cond:y}
\begin{split}
&(B_1 h_1 - B_2 h_2)^2 +8B_1BV_1V_2h_1h_2 -2B_1B_2V_1V_2(h_1^2 + h_2^2)+ 2(1-V_1V_2)\bigg[\frac{a_1-d_1I_1}{b_1+\hat{\nu}_{12}} +\frac{a_2-d_2I_2}{b_2+\hat{\nu}_{21}}\bigg]>0 \,, \\[10pt]
& \hspace{3cm} \text{with} \qquad B_i:=\frac{b_i}{b_i+\hat{\nu}_{ij}} \,, \qquad V_i:=\frac{\hat{\nu}_{ij}}{b_i+\hat{\nu}_{ij}} \qquad i,j=1,2 \;\;\; i\neq j \,.
\end{split}
\end{equation*}
%

\begin{figure}[htb!]
\centering
\includegraphics[width=1\linewidth]{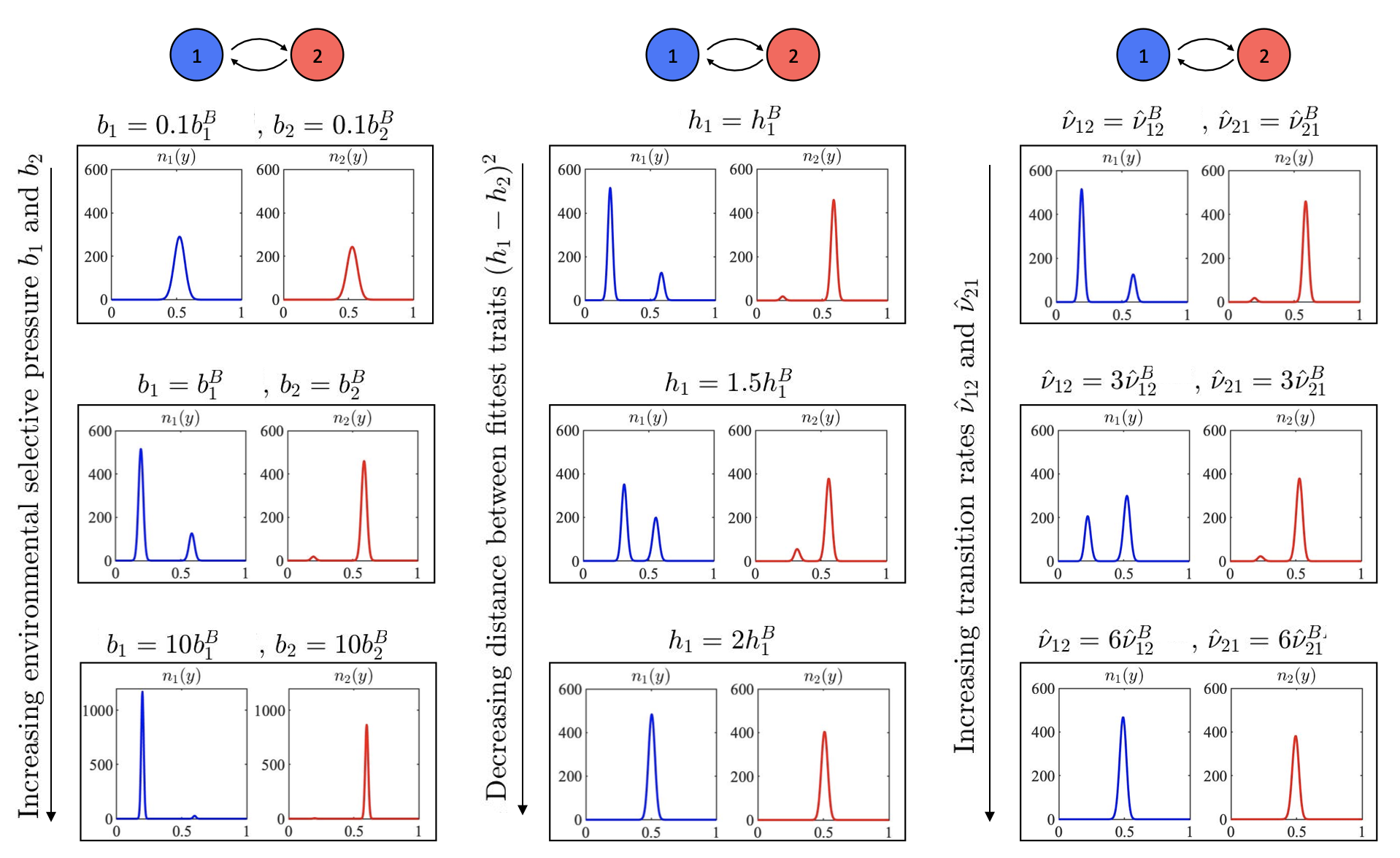}
\caption{\textbf{Outcome dependency on input factors in the secondary seeding scenario.} Illustrative example showing how the selection gradients $b_1$ and $b_2$, the fittest trait $h_1$ and the maximum migration rates $\hat{\nu}_{1,2}$ and $\hat{\nu}_{2,1}$ may affect the equilibrium distributions of the cancer cell populations of each site, in the secondary seeding case. 
We simulate system~(3), 
under definition~(7) 
for the fitness functions $R_i$ and definition~(9) 
for the migration rates $\nu_{i,j}(y)$,  under the initial conditions $n_{0,1}(y)>0$ and $n_{0,2}(y)=0$ $\forall y\in[0,1]$.  
Each column displays the equilibrium solutions of three simulations obtained by progressively varying the parameters $b_1$ and $b_2$ (first column), $h_1$ (second column) and $\hat{\nu}_{1,2}$ and $\hat{\nu}_{1,,12}$ (third column) form their baseline values $b_2^{B}=1$, $b_2^{B}=0.6$, $h_1^{B}=0.2$, $\hat{\nu}_{1,2}^{B}=0.1$ and $\hat{\nu}_{2,1}^{B}=0.05$. The remaining parameters are set to $\beta_1=\beta_2=10^{-7}$, $a_1=6$, $a_2=5$, $b_1=b_1^{B}$, $b_2=b_2^{B}$, $h_1=h_1^{B}$, $h_2=0.6$, $d_1=d_2=0.2$, $\hat{\nu}_{1,2}=\hat{\nu}_{1,2}^{B}$ and $\hat{\nu}_{2,1}=\hat{\nu}_{2,1}^{B}$. }
\label{plot_secondary_seeding}
\end{figure}

\noindent While this condition is quite complex, we may still consider the limit as $b_1,b_2\to\infty$, under which we have $B_i\to1$, $V_i\to 0$ ($i=1,2$) and therefore the condition reduces to $(h_1-h_2)^2>0$, which is satisfied as long as $h_1\neq h_2$. Therefore under extremely strong selective pressure from the two environments we can -- and we might expect to -- have dimorphism in both sites, which is in line with the results obtained in the case of constant transition rates. Figure \ref{plot_secondary_seeding} displays the possible model outcomes changing some of the parameters, and obtained simulating a nondimensional version of the model.

\newpage

\section{Supplementary figures}

\begin{figure}
    \centering
    \begin{subfigure}{0.9\textwidth}
        \centering
        \includegraphics[width=\textwidth]{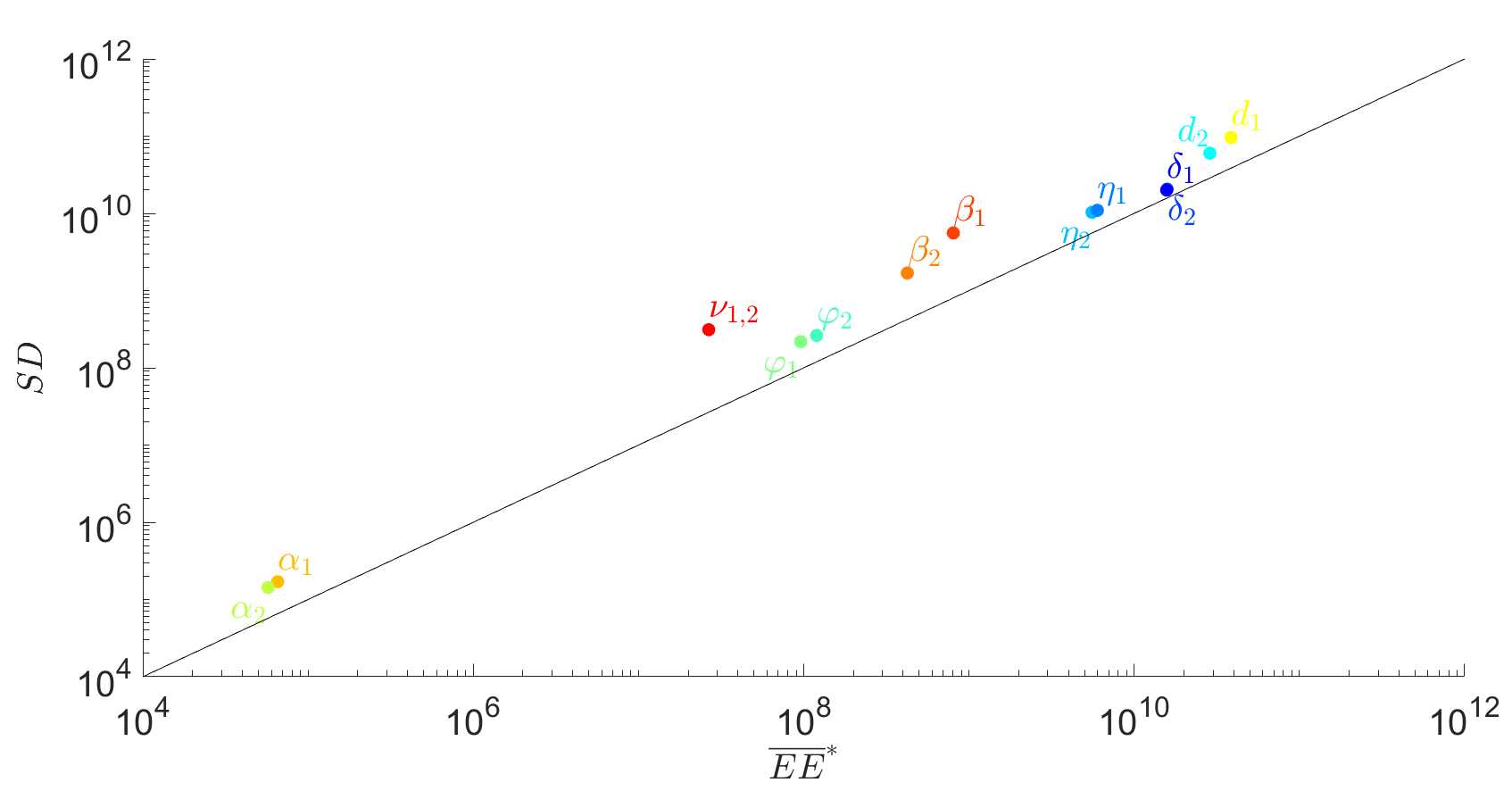}
        \caption{}
        \label{fig:EEligne_rho} 
    \end{subfigure}
    \begin{subfigure}{0.9\textwidth}
        \centering
        \includegraphics[width=\textwidth]{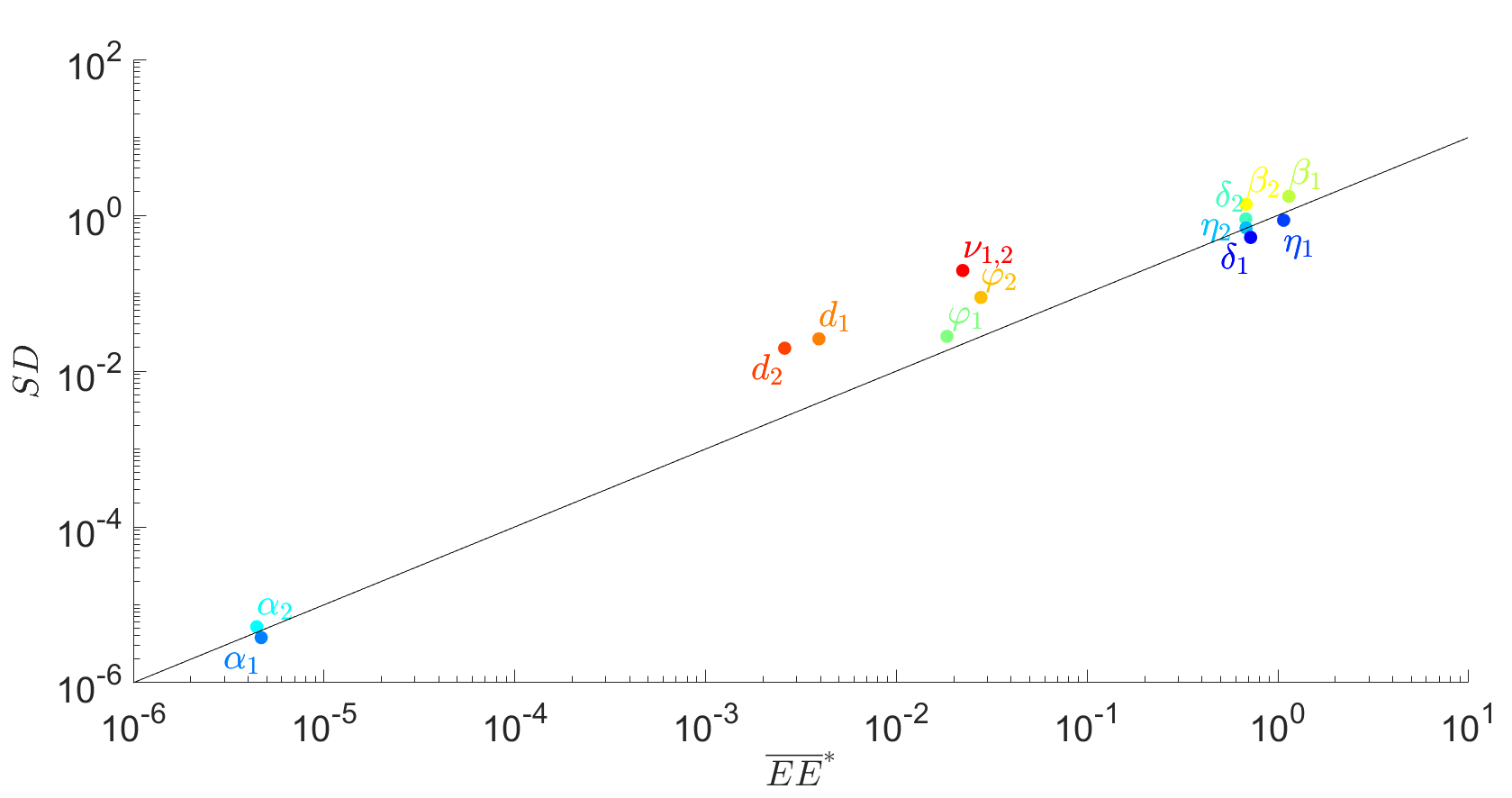}
        \caption{}
        \label{fig:EEligne_mu} 
    \end{subfigure}
    \label{fig:EEligne}
    \caption{\textbf{Results of the Elementary effects method for GSA: EE measures comparison.} 
    Comparison of the EE measures $\overline{EE}_i^*$ ($x$-axis) and $SD_i$ ($y$-axis), defined in~(37), 
    associated with the model output $Y_I$ (a) and $Y_{\mu}$ (b). The estimates are computed with $r=500$ evaluations of the elementary effect for each parameter.}
\end{figure}

\begin{figure}
    \renewcommand{\thesubfigure}{\alph{subfigure}.\arabic{row}}
    \centering
    \setcounter{row}{1}%
    \begin{subfigure}[b]{0.45\textwidth}
        \centering
        \includegraphics[width=0.95\textwidth]{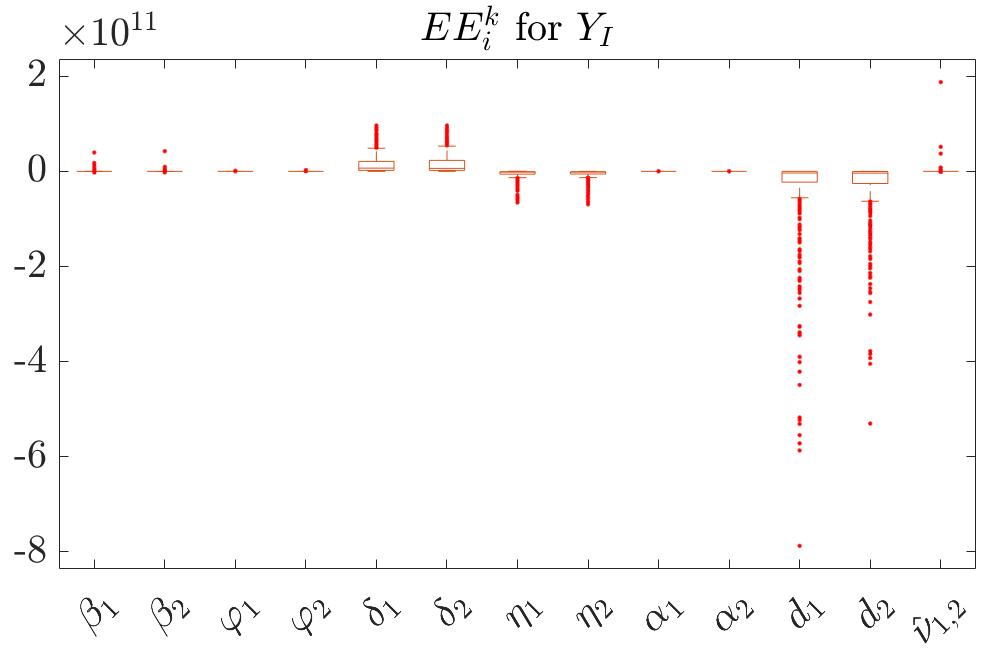}
        \caption{}
    \end{subfigure} \quad
    \begin{subfigure}[b]{0.45\textwidth}
        \centering
        \includegraphics[width=0.95\textwidth]{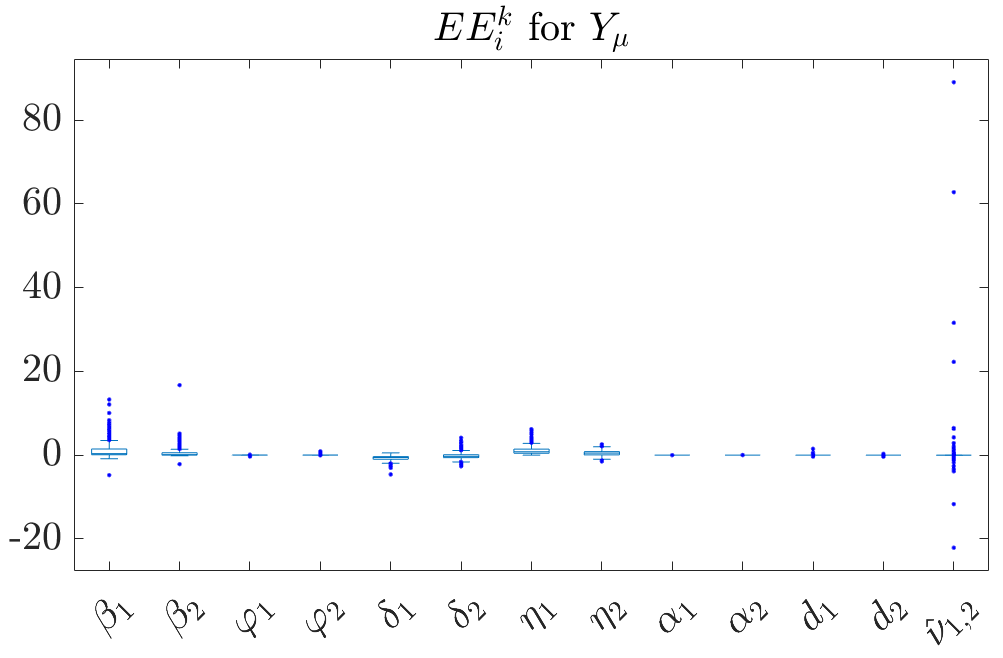}
        \caption{}
    \end{subfigure}
    \vskip\baselineskip
    \stepcounter{row}
    \setcounter{subfigure}{0}
    \begin{subfigure}[b]{0.45\textwidth}
        \centering
        \includegraphics[width=\textwidth]{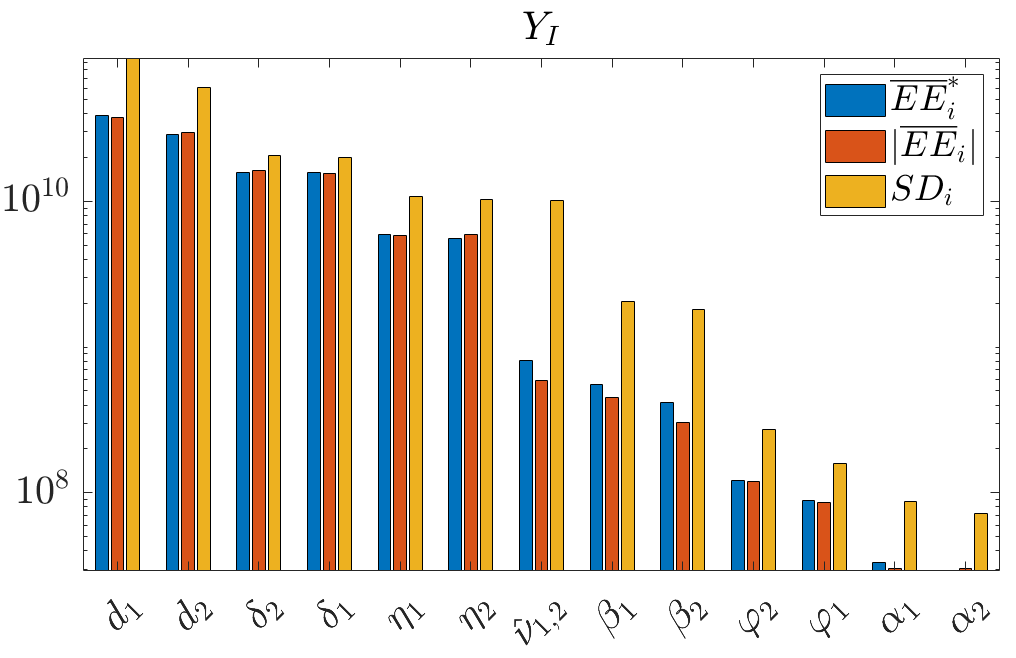}
        \caption{}
    \end{subfigure}  \quad
    \begin{subfigure}[b]{0.45\textwidth}
        \centering
        \includegraphics[width=\textwidth]{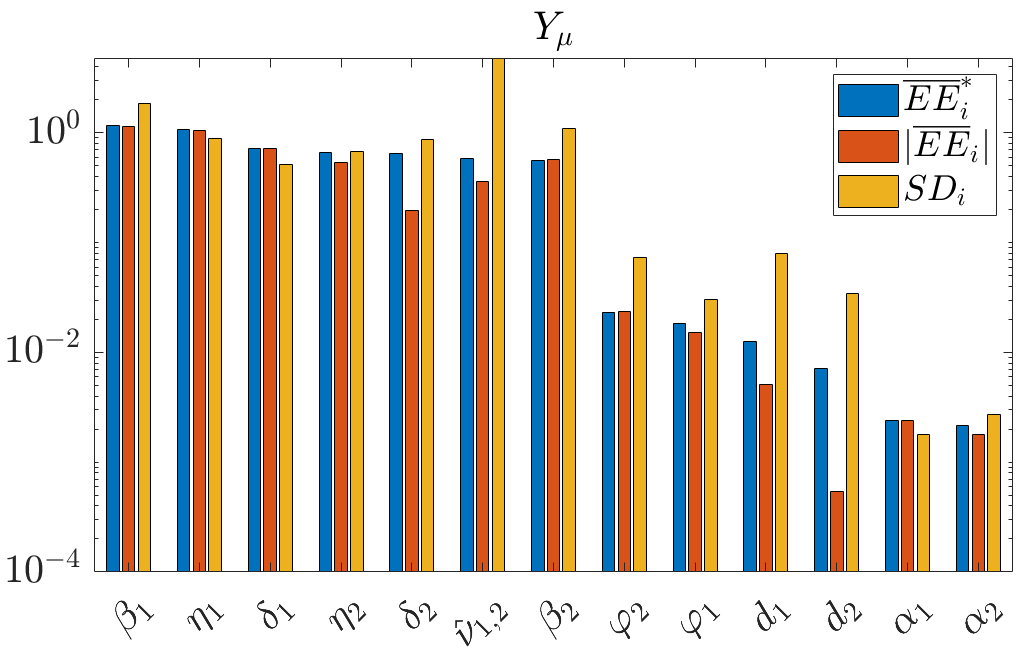}
        \caption{}
    \end{subfigure}
    \vskip\baselineskip
    \stepcounter{row}
    \setcounter{subfigure}{0}
    \begin{subfigure}[b]{0.45\textwidth}
        \centering
        \includegraphics[width=0.95\textwidth]{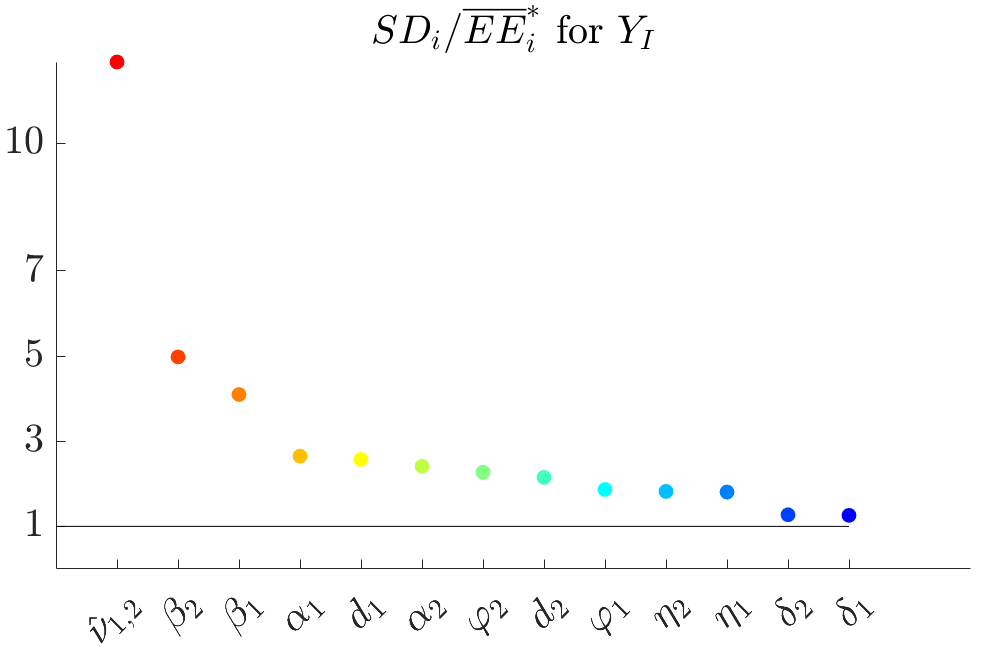}
        \caption{}
    \end{subfigure}  \quad
    \begin{subfigure}[b]{0.45\textwidth}
        \centering
        \includegraphics[width=0.95\textwidth]{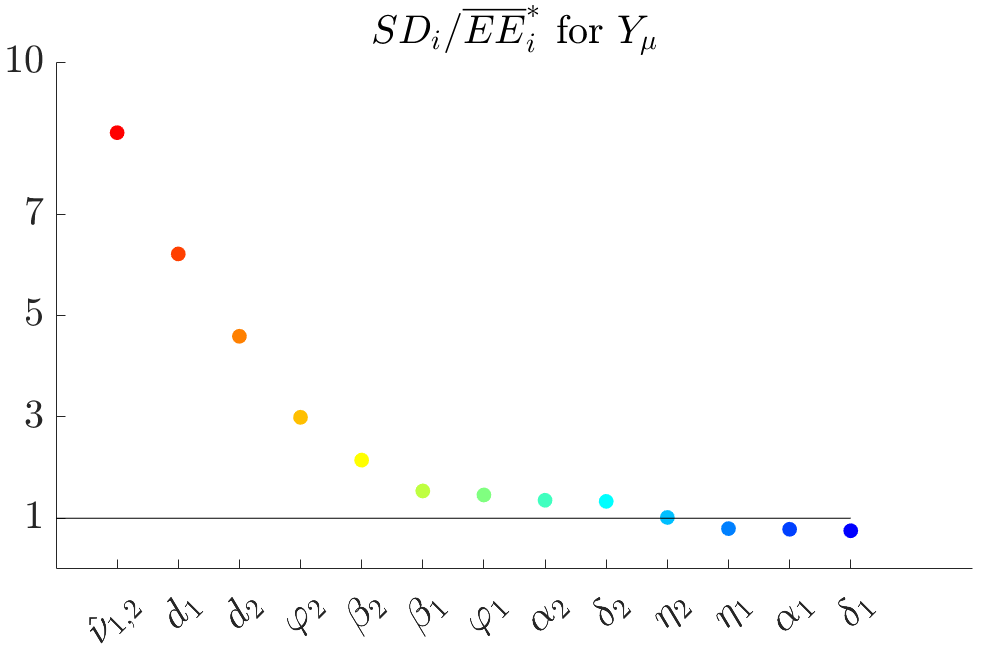}
        \caption{}
    \end{subfigure}
 \caption{\textbf{Results of the Elementary effects method for GSA under a wider range of values for the maximum migration rates.} EE measures associated with the model output $Y_{I}$ (a.$1$-$3$) and $Y_{\mu}$ (b.$1$-$3$), \textit{i.e.} the total tumour mass and average mean phenotypic state of the sites.  
  The estimates are computed with $r=500$ evaluations of the elementary effect for each parameter, and migration rates taking values in the range $\hat\nu_{1,2}\in[10^{-13},10^{-5}]$ 1/s. The elementary effects $EE_i$, defined in~(36), 
  calculated for each parameter are displayed in plots (a.$1$) and (b.$1$). The associated measures $\overline{EE}_i^*$, $|\overline{EE}_i^*|$ and $SD_i$, defined in~(37), 
  are shown in plots (a.$2$) and (b.$2$).  Plots (a.$3$) and (b.$3$) present the ratio between $SD$ and $\overline{EE}_i^*$.}
\label{EE_nu_wider_range}
\end{figure}

\begin{figure}
    \renewcommand{\thesubfigure}{\alph{subfigure}.\arabic{row}}
    \centering
    \setcounter{row}{1}%
    \begin{subfigure}[b]{0.45\textwidth}
        \centering
        \includegraphics[width=0.95\textwidth]{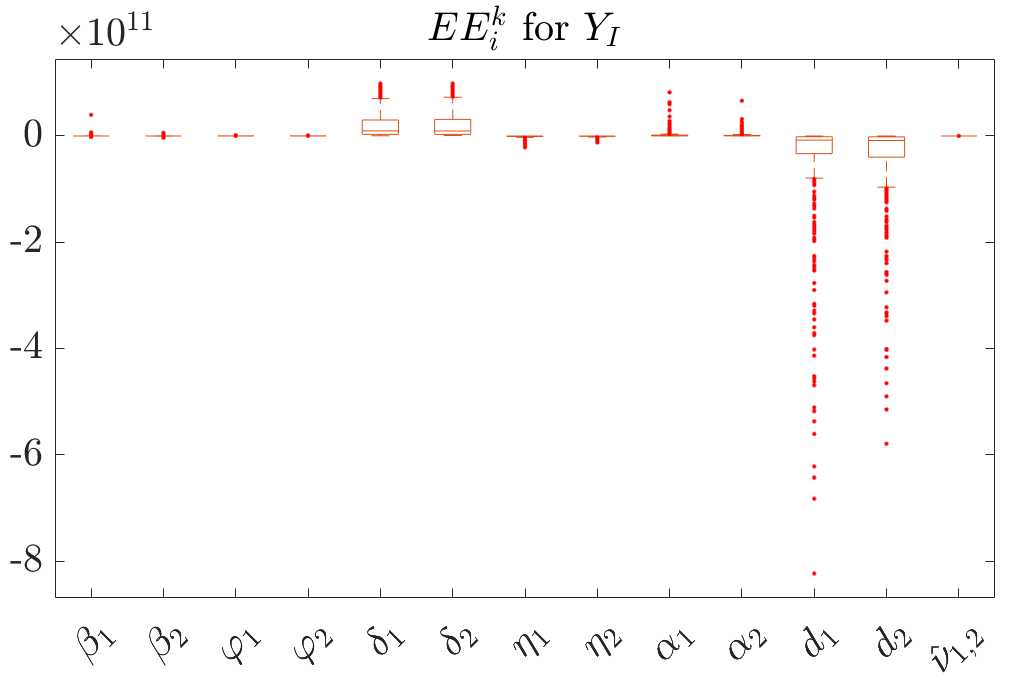}
        \caption{}
    \end{subfigure} \quad
    \begin{subfigure}[b]{0.45\textwidth}
        \centering
        \includegraphics[width=0.95\textwidth]{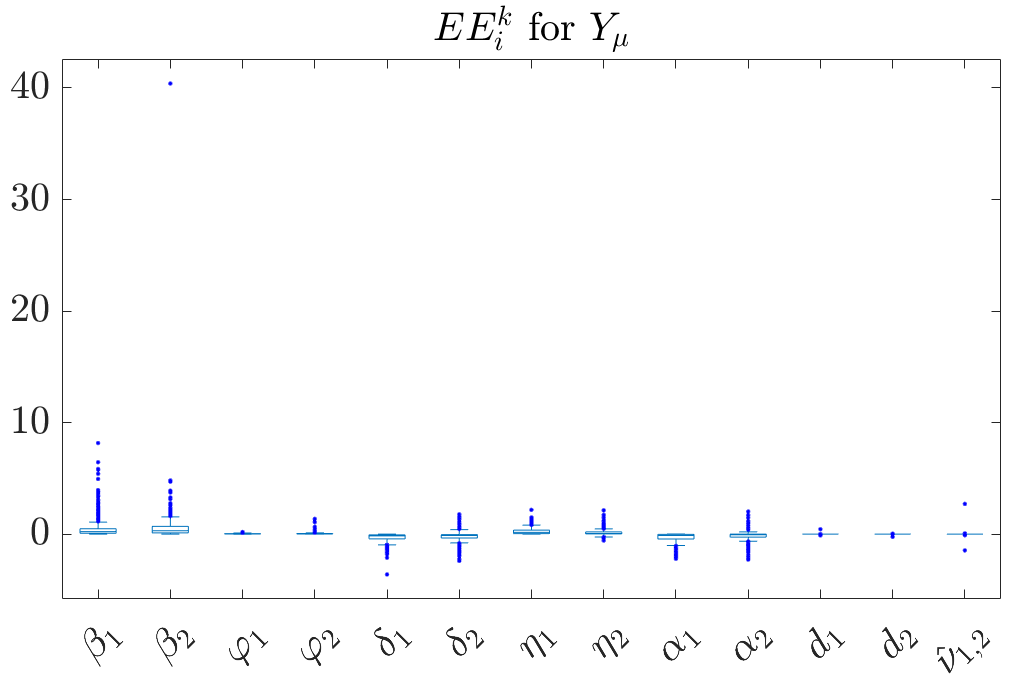}
        \caption{}
    \end{subfigure}
    \vskip\baselineskip
    \stepcounter{row}
    \setcounter{subfigure}{0}
    \begin{subfigure}[b]{0.45\textwidth}
        \centering
        \includegraphics[width=\textwidth]{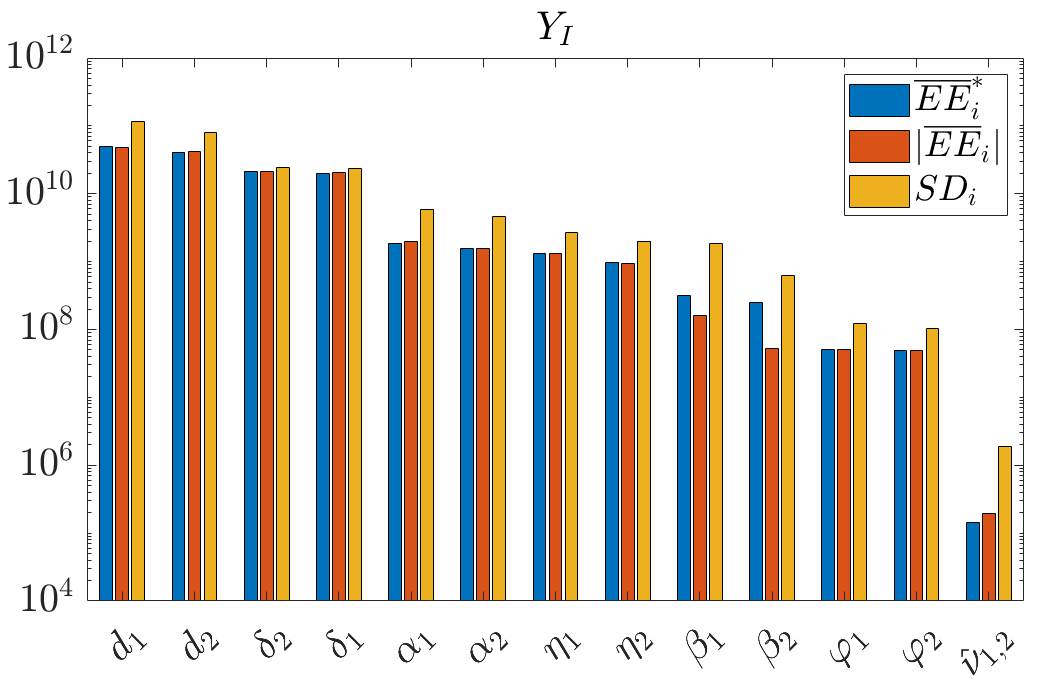}
        \caption{}
    \end{subfigure}  \quad
    \begin{subfigure}[b]{0.45\textwidth}
        \centering
        \includegraphics[width=\textwidth]{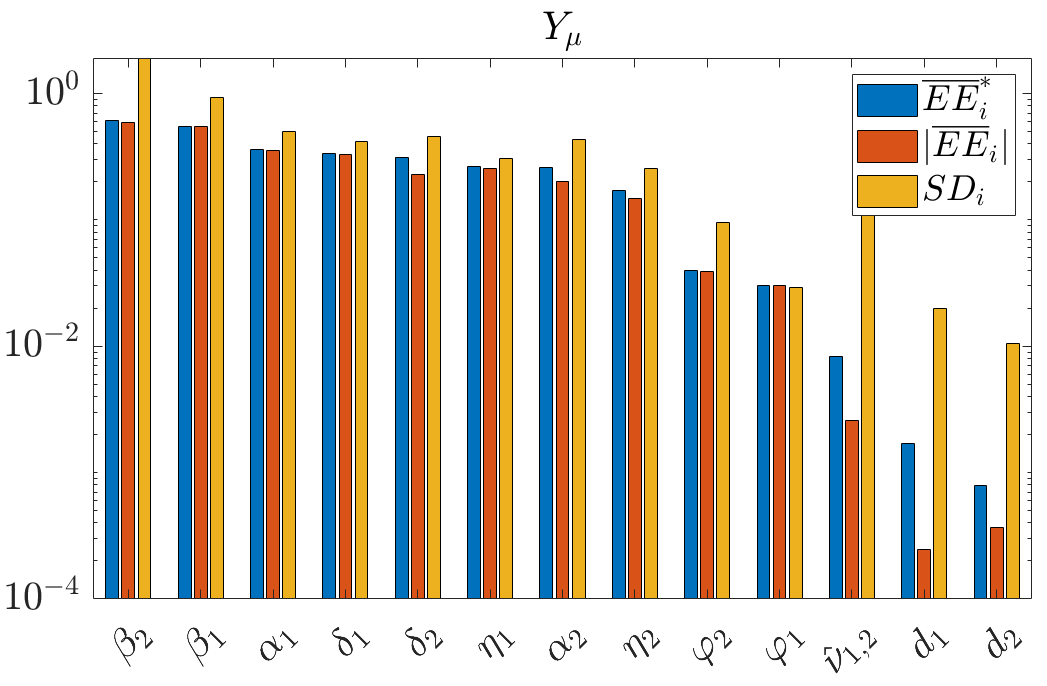}
        \caption{}
    \end{subfigure}
    \vskip\baselineskip
    \stepcounter{row}
    \setcounter{subfigure}{0}
    \begin{subfigure}[b]{0.45\textwidth}
        \centering
        \includegraphics[width=0.95\textwidth]{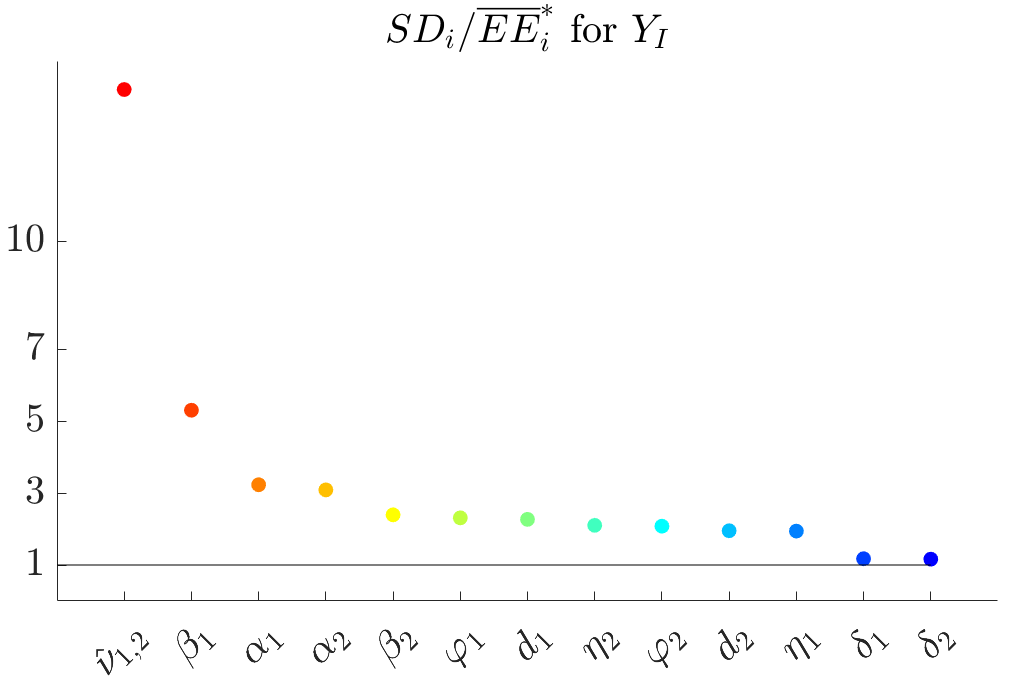}
        \caption{}
    \end{subfigure}  \quad
    \begin{subfigure}[b]{0.45\textwidth}
        \centering
        \includegraphics[width=0.95\textwidth]{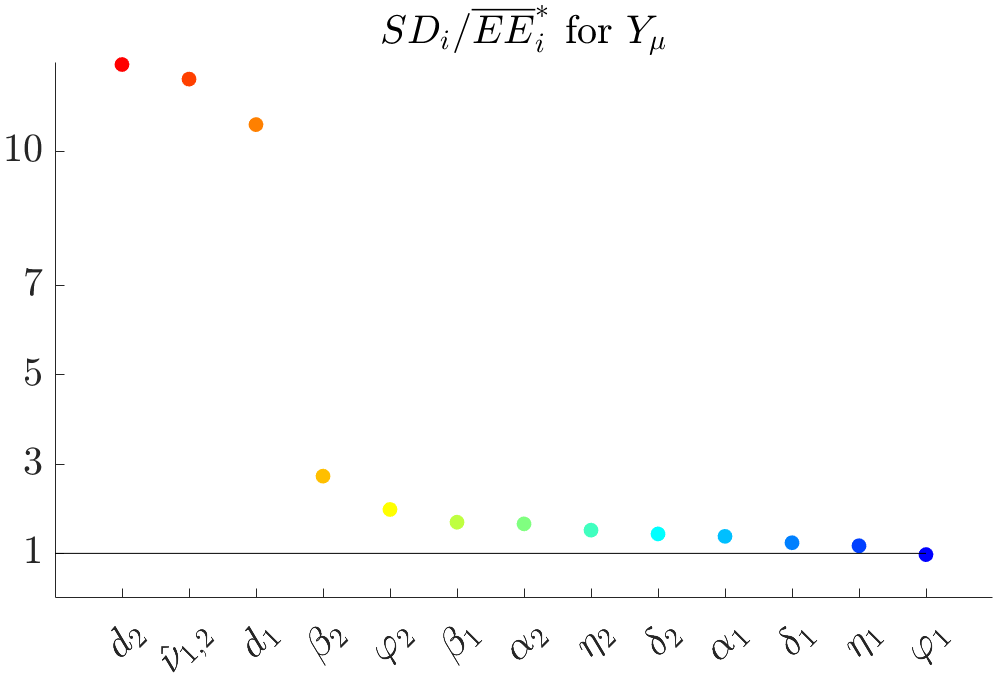}
        \caption{}
    \end{subfigure}
    \caption{\textbf{Results of the Elementary effects method for GSA under a lower drug dose.} EE measures associated with the model output $Y_{I}$ (a.$1$-$3$) and $Y_{\mu}$ (b.$1$-$3$), \textit{i.e.} the total tumour mass and average mean phenotypic state of the sites.  
  The estimates are computed with $r=500$ evaluations of the elementary effect for each parameter, and an intravenously-injected drug dose of $3.75\cdot10^{-4}\mbox{ }\mu g/s$. The elementary effects $EE_i$, defined in~(36), 
  calculated for each parameter are displayed in plots (a.$1$) and (b.$1$). The associated measures $\overline{EE}_i^*$, $|\overline{EE}_i^*|$ and $SD_i$, defined in~(37), 
  are shown in plots (a.$1$) and (b.$2$).  Plots (a.$3$) and (b.$3$) present the ratio between $SD$ and $\overline{EE}_i^*$.}
\end{figure}

\begin{figure}[h]
        \centering
        \begin{subfigure}[b]{0.45\textwidth}
            \centering
            \includegraphics[width=\textwidth]{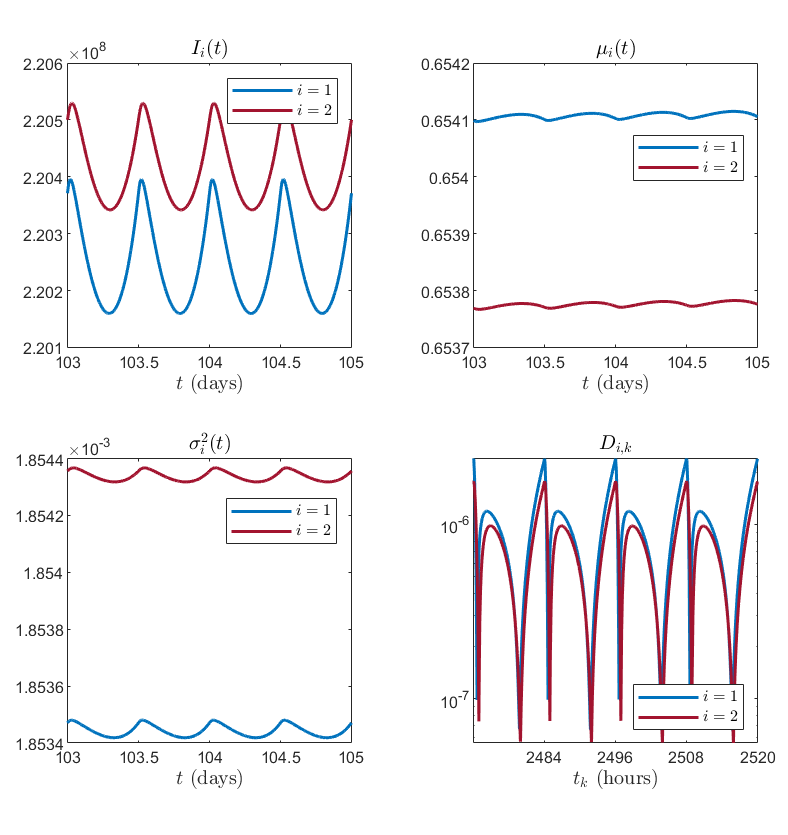}
        \caption{Baseline scenario}
        \label{fig:last2days_baseline}
        \end{subfigure}
          \qquad 
        \quad
        \begin{subfigure}[b]{0.45\textwidth}
            \centering
            \includegraphics[width=\textwidth]{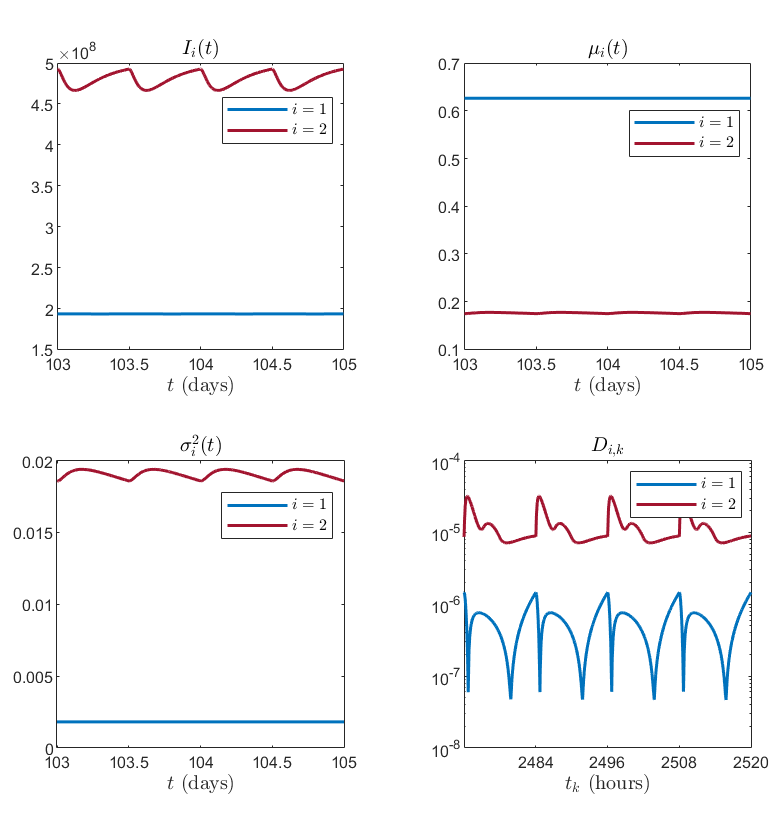} 
        \caption{Non baseline scenario}
        \label{fig:last2days_nonbaseline}
        \end{subfigure}
        \caption{\textbf{Cancer evolutionary dynamics under baseline and non-baseline scenarios: zoom-in analysis of the last two days of the simulations.} Cancer evolutionary dynamics results of the numerical simulations under the baseline (a) and non-baseline (b) scenarios with the drug schedule of 150 mg orally administered twice a day, and time period $t\in[103,105]$ days, \textit{i.e.} the last $26301$ time steps of the simulation results shown in Figure 7 of the Main Manuscript. For each scenario we plot the population size (top left panel), mean phenotypic trait (top right panel), the related variance of the phenotypic distribution (bottom left panel) and the step difference $D_{i,k}$ (bottom right panel), respectively defined in equations (1), (2) and (35). More details on the simulation set-up and numerical methods can be found in Section 4.1.}
        \label{fig:last2days}
\end{figure}

\begin{figure}
    \centering
    \includegraphics[scale=0.4]{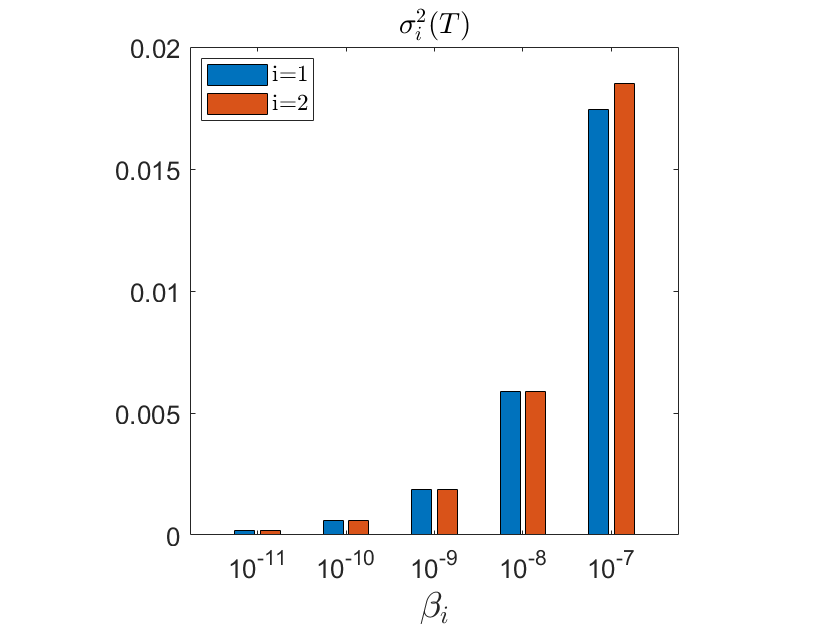}
    \caption{\textbf{Variance dependency on $\beta_i$.} 
Treatment outcomes obtained from numerical simulations of the model under the baseline scenario, for an intravenously-injected drug dose of $2.6915$ $\mu$g/s and a final time of $T=210$ days. The graph displays the variance $\sigma^2_i$, defined in equation~(2), of the phenotypic distribution $n_i$ at steady state, varying the epimutation rate $\beta_i$.  
The steady state time is the first time at which $D_{i,k}<tol$, with $D_{i,k}$ defined in~(35) and a tolerance of  $tol=10^{-6}$. 
 More details on the simulation set-up and numerical methods can be found in Section~4.1.}
    \label{fig:sensib_beta_variance}
\end{figure}